\documentclass[11pt,a4paper]{article}

\pdfoutput=1

\usepackage{jheppub}
\usepackage{multicol,bbm}
\usepackage{bm}
\usepackage{multirow}
\usepackage{xcolor}
\usepackage{mathrsfs}
\usepackage[T1]{fontenc}
\usepackage{lmodern}
\usepackage{tocloft}

\def\bnslash{\bar n\!\!\!\slash}

\newcommand{\nn}{\nonumber} 
\newcommand{\bn}{{\bar n}}

\newcommand{\be}{\begin{equation}}
\newcommand{\ee}{\end{equation}}
\newcommand{\bea}{\begin{eqnarray}}
\newcommand{\eea}{\end{eqnarray}}

\newcommand{\SCETb}{\mbox{${\rm SCET}_{\rm II}$ }}

\newcommand{\vect}[1]{\mathbf{#1}}
\newcommand{\abs}[1]{\left\lvert #1\right\rvert}
\newcommand{\bra}[1]{\left\langle #1\right\rvert}
\newcommand{\ket}[1]{\left\lvert #1\right\rangle}

\newcommand{\Lqcd}{\Lambda_{\text{QCD}}}

\newcommand{\p}{\partial}
\newcommand{\nnlo}{N$^n$LO\ }
\newcommand{\nkll}{N$^k$LL\ }
\newcommand{\nkllprime}{N$^k$LL$'$\ }

\newcommand{\LLF}{LL$_F$\ }

\newcommand{\NLLF}{NLL$_F$\ }

\newcommand{\as}{\alpha_s}
\newcommand{\minus}{\!-\!}
\newcommand{\plus}{\!+\!}
\newcommand{\Gcusp}{\Gamma_{\rm cusp}}
\newcommand{\dOmega}{\partial_{\Omega}}
\newcommand{\dOmegabar}{\partial_{\bar\Omega}}

\newcommand{\cO}{\mathcal{O}}
\newcommand{\cK}{\mathcal{K}}

\newcommand{\cC}{\mathcal{C}}

\newcommand{\cG}{\mathcal{G}}

\newcommand{\cF}{\mathcal{F}}
\newcommand{\cL}{\mathcal{L}}
\newcommand{\cA}{\mathcal{A}}

\newcommand{\cJ}{\mathcal{J}}
\newcommand{\cP}{\mathcal{P}}
\newcommand{\ca}[1]{\mathcal{#1}}

\newcommand{\wt}[1]{\widetilde{#1}}
\newcommand{\wh}[1]{\widehat{#1}}
\newcommand{\ord}[1]{\cO(#1)}

\newcommand{\eq}[1]{Eq.~\eqref{eq:#1}}
\newcommand{\eqs}[2]{Eqs.~\eqref{eq:#1} and \eqref{eq:#2}}
\newcommand{\eqss}[3]{Eqs.~\eqref{eq:#1}, \eqref{eq:#2}, and \eqref{eq:#3}}
\renewcommand{\sec}[1]{Sec.~\ref{sec:#1}}
\newcommand{\ssec}[1]{Sec.~\ref{ssec:#1}}
\newcommand{\appx}[1]{App.~\ref{app:#1}}
\newcommand{\fig}[1]{Fig.~\ref{fig:#1}}
\newcommand{\figs}[2]{Figs.~\ref{fig:#1} and \ref{fig:#2}}
\newcommand{\tab}[1]{Table~\ref{tab:#1}}

\newcommand{\red}[1]{{\color{red}#1}}
\newcommand{\blue}[1]{{\color{blue}#1}}
\newcommand{\green}[1]{{\color{green}#1}}

\newcommand{\purple}[1]{{\color{purple}#1}}

\newcommand{\LP}{\mathscr{L}}
\newcommand{\iLP}{\mathscr{L}^{-1}}
\DeclareMathOperator{\Tr}{Tr}

\title{Comparing and Counting Logs in Direct and Effective Methods of QCD Resummation}

 \author[a,b]{Leandro G. Almeida,}
 \author[c]{Stephen D. Ellis,}
 \author[d]{Christopher Lee,}
 \author[e]{George Sterman,}
 \author[f,g]{Ilmo Sung,}
 \author[h,i]{and Jonathan R. Walsh} 

\affiliation[a]{Laboratoire de Physique Th\'eorique, Universit\'e Paris-Sud 11 and CNRS, 91405 Orsay Cedex, France}
\affiliation[b]{Institut de Biologie de l'\'Ecole Normale Sup\'erieure \,(IBENS), Inserm 1024-CNRS 8197, 46 rue d'Ulm, 75005 Paris, France \footnote{Current Address}}
\affiliation[c]{Department of Physics, University of Washington, Seattle, WA  98195, USA}
\affiliation[d]{Theoretical Division, MS B283, Los Alamos National Laboratory, Los Alamos, NM  87544, USA}
\affiliation[e]{C.N. Yang Institute for Theoretical Physics, Stony Brook University, Stony Brook, NY  11794, USA}
\affiliation[f]{Department of Applied Physics, New York University, Brooklyn, NY 11201, USA}
\affiliation[g]{Queens College, City University of New York, Flushing, NY 11367, USA}
\affiliation[h]{Lawrence Berkeley National Laboratory, University of California, Berkeley, CA  94720, USA}
\affiliation[i]{Berkeley Center for Theoretical Physics, University of California, Berkeley, CA  94720, USA}

\emailAdd{leandro.g.almeida@gmail.com}
\emailAdd{sdellis@uw.edu}
\emailAdd{clee@lanl.gov}
\emailAdd{george.sterman@stonybrook.edu}
\emailAdd{ilmo.sung@nyu.edu}
\emailAdd{jwalsh@lbl.gov}

\abstract{We compare methods to resum logarithms in event shape distributions as they have been used in perturbative QCD directly and in effective field theory. We demonstrate that they are equivalent.  In showing this equivalence, we are able to put standard soft-collinear effective theory (SCET) formulae for cross sections in momentum space into a novel form more directly comparable with standard QCD formulae, and endow the QCD formulae with dependence on separated hard, jet, and soft scales, providing potential ways to improve estimates of theoretical uncertainty. We show how to compute cross sections in momentum space to keep them as accurate as the corresponding expressions in Laplace space.  In particular, we point out that that care is required in truncating differential distributions at \nkll accuracy to ensure they match the accuracy of the corresponding cumulant or Laplace transform. We explain how to avoid such mismatches at \nkll accuracy, and observe why they can also be avoided by working to \nkllprime accuracy.
}

\keywords{Resummation, Effective Field Theory, QCD, Soft Collinear Effective Theory}

\allowdisplaybreaks[1]

\begin{document}

{\flushright LA-UR-13-28363, YITP-SB-14-02\\[-9ex]}
\maketitle

%Main body of the paper

%%%%%%%%%%%%%%%%%%%%%%%%%%%%%%%%%%%%%%%%%%%%%%%%%%%%%%%%%%%%%%%%%%%%%%%%%%%%%%%%%%%%%%%%%%%%%%%%%%%%%%%%%
\section{Introduction}
\label{sec:intro}
%%%%%%%%%%%%%%%%%%%%%%%%%%%%%%%%%%%%%%%%%%%%%%%%%%%%%%%%%%%%%%%%%%%%%%%%%%%%%%%%%%%%%%%%%%%%%%%%%%%%%%%%%

The theory Quantum Chromodynamics (QCD) is remarkably successful in describing the strong interaction. Thanks to the phenomenon of asymptotic freedom \cite{Gross:1973id,Politzer:1973fx}, many inclusive strong interaction cross sections can be predicted to excellent accuracy in perturbation theory, expanding order by order in the strong coupling $\as(\mu)$. 

Cross sections that are not fully inclusive over final-state hadrons, however, require careful treatment in perturbation theory due to the appearance of large logarithms $\ln^k(\mu_1/\mu_2)$ at each order in fixed-order perturbation theory. The logarithms are of ratios of scales $\mu_{1,2}$ that are used to define the exclusivity of the measurement. If the scales are widely separated, the logs are large, and the expansion in $\as$ is poorly behaved. In particular, such logs appear in cross sections that count hadronic jets \cite{Sterman:1977wj}. They may depend on ratios of energy cuts $\Lambda$ on the final state to the total interaction energy $Q$. The logs may also depend on masses of jets that are measured in the final state. Methods to resum these logarithms to all orders in $\as$ may or may not yet exist, depending on the type of measurement.

Event shapes provide a large class of observables that can be used to measure the jet-like structure of the final state in collisions producing hadrons \cite{Dasgupta:2003iq}.    We consider those event shapes that are ``global'' observables, which sum over all the final-state hadrons with a single weight measure $e$ that introduces sensitivity of the cross section to a single collinear scale and a single soft scale. Non-global observables which induce sensitivity to multiple soft scales will not be considered here \cite{Dasgupta:2001sh,Hornig:2011iu}. For simplicity we consider event shapes in leptonic collisions $e^+ e^-\to X$ producing hadrons $X$. Our observations and conclusions  apply more generally to the resummation of logs in other types of collisions and measurements as well.  A very familiar event shape is the thrust $\tau = 1-T$ \cite{Farhi:1977sg}, defined here by
\be
\label{eq:thrust}
\tau \equiv 1-T = 1 - \frac{1}{Q}\max_{\vect{\hat t}}\sum_{i\in X} \abs{\vect{\hat t}\cdot \vect{p}_i}\,,
\ee
where $Q$ is the center-of-mass collision energy, $X$ the final hadronic state, $\vect{p}_i$ the 3-momentum of particle $i$, and $\vect{\hat t}$ the \emph{thrust axis}, defined as the axis that maximizes the sum over $i$ (and thus minimizes $\tau$) in \eq{thrust}. The thrust is one of a continuous set of event shapes that can be written in a more generic form,
\be
\label{eq:eventshape}
e = \frac{1}{Q}\sum_{i\in X}\abs{\vect{p}_T^i} f_e(\eta_i)\,,
\ee 
where the transverse momentum $\vect{p}_T$ and rapidity $\eta$ of each particle are measured with respect to the thrust axis $\vect{\hat t}$. (Equation~(\ref{eq:eventshape}) applies to massless hadrons $i$ in the final state. For the generalization to massive hadrons, see \cite{Salam:2001bd,Mateu:2012nk}.)  The function $f_e$ is taken to be continuous and to fall off sufficiently fast at large $\eta_i$ to preserve infrared and collinear safety \cite{Sterman:1978bj,Sterman:1981jc}. In the main discussion in this paper, we will consider the class known as \emph{angularities} $e = \tau_a$ \cite{Berger:2003iw,Berger:2003pk,Berger:2004xf}, defined by $f_a(\eta) = e^{-\abs{\eta}(1-a)}$, which are infrared safe for real numbers $a<2$, although our discussion of factorization and resummation below will be valid for $a<1$. 

Although our discussion will take place mainly within the context of angularity event shapes in $e^+e^-$ collisions, our conclusions will not limited to this application. Rather they provide a concrete arena in which to draw more general lessons about different techniques to resum logs in QCD perturbation theory and the relation between them.

Measurement of an event shape such as thrust induces sensitivity of the cross section on several different energy scales. Small thrust $\tau$ means that the final state is nearly two-jet-like. The cross section then depends on the hard collision energy $Q$, the typical jet mass, $Q\sqrt{\tau}$, and the energy of soft radiation from the jets, $Q\tau$. Ratios of these scales appear as logs of $\tau$ in the perturbative expansion of the thrust distribution. A method to resum two-jet event shapes was described in \cite{Catani:1992ua} (CTTW) based on the soft and collinear singularities of QCD and approximations to phase space in these regions. A unified derivation of the resummation of logarithms from the factorization properties of QCD cross sections was presented in Ref.~\cite{Contopanagos:1996nh}, which are also naturally derivable using the methods of effective field theory. We summarize these methods in parallel discussions below in the context of ``direct'' QCD (dQCD) and soft collinear effective theory (SCET) \cite{Bauer:2000ew,Bauer:2000yr,Bauer:2001ct,Bauer:2001yt,Bauer:2002nz} techniques to resum logs, applied to jet cross sections \cite{Bauer:2002ie,Bauer:2003di} and event shape distributions \cite{Bauer:2008dt}. 

Working directly in full QCD or going through the methods of EFT should theoretically be equivalent. The EFT, after all, is a systematic approximation to and method for computing in the full theory \cite{Manohar:1996cq}.  It is a primary goal of this paper to show that standard formulae derived in the two formalisms are, indeed, equivalent. However, some results for resummed event shapes published in the respective literature sometimes do not seem obviously equivalent and sometimes yield apparently different numerical results. 

We will illustrate one such previous discrepancy and its resolution by focusing on a comparison between angularity distributions reported in \cite{Berger:2003iw,Berger:2003pk} using full QCD and \cite{Hornig:2009kv,Hornig:2009vb} using SCET. (QCD NLL results for $\tau_a$ and many other event shapes are also given in \cite{Banfi:2004yd}.) In \cite{Hornig:2009vb} a comparison of the two sets of results at next-to-leading logarithmic (NLL) accuracy was performed, and significant numerical differences were found. In this paper, we resolve the source of this discrepancy. The resolution will involve a careful consideration of the meaning of \nkll accuracy for various ways to write the cross section, e.g. as a differential distribution, the cumulative distribution (cumulant) or in Laplace or position space. We find that the results in \cite{Berger:2003iw,Berger:2003pk} and  \cite{Hornig:2009vb} were not evaluated according to a consistent definition of NLL accuracy. Once we implement such a consistent scheme, we find that results in the two formalisms yield numerically equivalent results.

In \sec{orders} we will review a standard definition of \nkll accuracy, namely, which terms in the exponent of the Laplace transform $\wt\sigma(\nu_a)$ of the event shape distribution $d\sigma/d\tau_a$ are accurately predicted. We will compare it to the definition given by CTTW in terms of the cumulative distribution (or ``cumulant'' or ``radiator'')  $R(\tau_a)$. We will observe that the two descriptions of \nkll are not precisely equivalent. This sets the stage for a more careful prescription for computing to \nkll accuracy in the subsequent sections.

In \sec{compare} we consider factorized and resummed event shape distributions predicted in \cite{Bauer:2008dt,Hornig:2009vb} using the formalism of SCET and compare to a form based on the predictions of \cite{Berger:2003iw,Berger:2003pk} derived directly in full QCD. These distributions look similar but not identical at first, but we will put the two into forms that are precisely and transparently equivalent. This comparison serves not only a pedagogical purpose, but improves both versions in real ways. The SCET version is put in a form in which more logarithmic terms are explicitly exponentiated and thus sums a larger number of terms to all orders in $\as$ at a given level of logarithmic accuracy.  In the QCD version, we generalize the jet and soft scales from the fixed values they usually take to generic, variable forms, whose variation can provide an accurate estimation of theoretical uncertainty at finite \nkll accuracy. This restoration of variable scales to the traditional QCD form is a direct byproduct of our proof of its equivalence with the SCET form. This proof in \ssec{resumequiv} is the technical heart of the paper.

\sec{schemes} contains a detailed pedagogical review of the orders to which one must compute various ingredients of the factorization theorem (hard, jet, soft functions and their anomalous dimensions) to achieve \nkll accuracy in event shape cross sections. The counting scheme for the Laplace transform $\wt\sigma(\nu)$ is standard and straightforward. If we base the definition of \nkll accuracy on this object, however, then we argue that in order to evaluate cross section in momentum ($\tau_a$) space to the equivalent accuracy, some additional care is warranted in evaluating typical formulae for the cumulant $R(\tau_a)$, and even greater care in evaluating the differential distribution $\sigma(\tau_a)$. Na\"{i}vely applying the same procedures that one uses to truncate ingredients of the resummed Laplace transform at a given order of accuracy directly to some existing forms for the resummed momentum-space distributions can cause the latter to predict fewer terms at \nkll accuracy than one might expect them to. We will give clear prescriptions on how to evaluate the cross sections in momentum space to the equivalent accuracy as the Laplace transforms. Namely, it is preferable to evaluate some quantities in $R(\tau_a)$ to what is known as \nkllprime accuracy (keeping fixed-order jet and soft functions to one higher power in $\as$ than at \nkll accuracy) to maintain an equivalent level of accuracy with $\wt\sigma(\nu)$, and it is \emph{essential} to evaluate some terms in $\sigma(\tau_a)$ to \nkllprime accuracy% (and some even higher)
, so that an equivalent level of accuracy between $\sigma(\tau_a)$ and the result of taking the derivative of $R(\tau_a)$ is maintained. We will provide a new formula for the resummed differential distribution $\sigma(\tau_a)$ defined in terms of the derivative of the cumulant and to which simple truncation rules can be applied.

\sec{summary} contains in summary form the results of the study in \sec{schemes}, and the reader who wishes to skip to the final formulae and prescriptions for evaluating $\wt\sigma(\nu)$, $R(\tau_a)$ and $\sigma(\tau_a)$ to \nkll or \nkllprime accuracy without working through the details of their development in \sec{schemes} should turn directly to \sec{summary}.

In \sec{numerical} we will provide some numerical comparisons of formulae in previous literature and in the current paper to illustrate the effect of consistent counting of logarithms and the implementation of the improved procedures we advocate.

Much of the information in this paper is a review of known, existing procedures, and even some of the comparisons we perform have been presented in different contexts in the past (e.g. \cite{Manohar:2003vb,Becher:2006mr,Becher:2006nr,Bonvini:2012az,Bonvini:2013td}, and more recently \cite{Sterman:2013nya}). For these parts, we hope the reader finds pedagogical value in the unified, coherent explanations we attempt to provide here. However, we believe the new observations we make in comparing SCET and QCD event shape resummation add new practical value as well, increasing the number of resummed terms in one case and scale variability/uncertainty estimation in the other. Moreover, the observations we make about how to calculate the momentum-space cumulant to keep it as accurate as the resummed Laplace transform and, more importantly, how to calculate the differential distribution to an accuracy equivalent to that of the cumulant, have not to our knowledge been made clearly and explicitly before. It is our hope that future studies in this direction may avoid the counting complications we identify, so that with knowledge of certain ingredients to a given order in $\as$, they may achieve the greatest accuracy and most realistic uncertainty estimates that are possible. We also hope that the issues resolved here will help ensure that the most precise QCD predictions will be available at the LHC and future facilities.

%%%%%%%%%%%%%%%%%%%%%%%%%%%%%%%%%%%%%%%%%%%%%%%%%%%%%%%%%%%%%%%%%%%%%%%%%%%%%%%%%%%%%%%%%%%%%%%%%%%%%%%%%
\section{Orders of Resummed Perturbative Accuracy}
\label{sec:orders}
%%%%%%%%%%%%%%%%%%%%%%%%%%%%%%%%%%%%%%%%%%%%%%%%%%%%%%%%%%%%%%%%%%%%%%%%%%%%%%%%%%%%%%%%%%%%%%%%%%%%%%%%%

\subsection{Definition of \nkll Accuracy: Counting in the Laplace Exponent}

In this paper we will review and compare methods to resum large logarithms in event shape distributions, expressed in three ways: as a differential cross section, as its Laplace transform, and as a cumulant cross section (also called the \emph{radiator}, e.g. in \cite{Berger:2003pk}):
\begin{align}
\label{eq:sigmadef}
\sigma(\tau) &\equiv \frac{1}{\sigma_0}\frac{d\sigma}{d\tau} \,, \\
\label{eq:Laplacedef}
\wt\sigma(\nu) &\equiv \int_0^\infty d\tau \, e^{-\nu\tau} \sigma(\tau) \\
\label{eq:cumulantdef}
R(\tau) &\equiv \int_0^\tau d\tau' \sigma(\tau')\,,
\end{align}
where $\sigma_0$ is the Born cross section.
Each of the three ways of writing the cross section in \eqss{sigmadef}{Laplacedef}{cumulantdef} exhibits logarithms that become large in the two-jet endpoint region $\tau\ll 1$. The differential distribution can be computed as a perturbative expansion in the strong coupling $\as$, for sufficiently large $\tau$. For very small $\tau\lesssim\Lqcd/Q$, the distribution must be convolved with a nonperturbative shape function (e.g. \cite{Korchemsky:1999kt,Korchemsky:2000kp,Berger:2003pk,Hoang:2007vb,Hornig:2009vb}) whose effects dominate in this region, but which we do not consider in this paper. The distribution predicted by perturbation theory takes the form,
\be
\label{eq:distributionexpansion}
\begin{split}
\sigma(\tau) = \delta(\tau) +  \frac{\as}{4\pi}\Bigl[ a_{12} \cL_1 &+ a_{11} \cL_0 + a_{10}\delta(\tau)\Bigr] \\
 + \Bigl(\frac{\as}{4\pi}\Bigr)^{\!2} \Bigl[ a_{24} \cL_3 &+ a_{23} \cL_2 + a_{22} \cL_1+ a_{21}\cL_0 + a_{20}\delta(\tau) \Bigr] \\
 + \Bigl(\frac{\as}{4\pi}\Bigr)^{\!3} \Bigl[ a_{36} \cL_5 &+ a_{35} \cL_4 + a_{34}\cL_3 + a_{33} \cL_2 + a_{32} \cL_1 + a_{31} \cL_0 + a_{30}\delta(\tau)\Bigr]  \\
  +\quad\quad\,\,\,    \cdots  \qquad  &+  d[\as](\tau)\,,
\end{split}
\ee
where $\cL_n\equiv\cL_n(\tau) = [\theta(\tau)\ln^n\tau/\tau]_+$ is a plus distribution defined in \appx{plus}, and where $d$ is an integrable function of $\tau$, no more singular than $\ln(1/\tau)$. Meanwhile the Laplace transform and the cumulant both take the form
\be
\label{eq:Laplaceexpansion}
\begin{split}
\{\tilde \sigma(\nu),R(\tau)\}  = 1+ \frac{\as}{4\pi}\Bigl[ b_{12} L^2 &+ b_{11} L + b_{10} \Bigr] \\
 + \Bigl(\frac{\as}{4\pi}\Bigr)^{\!2} \Bigl[ b_{24} L^4 &+ b_{23} L^3 + b_{22} L^2 + b_{21} L + b_{20} \Bigr] \\
 + \Bigl(\frac{\as}{4\pi}\Bigr)^{\!3} \Bigl[ b_{36} L^6 &+ b_{35} L^5 + b_{34} L^4 + b_{33} L^3 + b_{32} L^2 + b_{31} L + b_{30} \Bigr] \\
 & + \;\;  \cdots \;\;  + D(\as)\,,
\end{split}
\ee
where for the Laplace transform $L = \ln\nu$ while for the cumulant $L = \ln (1/\tau)$. The remainder function $D$ is either the Laplace transform or the integral of the integrable function $d(\tau)$, with the property $D\!\to\!0$ as $\tau\!\to\!0$ or $\nu\!\to\! \infty$.  Of course, the coefficients $b_{ij}$ differ between the Laplace transform and the cumulant. 
\begin{table}[t]
\begin{center}
$
\begin{array}{ | c | c | c |}
\hline
\text{accuracy} & c_{nk} & C_n \\  \hline
\text{LL} & k=n+1 & n=0 \\ \hline
\text{NLL} & k\geq n & n=0 \\ \hline
\text{NNLL} & k\geq n-1 & n\leq 1 \\ \hline
\text{N$^3$LL} & k\geq n-2 & n\leq 2 \\ \hline 
\end{array}
\qquad
\begin{array}{ | c | c | c |}
\hline
\text{accuracy} & c_{nk} & C_n \\  \hline
\text{LL} & k=n+1 & n=0 \\ \hline
\text{NLL}' & k\geq n & n=1 \\ \hline
\text{NNLL}' & k\geq n-1 & n\leq 2 \\ \hline
\text{N$^3$LL}' & k\geq n-2 & n\leq 3 \\ \hline 
\end{array}
$
\end{center}
\vspace{-1em}
{\caption{Conventions for order counting in $\wt\sigma(\nu)$. At \nkll accuracy, all $c_{nk}$ for all $n$ and $m\geq n+k-1$ are included, as are all $C_n$ for $n\leq k-1$. In the primed \nkllprime counting, the fixed-order coefficients $C_n$ are computed to one higher order in $\as$.}
\label{tab:Laplace} }
\end{table}

Fixed-order perturbation theory calculates the cross sections in \eqs{distributionexpansion}{Laplaceexpansion} row-by-row, order-by-order in $\as$, which in \eqs{distributionexpansion}{Laplaceexpansion} means $\as(Q)$, where $Q$ is the CM $e^+e^-$ collision energy. At this high scale, $\as\ll 1$. However when the logs $\cL_n$ or $L^n$ become large, this expansion breaks down. Instead, the expansions should be reorganized so that sets of large logarithms to all orders in $\as$ are summed one set at a time. One might be tempted to sum \eqs{distributionexpansion}{Laplaceexpansion} column-by-column, capturing the largest logs first, then the next-to-largest, etc. However, it turns out that this is not the most systematic way to reorganize the summation. Instead, systematic methods of resummation prefer to sum the \emph{logarithm} of $\wt\sigma(\nu)$, which exponentiates in a simple fashion:
\be
\label{eq:Laplaceexponent}
\begin{split}
\tilde \sigma(\nu)  = C(\as)\exp\biggl\{\frac{\as}{4\pi}\Bigl[ c_{12} L^2 &+ c_{11} L \Bigr] \\
 + \Bigl(\frac{\as}{4\pi}\Bigr)^2 \Bigl[ c_{23} L^3 &+ c_{22} L^2 + c_{21} L \Bigr] \\
 + \Bigl(\frac{\as}{4\pi}\Bigr)^3 \Bigl[ c_{34} L^4 &+ c_{33} L^3 + c_{32} L^2 + c_{31} L \Bigr] + \cdots\biggr\} + D(\as)\,, \\
 \small \text{LL} \ &+   \small\text{NLL} \ \ + \text{NNLL} + \text{N$^3$LL} + \cdots 
\end{split}
\ee
where $C$ has an expansion in $\as$ independent of the logarithms,
\be
\label{eq:Cexpansion}
C(\as) = 1 + \frac{\as}{4\pi} C_1 + \Bigl(\frac{\as}{4\pi}\Bigr)^2 C_2+\cdots\,.
\ee
In the expansion in \eq{Laplaceexponent}, the highest power of logs in the exponent at each order $\as^n$ is $L^{n+1}$. It turns out that $\wt\sigma(\nu)$ exponentiates more simply than $\sigma(\tau)$ or $R(\tau)$, and so we use $\wt\sigma(\nu)$ written as in \eq{Laplaceexponent} as the benchmark for defining orders of logarithmic accuracy. Systematic methods of resummation give a well-defined, simple definition of this exponent in terms of a limited number of functions in QCD/SCET: 
\be
\label{eq:Laplaceexpsimple}
\wt\sigma(\nu) = C(\as) e^{\bar E(\ln\nu)} + \wt D(\as,\nu)\,,
\ee
where $\wt D$ is the non-singular part of $\wt\sigma$ in Laplace space.
We will give a precise definition of $\bar E$ in terms of anomalous dimensions and other functions in \sec{compare}. 
We use \eq{Laplaceexponent} to define the order of accuracy to which we resum logarithms in the cross section: the first column contains all the terms at \emph{leading-log} (LL) accuracy, the second column next-to-leading log (NLL), the third next-to-next-to-leading log (NNLL), etc. This counting is natural in a regime where the logs are large enough so that $L\sim 1/\as$. Then LL sums all terms of order $\as^{-1}$, NLL all terms of order 1, NNLL all terms of order $\as$, etc. In this counting, the fixed-order terms in $C$ are counted as $C_1$ being NNLL (i.e. it is explicitly order $\as$), $C_2$ being N$^3$LL, etc. Since the terms in the prefactor $C$ in \eq{Laplaceexpsimple} do organize themselves differently than the exponent $\bar E$, though, one can adopt an alternative convention for them. Often, the fixed-order coefficients are included to one higher order in $\as$, yielding the so-called ``primed'' counting \cite{Abbate:2010xh}, summarized in \tab{Laplace}, so $C_1$ would be NLL$'$, $C_2$ NNLL$'$, etc. This counting is more useful when one wishes to compute the transition region between small and large $\tau$ to N$^k$LO accuracy where the fixed-order terms in the non-singular $D$ are the same size as the logs. In addition, we will find below that in a number of ways the primed counting is a more consistent scheme for counting the accuracy of the resummed logs themselves, at least when computing $R(\tau)$ or $\sigma(\tau)$ in momentum space. When the non-singular terms $D$ are calculated in an ordinary fixed-order expansion to \nnlo accuracy, etc. we speak of the cross section resummed to \nkll\!\!$(')$ accuracy and matched onto fixed-order at \nnlo\!, or, for short, \nkll\!\!$(')$+\nnlo\!\!.

We will define \nkll accuracy for the cumulant $R(\tau)$ or the distribution $\sigma(\tau)$ similarly to $\wt\sigma(\nu)$ in \eq{Laplaceexponent}. However, because they do not exponentiate as simply as \eq{Laplaceexpsimple}, the prescriptions for calculating $R(\tau)$ or $\sigma(\tau)$ to \nkll (or \nkllprime\!\!) accuracy are a bit more involved. For example, the most compact form for systematic resummation of $R(\tau)$ that we will find below is:
\be
\label{eq:Rexpsimple}
R(\tau) = C(\as) \exp\biggl[\bar E(\ln1/\tau) + \sum_{n=2}^\infty \frac{1}{n!} \bar E^{(n)}\partial_{\bar E'}^{n}\biggr] \frac{1}{\Gamma(1-\bar E')} + D(\as,\tau)\,,
\ee
where $D$ is the non-singular part of $R$ in $\tau$-space.
The exponents $\bar E$ in \eqs{Laplaceexpsimple}{Rexpsimple} take the same form, while the terms generated by the gamma function and its derivatives with respect to $\bar E'$ [which itself is the derivative $d\bar E/d(\ln1/\tau)$] compensate for the additional terms generated due to the Laplace transform of $\ln^n(1/\tau)$ being not exactly equal to $\ln^n(\nu e^{\gamma_E})$ (lower order logs are also generated in the transform, as reviewed in \appx{Laplace}). Thus, to make \eq{Rexpsimple} reproduce the entire \nkll accurate $\wt\sigma(\nu)$ in \eq{Laplaceexpsimple} \emph{exactly} requires evaluating the differential operators in \eq{Rexpsimple} acting on the gamma function to infinitely high order, for which a closed-form algebraic expression cannot be obtained.  However, we will find that keeping the differential operators to sufficiently high finite order, all the terms in $\wt\sigma(\nu)$ that should be correct at \nkll accuracy are indeed reproduced by Laplace transforming $R(\tau)$ in \eq{Rexpsimple}. We will describe the appropriate procedure and counting in \ssec{Laplacecomparison}. Additional similar considerations apply to $\sigma(\tau)$, which we will explain in \ssec{distributionaccuracy}.

\subsection{Original CTTW Convention: Counting in the Cumulant}

Resummation of large logarithms in event shape distributions was described by CTTW \cite{Catani:1992ua} in terms of the cumulant cross section (also called the ``radiator'') $R(\tau)$, which can be organized in the form
\begin{equation}
\label{eq:CTTWradiator}
R(\tau) = C(\alpha_s)\Sigma(\tau,\alpha_s) + D(\tau,\alpha_s)\,,
\end{equation}
where 
\begin{subequations}
\label{eq:CTTWdefs}
\begin{align}
\label{eq:CTTWdefC}
C(\alpha_s) &= 1+ \sum_{n=1}^\infty \left(\frac{\as}{2\pi}\right)^n C_n \,, \\
\label{eq:CTTWdefSigma}
\begin{split}
\ln\Sigma(\tau,\as) &= \sum_{n=1}^\infty \sum_{m=1}^{n+1} \left(\frac{\as}{2\pi}\right)^n G_{nm}  \ln^m\frac{1}{\tau}  \\
&= \quad \Bigl( \frac{\as}{2\pi} \Bigr) \; \biggl( G_{12} \ln^2 \frac{1}{\tau} + G_{11} \ln\frac{1}{\tau}  \biggr) \\
&\quad+ \Bigl(\frac{\as}{2\pi}\Bigr)^2 \biggl( G_{23} \ln^3 \frac{1}{\tau} + G_{22} \ln^2\frac{1}{\tau} + G_{21} \ln\frac{1}{\tau} \biggr) \\
&\quad+ \Bigl(\frac{\as}{2\pi}\Bigr)^3 \biggl( G_{34} \ln^4 \frac{1}{\tau} + G_{33} \ln^3\frac{1}{\tau} + G_{32} \ln^2\frac{1}{\tau} +G_{31} \ln\frac{1}{\tau} \biggr) + \cdots \,,
\end{split}
\end{align}
\end{subequations}
and $D(\tau,\as)$ is a remainder function that vanishes as $\tau\to 0$.  In \eqs{CTTWradiator}{CTTWdefs}, $\as$ is again evaluated at the scale $Q$.  
Expanding $R(\tau)$ in powers of $\as$ explicitly,
\begin{equation}
\label{eq:radiatorFO}
R(\tau) = \sum_{n=0}^\infty \sum_{m=0}^{2n} R_{nm}\left(\frac{\as}{2\pi}\right)^n \ln^m\frac{1}{\tau} + D(\tau, \as) \,.
\end{equation}
In \tab{CTTWFO} we give the coefficients $R_{nm}$ in terms of the coefficients $G_{nm}$ in the exponent of $\Sigma$ and coefficients $C_n$ in the multiplicative prefactor up to $\ord{\as^3}$.
\begin{table}
\begin{center}
\scriptsize
$
\begin{array}{|c|c|c|c|}
\hline
R_{nm} & n=1 & n=2 & n=3  \\ \hline
m=2n & R_{12} = G_{12} & R_{24}= \frac{1}{2}G_{12}^2 &  R_{36} = \frac{1}{6}G_{12}^3 \\ \hline
m=2n-1 & R_{11} = G_{11} & R_{23} = G_{23} + G_{12}G_{11} & R_{35} = G_{23}G_{12} + \frac{1}{2}G_{12}^2 G_{11} \\ \hline  
m={2n-2} & R_{10} = C_1 &  R_{22} = G_{22} + \frac{1}{2}G_{11}^2  + C_1 G_{12} & R_{34} = G_{34} + G_{23}G_{11} + G_{22}G_{12} + \frac{1}{2}G_{12}G_{11}^2 + \frac{1}{2} C_{1}G_{12}^2 \\ \hline
m=2n-3 & \text{---} & R_{21} = G_{21} + C_1 G_{11} & R_{33} = G_{33} + G_{22}G_{11} + G_{21}G_{12} + \frac{1}{6}G_{11}^3 + C_1(G_{23}+G_{12}G_{11}) \\ \hline
m = 2n-4 & \text{---} & R_{20} = C_2 & R_{32} = G_{32} + G_{21}G_{11} + C_2G_{12} + C_1(G_{22}+\frac{1}{2}G_{11}^2)  \\ \hline
m = 2n-5 & \text{---} & \text{---} & R_{31} = G_{31} + C_2 G_{11} + C_1 G_{21} \\ \hline
m = 2n-6 & \text{---} & \text{---} & R_{30} = C_3  \\ \hline
\end{array}
$
\end{center}
\vspace{-1em}
\caption{Coefficients in the fixed-order expansion of radiator $R(\tau)$. $R_{nm}$ is the coefficient of $(\as/2\pi)^n \ln^{m}(1/\tau)$ in the expansion of $R(\tau)$, given here up to $\cO(\as^3)$ in terms of the coefficients $C_n$ and $G_{nm}$ in the exponentiated form of the radiator \eq{CTTWradiator}.}
\label{tab:CTTWFO}
\end{table}

Comparing \eqs{CTTWradiator}{CTTWdefs} to \eq{Rexpsimple}, we note that the CTTW exponent $\ln\Sigma$ is not simply equal to $\bar E$, but also contains terms generated by the gamma function $\Gamma(1-\bar E')$ and the derivatives in \eq{Rexpsimple} acting on it. Although \eq{Rexpsimple} is thus in a slightly more involved mathematical form, in practice it is actually simpler in the sense that \eq{Rexpsimple} can be computed from the single object $\bar E$ and its derivatives and shows how the CTTW coefficients $G_{nm}$ would be computed systematically to arbitrarily high accuracy. Because \eq{Rexpsimple} is not simply an exponential of $\bar E$ as in \eq{Laplaceexpsimple}, we must take care in defining what \nkll accuracy means in evaluating each part of \eq{Rexpsimple} in a manner consistent with the prescription for \eq{Laplaceexpsimple}. This will be the subject of \ssec{Laplacecomparison}. 

In the presentation of CTTW, \nkll accuracy describes the number of terms in the exponent $\ln\Sigma$ of $R$ in \eq{CTTWdefs} that are known. 
We have instead defined \nkll accuracy by the number of terms known in the exponent of the Laplace transform $\wt\sigma(\nu)$ in \eq{Laplaceexponent}. In principle, knowing one set of terms in either of \eqs{Laplaceexponent}{CTTWdefSigma} to \nkll accuracy would allow obtaining the other one to the same accuracy by (inverse) Laplace transformation. However, unlike for $R(\tau)$, it is possible to give a closed algebraic form \eq{Laplaceexpsimple} for $\wt\sigma(\nu)$ that predicts all the \nkll terms to arbitrarily high order in $\as$ in \eq{Laplaceexponent}. In contrast, \eq{Rexpsimple} for $R(\tau)$ does so in principle, but not in a closed algebraic form---the exponential of derivatives $\partial_{\bar E'}^n$ cannot actually be evaluated to infinitely high order in practice. So, one has to decide at what finite order in $\as$ to truncate the exponential differential operator to finite order at \nkll accuracy. Because of this additional ambiguity or complexity in defining \nkll accuracy in terms of $R(\tau)$ (similarly $\sigma(\tau)$) in momentum space, we define \nkll accuracy instead more simply by the number of terms accurately predicted in the exponent of $\wt\sigma(\nu)$. We will explain in \ssec{Laplacecomparison} below how to truncate \eq{Rexpsimple} for $R(\tau)$ at \nkll accuracy properly so that it does in fact match the accuracy of $\wt\sigma(\nu)$.

For the distribution $\sigma(\tau)$, \nkll accuracy will be similarly defined, namely, the Laplace transform of the \nkll $\sigma(\tau)$ must match the accuracy of the \nkll $\tilde\sigma(\nu)$. This can be accomplished by simply differentiating $R(\tau)$ computed properly to \nkll accuracy or by appropriate truncation of a formula derived for $\sigma(\tau)$ directly.  In \ssec{countingdistribution}, we will  show that care must be exercised in making such truncations of some standard formulae for the resummed distribution. Na\"{i}ve truncation according to standard rules for \nkll accuracy can make those formulae for $\sigma(\tau)$ yield less accurate results than directly differentiating the \nkll cumulant. This problem does not exist at \nkllprime accuracy.

To close this section, let us comment on counting logs in the fixed-order expansion \eq{radiatorFO} of the cumulant $R(\tau)$. Note that knowing all the coefficients $G_{nk}$ and $C_n$ at a given order  in \nkll accuracy also captures all the terms in the $(k\plus 1)$-th row of the fixed-order expansion of $R(\tau)$ shown in \tab{CTTWFO}. In the past, this is what was sometimes meant by \nkll  when applied directly to the expansion of $R(\tau)$ in powers of $\as$ (this is what is called ``\nkll$_{\!\!\!\text{F}}$'' accuracy in \cite{Chiu:2007dg}). That is, \LLF accuracy means the first row of \tab{CTTWFO}, \NLLF accuracy means the second row, etc. However, \nkll accuracy in the exponent of $R(\tau)$ resums many more terms than this. The number of rows in \tab{CTTWFO} do not stand in one-to-one correspondence with the orders of accuracy \nkll in \tab{Laplace}  \cite{Chiu:2007dg}. For resummed calculations,  nowadays \nkll accuracy invariably refers to the number of terms resummed in the exponent $\ln\Sigma$ (and prefactor $C$) of the radiator $R$ (or, rather, in the exponent $\bar E$ of the Laplace transform $\wt\sigma(\nu)$). This counting is also much more natural and consistent with systematic calculation using the renormalization group methods that are reviewed in \sec{compare}.

%%%%%%%%%%%%%%%%%%%%%%%%%%%%%%%%%%%%%%%%%%%%%%%%%%%%%%%%%%%%%%%%%%%%%%%%%%%%%%%%%%%%%%%%%%%%%%%%%%%%%%%%%
\section{Comparison of Direct QCD and SCET resummation}
\label{sec:compare}
%%%%%%%%%%%%%%%%%%%%%%%%%%%%%%%%%%%%%%%%%%%%%%%%%%%%%%%%%%%%%%%%%%%%%%%%%%%%%%%%%%%%%%%%%%%%%%%%%%%%%%%%%

In this section we review techniques for achieving the resummation of logarithms in \eq{Laplaceexponent} to all orders in $\as$. On the one hand, none of the techniques reviewed here are new. On the other hand, by gathering them in one place, we will notice some new connections that help generalize or simplify existing results into more illuminating, useful forms, and even improve the accuracy of some resummation formulae. We hope the reader finds new pedagogical and practical value in the combined results we collect and review here.

A variety of techniques have been employed to perform resummation for processes with large logarithms, both within the context of full QCD and also using the tools of effective field theory. For jets, the EFT is SCET.  For threshold resummation, these methods have been compared in detail from the perspective of both fields \cite{Becher:2006mr,Becher:2006nr,Bonvini:2012az,Bonvini:2013td,Sterman:2013nya}).  However, for event shapes, such a detailed comparison has not been performed.  These techniques are applicable to more complex jet observables \cite{Ellis:2009wj,Ellis:2010rw,Stewart:2010tn,Thaler:2010tr,Bauer:2011uc}, and will find use in studies of jet substructure \cite{Abdesselam:2010pt,Altheimer:2012mn,Altheimer:2013yza}.

In the title of this section we refer to ``direct QCD'' (dQCD) by which we mean the method of resummation derived directly from the properties of scattering amplitudes in perturbative QCD. This is in contrast to SCET methods derived using the techniques of effective field theory. The two methods differ in details but both describe in a controlled series of approximations the same full theory of QCD.   The EFT can be viewed as a systematic method to organize the approximations made in order to factorize cross sections in full QCD rather than being a different theory. The tools of EFT then make systematic the methods of resumming logs via RG evolution. Whether working directly in the full theory or going through the machinery of EFT, the two methods lead to the same resummation results because both arrive at similar factorization of hard, collinear and soft scales for the cross section. In SCET the collinear (jet) and soft functions in the factorization theorem are matrix elements of operators built out of effective theory fields. In direct QCD they rely on perturbative techniques to separate out the subleading contributions that violate the factorization, but can also be expressed as matrix elements of operators similar to those in SCET (see, \emph{e.g.}, \cite{Berger:2003iw,Lee:2006nr}). We use the adjective ``direct'' simply to distinguish whether one uses explicitly the construct of an EFT or not to resum logs. We will, however, often drop the adjective and simply refer to the direct method as ``QCD'' below.

We begin this section with a brief overview of  factorization in both QCD and SCET, and provide separate short reviews of common resummation techniques in the two formalisms.  We then show some rather compact and remarkable ways to rewrite the all-orders resummation formalism that underscore the connection between QCD and SCET.  We show that the two methods are equivalent at all orders, and furthermore that, if the resummation is truncated at a given order, the two methods still give the same result.  Additionally, we use the connection between the two resummation formalisms to show how non-canonical scale choices can be made in the QCD resummation. This connection also leads us to a form for event shape distributions resummed in SCET that resums more terms than most standard formulae used in the literature to date.

While there is a wide variety of event shapes one can consider (e.g. thrust \cite{Farhi:1977sg}, jet mass \cite{Clavelli:1979md,Chandramohan:1980ry,Clavelli:1981yh}, broadening \cite{Catani:1992jc}, $C$-parameter \cite{Ellis:1980wv}, $N$-jettiness \cite{Stewart:2010tn}, etc.), a much larger class of event shapes that also encompasses some of these are the \emph{angularities}
\cite{Berger:2003iw,Berger:2003pk,Berger:2004xf}, defined for events in $e^+ e^-\to \text{hadrons}$  as
\be
\tau_a = \frac{1}{Q} \sum_{i\in X} E_i \sin^a \theta_i (1-\cos\theta_i)^{1-a} \overset{m_i = 0} {\longrightarrow} \frac{1}{Q}\sum_{i\in X} \bigl\lvert \vect{p}_i^\perp \bigr\rvert e^{-\abs{\eta_i}(1-a)}\,,
\ee
where $Q$ is the center-of-mass collision energy, $X$ the hadronic final state, $E_i$ the energy of particle $i$, and where the angle $\theta_i$, transverse momentum $\vect{p}_i^\perp$ and pseudorapidity $\eta_i$ are measured with respect to the thrust axis $\vect{\hat t}_X$ of the final state $X$. The second expression for $\tau_a$ holds for massless hadrons; for generalization to nonzero mass, see \cite{Mateu:2012nk}. The angularities interpolate between thrust at $a=0$ and broadening at $a=1$, although $a$ is allowed to vary between $-\infty<a<2$. The factorization theorem and resummation we discuss below hold for $a<1$. For the case $a=1$ see \cite{Dokshitzer:1998kz,Becher:2011pf,Chiu:2011qc,Chiu:2012ir,Becher:2012qc}. For ``recoil-free'' versions of angularities and other event shapes with nice properties at $a=1$, see \cite{Larkoski:2014uqa}. Thus, framing our discussion in terms of angularities is not a narrow specialization to a particular variable, but rather a way to encompass many of the above-mentioned event shapes at once in a generic way, and also illustrates how to modify the resummation formulae for observables of different mass dimensions. The discussion below for angularities should be read in this generic light.

%%%%%%%%%%%%%%%%%%%%%%%%%%%%%%%%%%%%%%%%%%%%%%%%%
\subsection{Factorization and Resummation of Event Shape Distributions}
\label{ssec:factorizationoverview}
%%%%%%%%%%%%%%%%%%%%%%%%%%%%%%%%%%%%%%%%%%%%%%%%%

\subsubsection{The Factorization Theorem}

The cross section for $e^+ e^- \to \text{hadrons}$ differential in the angularity $\tau_a$ can be shown to factorize in the two-jet region $\tau_a\ll 1$. In this regime, soft and collinear degrees of freedom dominate the final state, and the cross section factorizes into hard, jet, and soft functions \cite{Berger:2003iw,Lee:2006nr,Bauer:2008dt,Hornig:2009vb}:
\begin{align}
\label{eq:factorization}
\frac{d\sigma}{d\tau_a} &= \sigma_0 \, H_2(Q^2,\mu)\int dt_a^n \, dt_a^{\bn} \, dk_s \, \delta\Bigl( \tau_a -\frac{ t_a^n + t_a^{\bn}}{Q^{2-a}} - \frac{k_s}{Q} \Bigr)   J_n^a \big( t_a^n , \mu) J_{\bn}^a \big( t_a^{\bn} , \mu) S_2^a \big( k_s , \mu\big) \,.
\end{align}
For larger $\tau_a\sim 1$, non-singular terms which are power-suppressed in the two-jet region become leading order and must be added to this expression. The factorization theorem \eq{factorization} has been derived both using the methods of direct QCD \cite{Berger:2003iw,Lee:2006nr} and SCET \cite{Lee:2006nr,Bauer:2008dt,Hornig:2009vb}.
We have changed the notation from that used in \cite{Berger:2003iw,Lee:2006nr,Bauer:2008dt,Hornig:2009vb} so the arguments of the jet and soft functions are dimensionful, reflecting the natural quantities on which they depend. As alluded to in \sec{orders}, often it is advantageous to study the Laplace transform of \eq{factorization}, for which the factorization theorem is
\be
\label{eq:factorizationx}
\wt\sigma(\nu_a)  = H_2(Q^2,\mu) \wt J_n^a\Bigl( \frac{\nu_a}{Q^{2-a}},\mu\Bigr)  \wt J_\bn^a\Bigl( \frac{\nu_a}{Q^{2-a}},\mu\Bigr) \wt S_2^a\Bigl(\frac{\nu_a}{Q},\mu\Bigr)\,,
\ee
where 
\begin{subequations}
\begin{gather}
\wt\sigma(\nu_a) = \frac{1}{\sigma_0}\int_0^\infty d\tau_a\,e^{-\nu_a\tau_a}\frac{d\sigma}{d\tau_a}\,,\\
 \wt J_n^a(x_a,\mu) = \int_0^\infty dt_a\, e^{-x_a t_a} J_n^a(t_a,\mu) \,,\quad \wt S_2^a(x_a,\mu) = \int_0^\infty dk_s \,e^{-x_a k_s} S_2^a(k_s,\mu)\,, \\
 x_a = \frac{\nu_a}{Q^{2-a}} \ {\rm for}\ J\, , \quad x_a = \frac{\nu_a}{Q}\ {\rm for}\ S\, ,
\end{gather}
\end{subequations}
where for simplicity of notation we let $x_a$ take the form appropriate to the function in which it appears.
Below, we will generally use $\nu_a$ for dimensionless transformed variables and $x_a$ for dimensionful ones (with the dimension depending on whether it appears in $J$ or $S$). Alternatively one could use the Fourier transforms, for which \eq{factorizationx} would look similar.  We will find, as in previous literature, that the Laplace (or Fourier) transform offers the most straightforward path to defining orders of resummed logarithmic accuracy.

The jet and soft functions in \eqs{factorization}{factorizationx} encode the collinear and soft limits of QCD scattering amplitudes and phase space constraints used in \cite{Catani:1992ua} to separate jet and soft contibutions to event shape distributions and achieve resummation of logarithms arising from the collinear and soft divergences of QCD. The factorization approach that starts from \eqs{factorization}{factorizationx} provides matrix element definitions of these jet and soft functions and is thus very powerful to organize the computation of higher-order perturbative corrections and subleading power corrections \cite{Freedman:2013vya}, and the derivation of general properties of the cross section, such as universality of the leading nonperturbative corrections \cite{Lee:2006fn,Lee:2006nr,Mateu:2012nk}.  

We collect here only the basic definitions of the jet and soft functions in \eq{factorization}, leaving discussions of their derivation and calculation to the relevant references. The presentation here, which aims at comparing the two formalisms of direct QCD and SCET factorization of the cross section in \eqs{factorization}{factorizationx}, is very much parallel to that given in \cite{Lee:2006nr}.

The hard coefficient $H_2(Q^2,\mu)$ is a short-distance or hard function containing the underlying partonic hard-scattering diagrams, prior to the collinear branching and soft radiation encoded in the jet and soft functions. It is computed perturbatively and depends on dynamics only at the large energy scale $Q$.

\subsubsection{Jet and Soft Functions in Direct QCD}

In direct QCD, the quark jet functions in \eq{factorization} are defined in terms of matrix elements \cite{Berger:2003iw,Lee:2006nr}
\be
\label{eq:QCDjet}
\begin{split}
{J'_c}^{\mu}(t_a^n,a,\mu) &= \frac{2}{Q^2} \frac{(2\pi)^6}{N_C} \sum_{N_{J_c}} \Tr\Bigl[ \gamma^\mu \bra{0} \Phi_{\xi_n}^{(q)\dag}(0) q(0) \ket{N_{J_c}} \bra{N_{J_c}} \bar q(0) \Phi_{\xi_c}^{(q)}(0)\ket{0}\Bigr] \\
&\quad\times \delta(t_a^n- Q^{2-a}\tau_a(N_{J_c})) \delta (Q-\omega(N_{J_c}) ) \delta^2(\hat n_{J_c} - \hat n(N_{J_c}))\,,
\end{split}
\ee
where $c=n,\bn$ labels the direction of the jet, $N_C$ is the number of colors, the sum is over a set of collinear states $N_{J_n}$, $\tau_a(N_{J_c})$ is the measured angularity of state $N_{J_c}$, and $\omega(N_{J_c})$ its total energy. These jet functions are, in the notation of \cite{Berger:2003iw}, also differential in the direction $\hat n_{J_{c}}$ of the final state jet. In the inclusive event shape distribution \eq{factorization} we integrate over these directions. The scalar jet functions in \eq{factorization} are the projection of \eq{QCDjet} along the lightlike direction $\beta_c$ of the jet, $J_n^a(t_a,\mu) = \bar\beta_c\cdot J_c'(t_a^n,a,\mu)$. The jet operators in \eq{QCDjet} contain the Wilson lines
\be
\label{eq:QCDWilson}
\Phi_{\xi_c}^{(f)}(z) = P\exp \biggl[ ig\int_{-\infty}^0 d\lambda \, \xi_c\cdot \cA^{(f)} (\lambda\xi_c +z)\biggr]\,,
\ee
which are path-ordered exponentials of gluons in the color representation $(f)$ along a 4-vector direction $\xi_c$, which in \cite{Berger:2003iw} was taken to be off the light-cone, at least to start. 

The soft function, meanwhile, is defined in direct QCD by starting with an ``eikonal'' cross section reflecting the eikonal Feynman rules for emissions of soft gluons from energetic partons,
\be
\label{eq:QCDeikonal}
\bar\sigma^{\text{(eik)}} (k_s,\mu) = \frac{1}{N_C} \sum_{N_{\text{eik}}} \bra{0} \Phi_{\bn}^{(\bar q)\dag}(0) \Phi_n
^{(q)\dag}(0)\ket{N_{\text{eik}}}\bra{N_{\text{eik}}} \Phi_{n}^{(q)}(0) \Phi_\bn^{(\bar q)}(0)\ket{0}\delta(k_s - Q\tau_a(N_{\text{eik}}))\,,
\ee
where the final states $N_{\text{eik}}$ are those produced by Wilson lines in the directions $n,\bn$. Again $\tau_a(N_{\text{eik}})$ is the value of the angularity measured in state $N_{\text{eik}}$. The eikonal cross section provides a good approximation to the soft radiation at large angles from $n,\bn$ but double counts soft radiation along these jet directions that are already in the jet functions \eq{QCDjet}. To avoid this double-counting in the factorization theorem \eq{factorization}, one defines a set of eikonal jet functions to be subtracted out of the eikonal cross section:
\be
\begin{split}
\label{eq:QCDeikonaljet}
\bar J_c^{\text{(eik)}}(k_s,\mu) = \frac{1}{N_C} \sum_{N_c^{\text{eik}}} \bra{0} \Phi_{\xi_c}^{(f_c)\dag}(0) \Phi_{\beta_c}
^{(f_c)\dag}(0)\ket{\smash{N_c^{\text{eik}}}} & \bra{\smash{N_c^{\text{eik}}}} \Phi_{\beta_c}^{(f_c)}(0) \Phi_{\xi_c}^{(f_c)}(0)\ket{0} \\
&\times \delta(k_s - Q\tau_a(N_c^{\text{eik}}))\,,
\end{split}
\ee
where the roles of the quarks in \eq{QCDjet} are replaced by lightlike Wilson lines. Defining the soft function in Laplace transform space,
\be
\label{eq:QCDsoft}
\wt S(x_s) = \frac{\wt \sigma^{\text{(eik)}}(x_s)}{\tilde J_n^{\text{(eik)}}(x_s) \tilde J_\bn^{\text{(eik)}}(x_s)}\,,
\ee
avoids double counting in the factorized cross section, and leads to the correct factorization theorem \eqs{factorization}{factorizationx}. In \cite{Lee:2006nr} it was argued that subtracting the double-counted contributions out of the jet functions instead leads to definitions of jet and soft functions more parallel to SCET:
\be
\label{eq:QCDfactlikeSCET}
\wt\sigma(\nu) = \sigma_0(Q) \wt \cJ_n(x_a)\wt\cJ_\bn(x_a) \wt\sigma^{\text{(eik)}}(x_a)\,,
\ee
where 
\be
\label{eq:QCDjetsubtracted}
\wt\cJ_c(x_a) = \frac{\tilde J_c(x_a)}{\tilde J_c^{\text{(eik)}}(x_a)}\,.
\ee
This organization is directly related to the method of ``zero-bin subtraction'' in SCET \cite{Manohar:2006nz}, a relation which was discussed in some detail in \cite{Lee:2006nr,Idilbi:2007ff,Idilbi:2007yi}

\subsubsection{Jet and Soft Functions in SCET}

The jet and soft functions in SCET have similar definitions as above. Some differences in approach are that SCET begins with a Lagrangian built out of collinear and soft quark and gluon fields, formed after integrating out hard modes at the scale $\mu\sim Q$. This Lagrangian encodes the Feynman rules for evaluating matrix elements of collinear or soft fields. Details of the derivation of the leading-order Lagrangian can be found in  \cite{Bauer:2000yr,Bauer:2001ct}, the decoupling of soft and collinear modes in the theory in \cite{Bauer:2001yt}. The factorization in \eq{factorization} of event shape distributions in the two-jet region proceeds by matching the quark electroweak current in QCD onto currents of collinear and soft operators in SCET, the details of which can be found to $\cO(\as)$ accuracy in \cite{Bauer:2003di,Manohar:2003vb}. Details of the proof of the factorization are found in  \cite{Bauer:2003di,Bauer:2008dt}. 

It is not our intent to review all of these details here, but only to give the definitions of the jet and soft functions in SCET arising from the proof of \eq{factorization} found in the above references. The collinear quark jet functions appearing in \eq{factorization} are defined in SCET in terms of matrix elements of collinear jet operators,
\be
\label{eq:SCETjet}
J_n^a(t_a^n, \mu) = \int\frac{dl^+}{2\pi} \frac{1}{2N_C}\Tr \int d^4x \, e^{il\cdot x} \bra{0} \frac{\bnslash}{2}\chi_n(x)\delta(t_a^n - Q^{2-a}\hat\tau_a^n)\delta(Q + \bn\cdot\cP)  \delta^2(\cP_\perp)\bar\chi_n(0)\ket{0}\,,
\ee
where the trace is over colors and Dirac indices, $\cP^\mu$ is a ``label'' momentum operator picking out the large components of the momentum of the collinear modes $\chi_n$ \cite{Bauer:2001ct}, and $\hat\tau_a^n$ is an operator measuring the angularity of final states in the cut diagrams that must be evaluated to compute the matrix element in \eq{SCETjet} \cite{Bauer:2008dt}. The collinear jet fields $\chi_n$ are themselves built out of collinear quark fields and collinear Wilson lines in SCET:
\be
\label{eq:SCETchi}
\chi_n(x) = \sum_{\tilde p} \chi_{n,\tilde p}(x)\,,\quad \chi_{n,\tilde p}  = [\delta(\omega - \bn\cdot\cP)\delta^2(\tilde p_\perp - \cP_\perp) W_n^\dag\xi_n]\,,
\ee 
where $\tilde p^\mu = \omega n^\mu/2 + \tilde p_\perp^\mu$ is the large label momentum of the jet field $\chi_n$, and $\xi_n$ is a collinear quark field, and $W_n$ is the Wilson line of collinear gluons,
\be
W_n(x) = \sum_{\text{perms}} \exp\biggl[ - \frac{g}{\bn\cdot\cP}\bn\cdot A_n(x) \biggr] \,,
\ee
where $A_n^\mu(x) = \sum_{\tilde p}A_{n,p}^\mu(x)$ is an $n$-collinear gluon field.

Meanwhile the soft function in \eq{factorization} is a matrix element of soft gluon Wilson lines,
\be
\label{eq:SCETsoft}
S(k_s,\mu) = \frac{1}{N_C} \Tr\bra{0}\overline Y_{\bn}^\dag (0) Y_n^\dag(0) \delta(k_s - Q\hat\tau_a^s) Y_n(0) \overline Y_{\bn}(0)\ket{0}\,,
\ee
where the trace is over colors, $\hat \tau_a^s$ is an operator \cite{Bauer:2008dt} measuring the angularity $\tau_a$ of soft final states, and the Wilson lines are defined
\be
Y_n(x) = P\exp\biggl[ig \int_0^\infty ds\,n\cdot A_s(ns+x) \biggr] \,, \quad \overline Y_{\bn}(x) = P\exp\biggl[ ig\int_0^\infty ds\,\bn\cdot \bar A_s(\bn s+x)\biggr]\,,
\ee
where $A_s$ and $\bar A_s$ are soft gluons in the fundamental and anti-fundamental representation, respectively \cite{Bauer:2003di}. The Wilson lines $Y_{n,\bn}$ in \eq{SCETsoft} arise by a field redefinition of the collinear fields in the SCET Lagrangian that achieves a decoupling of soft and collinear interactions  \cite{Bauer:2001yt}. 

The jet and soft functions in SCET must also be carefully defined to avoid double counting of soft radiation that happens to go along the collinear directions $n,\bn$. In SCET this is achieved by a ``zero-bin subtraction'' \cite{Manohar:2006nz} that accounts for the fact that the sums over collinear label momenta in equations like \eq{SCETchi} do not include the momentum bin $\tilde p =0$. Since this subtraction is performed out of the collinear functions, it is akin to the direct QCD scheme in \eqs{QCDfactlikeSCET}{QCDjetsubtracted}. This equivalence was discussed in \cite{Lee:2006nr,Idilbi:2007ff,Idilbi:2007yi}.

\bigskip

The factorization theorems \eqs{factorization}{factorizationx} separate the dependence on the hard, jet, and soft scales whose ratios appear in the arguments of the large logarithms of $\tau_a$ in the QCD cross section. In the hard, jet, and soft functions in dimensional regularization, the logs are of ratios of the scale $\mu$ over a single scale, the hard scale $Q$, the jet scale $Q\tau_a^{1/(2-a)}$, or the soft scale $Q\tau_a$. The RG evolution of each of these functions with $\mu$ is what allows for systematic resummation of the large logarithms in the cross section. We review below how this is done in both methods, first in SCET and then in direct QCD.

%%%%%%%%%%%%%%%%%%%%%%%%%%%%%%%%%%%%%%%%%%%%%%%%%
\subsection{Resummed Event Shape Distributions in SCET}
\label{ssec:SCEToverview}
%%%%%%%%%%%%%%%%%%%%%%%%%%%%%%%%%%%%%%%%%%%%%%%%%

\subsubsection{Perturbative Expansions and Evolution of Hard, Jet, and Soft Functions}

The hard function $H_2$ in \eq{factorization} is given by $H_2(Q^2,\mu) = \lvert C_2(Q^2,\mu)\rvert^2$, where $C_2$ is the Wilson coefficient describing the matching from QCD onto the 2-jet operator $O_2$ in SCET, which comes from integrating out the short distance, energetic modes in QCD capable of creating energetic jets, and is observable independent \cite{Bauer:2002nz}.  To $\cO(\as)$, the hard function is given by \cite{Bauer:2003di,Manohar:2003vb}
\be
\label{eq:hardoneloop}
H_2(Q^2,\mu) = 1 + \frac{\as(\mu)C_F}{4\pi} \Bigl( -16 + \frac{7\pi^2}{3} - 12\ln\frac{\mu}{Q} - 8\ln^2 \frac{\mu}{Q}\Bigr)\,,
\ee
and is known to three loops \cite{Moch:2005id,Idilbi:2006dg, Becher:2006mr}.
Each jet function $J_n^a$ comes from collinear radiation in a jet, and is given to $\cO(\as)$ by \cite{Hornig:2009vb}:
\begin{align}
\label{eq:jetoneloop}
J_{n,\bn}^a(t_a,\mu) &= \delta(t_a) \Bigl[ 1+ \frac{\as(\mu) C_F}{4\pi} f(a)\Bigr]  \\
& \quad + \frac{\as(\mu)C_F}{4\pi} \frac{1}{2-a} \biggl[ - \frac{6}{\mu^{2-a}} \cL_0\Bigl( \frac{t_a}{\mu^{2-a}}\Bigr) + \frac{8}{1-a}\frac{1}{\mu^{2-a}} \cL_1\Bigl( \frac{ t_a}{\mu^{2-a}}\Bigr) \biggr]\,, \nn
\end{align}
where 
\be
f(a) = \frac{1}{2-a} \biggl( 14 - 13a - \frac{\pi^2}{6} \frac{12-20a+9a^2}{1-a} - 4\int_0^1 dx\,\frac{2-2x+x^2}{x} \ln[(1-x)^{1-a}+x^{1-a}]\biggr)\,,
\ee 
which reduces to $f(0) = 7 - \pi^2$ for $a=0$, agreeing with the one-loop result for the standard jet function $J^0(t_0)$ \cite{Bauer:2003pi,Becher:2006qw}. 
The jet function depends naturally on a $(2-a)$-dimensional variable $t_a$. For $a=0$ the jet function is known to two loops \cite{Becher:2006qw} and the anomalous dimension to three loops \cite{Becher:2006mr}. Meanwhile, the soft function $S_2^a$ describes the global soft radiation over the entire event, and is given to $\cO(\as)$ by \cite{Hornig:2009vb}:
\be
\label{eq:softoneloop}
S_2^a(k,\mu) = \delta(k) \Bigl( 1 + \frac{\as(\mu) C_F}{4\pi} \frac{1}{1-a}\frac{\pi^2}{3}\Bigr) - \frac{\as(\mu) C_F}{4\pi} \frac{16}{1-a} \frac{1}{\mu} \cL_1\Bigl(\frac{k}{\mu}\Bigr)\,.
\ee
$S_2^a$ is a function of the sum of momenta $k = n\cdot k_S^A + \bn\cdot k_S^B$, the sum of soft momenta in the hemispheres $A,B$ projected onto the $n,\bn$ directions that determine the two hemispheres. For $a=0$ the soft function is known to two loops \cite{Hornig:2011iu,Kelley:2011ng,Monni:2011gb}, and the three-loop anomalous dimension can be obtained from the three-loop hard and jet anomalous dimensions by the requirement of RG invariance of the cross section \eq{factorization} \cite{Becher:2008cf}.
The distributions $\cL_n$ in \eqs{jetoneloop}{softoneloop} are ``plus distributions'':
\be
\cL_n(x) \equiv \left[ \frac{\theta(x) \ln^n x}{x}\right]_+\,,
\ee
and are defined in \appx{plus}.

\begin{figure}[t]
\begin{center}
\hspace{3em}\includegraphics[width=.75\textwidth]{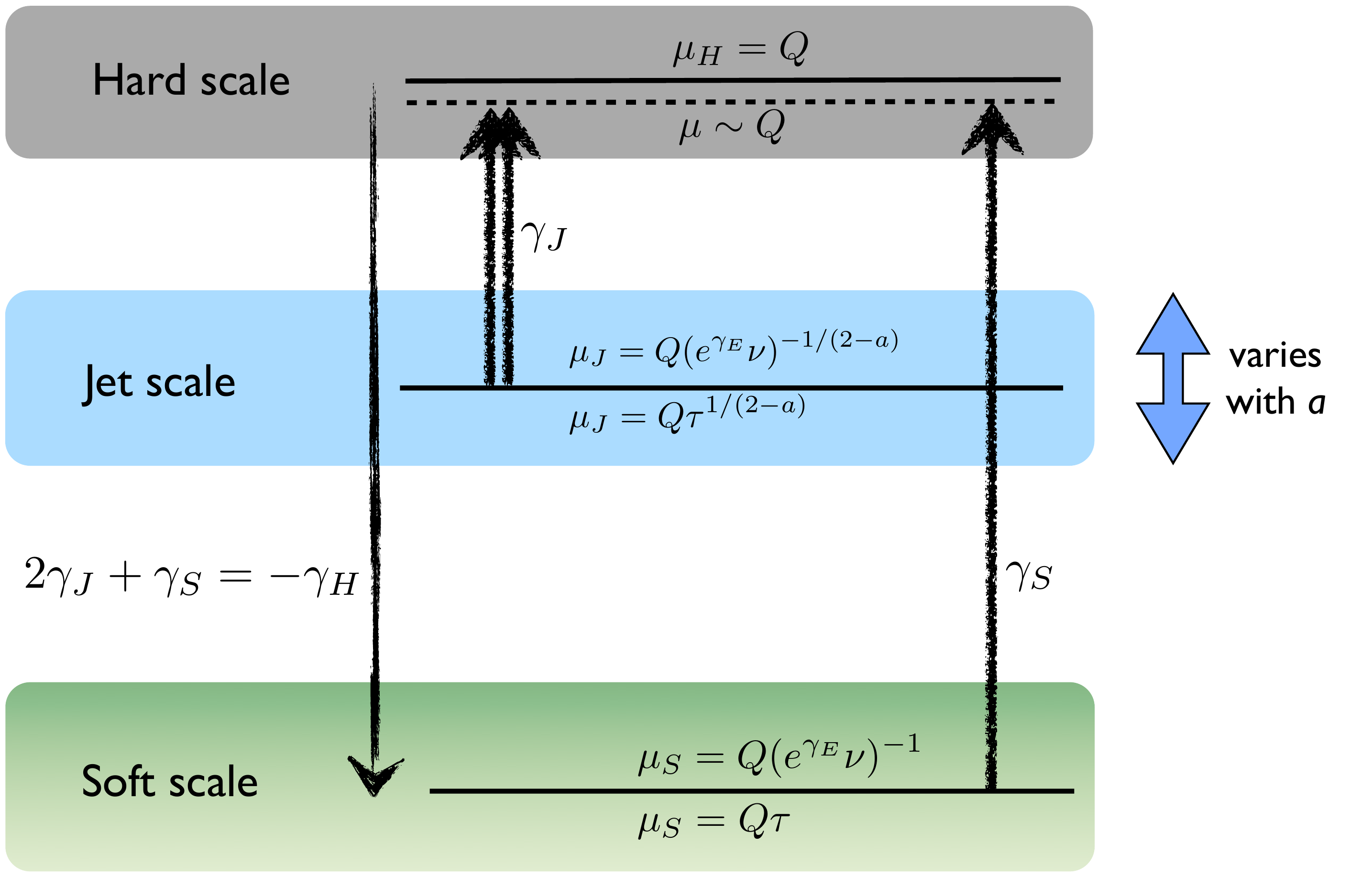}
\vspace{-1em}
\end{center}
\caption{Natural scales for angularity event shapes in the factorization theorem \eq{factorization}. The hard, jet, and soft scales in this ladder minimize logs in the hard, jet, and soft functions in the angularity cross sections \eq{SCETresummedcumulant} in $\tau_a$ space or \eq{sigmanuresummed} in Laplace space. The common factorization scale $\mu$ in \eq{factorization} can be chosen anywhere, but is commonly chosen near $\mu\sim Q$. The functions are evolved via RGEs from their natural scales to $\mu$. The anomalous dimensions satisfy the consistency condition $\gamma_H+2\gamma_J+\gamma_S = 0$. Note that the jet scale varies with $a$. At $a=1$ it coincides with the soft scale, and a new factorization theorem is required to sum logs in the jet and soft functions, e.g. using \SCETb \cite{Becher:2011pf,Becher:2012qc,Chiu:2011qc,Chiu:2012ir} or using ``recoil-free'' versions of angularities \cite{Larkoski:2014uqa}.
\label{fig:evolution}}
\end{figure}
The different functions in the factorization theorem each describe physics at different scales, and the natural scales associated with these functions are well separated. Ratios of these scales produce the logs of $\tau_a$ in the QCD cross section. The factorization theorem \eq{factorization} splits up these logs into logs of the factorization scale $\mu$ over only one of these natural scales at a time. The one-loop results in \eqss{hardoneloop}{jetoneloop}{softoneloop} for the hard, jet and soft functions display this natural scale dependence illustrated in \fig{evolution}. After plugging them into the factorized cross section \eq{factorization}, these scales are identified to be:
\begin{align} \label{eq:SCETnatscales}
\mu_H^{\rm nat} = Q \,,\qquad
\mu_J^{\rm nat} = Q\tau_a^{1/(2-a)} \,, \qquad
\mu_S^{\rm nat} = Q\tau_a \,. 
\end{align}
Large logarithms of the ratios of these scales to the factorization scale $\mu$ are produced in the fixed-order expansions of these functions, and these logs must be resummed to obtain accurate predictions in the small $\tau_a$ regime.  The renormalization group (RG) allows us to independently evolve each function in $\mu$ and sum these logs.  Each function has an associated RG equation relating the renormalized function to its anomalous dimension:
\begin{align} \label{eq:anomdimdef}
\frac{d}{d\ln\mu} H_2 (\mu) &= \gamma_H (\mu) \, H_2 (\mu) \,, \nn \\
\frac{d}{d\ln\mu} J_i^a \big( t_a, \mu \big) &= \int dt_a' \, \gamma_J^a (t_a - t_a', \mu) \, J_i^a (t_a', \mu) \,, \\
\frac{d}{d\ln\mu} S_2^a \big( k, \mu \big) &= \int dk' \, \gamma_S^a (k - k', \mu) \, S_2^a (k', \mu) \,. \nn
\end{align}
The hard function anomalous dimension takes the form
\be
\label{eq:gammaH}
\gamma_H (\mu) = \kappa_H \Gcusp^q [\as] \ln\Bigl( \frac{Q}{\mu} \Bigr) + \gamma_H [\as] \,,
\ee
dependent on ``cusp'' and ``non-cusp'' anomalous dimensions $\Gcusp^q[\as],\, \gamma_H[\as]$.
The jet and soft functions have a common form for the anomalous dimension:
\be
\gamma_F (t_F,\mu) = \kappa_F \Gcusp^q [\as] \frac{1}{\mu^{j_F}} \cL_0\Bigl(\frac{t_F}{\mu^{j_F}}\Bigr) + \gamma_F [\as] \delta(t_F) \,,
\ee
where $F = J,S$ for the jet and soft functions, and $t_F$ is a variable of mass dimension $j_F$, equal to $t_a$ for the jet function and $k$ for the soft function.

The cusp anomalous dimension $\Gcusp^q [\as]$ is a universal series in $\as$,
\be
\label{eq:cuspexpansion}
\Gcusp^q [\as] = \sum_{n=0}^{\infty} \Bigl( \frac{\as}{4\pi} \Bigr)^{\! n+1} \Gamma^q_{n} \,, 
\ee
where the coefficients are given up to $n=2$, or $\cO(\as^3)$, by \cite{Korchemsky:1987wg, Moch:2004pa}:
\begin{align}
&\Gamma^q_0 = 4C_F \,, \qquad \Gamma^q_1 = \Gamma^q_0 \Bigl[ \Bigl( \frac{67}{9} - \frac{\pi^2}{3} \Bigr) C_A - \frac{20}{9} T_F n_f \Bigr] \,, \\
&\Gamma^q_2 =  \Gamma^q_0\Bigl[
\Bigl(\frac{245}{6} -\frac{134 \pi^2}{27} + \frac{11 \pi ^4}{45}
  + \frac{22 \zeta_3}{3}\Bigr)C_A^2 
  + \Bigl(- \frac{418}{27} + \frac{40 \pi^2}{27}  - \frac{56 \zeta_3}{3} \Bigr)C_A\, T_F\,n_f
\nn\\* 
& \hspace{8ex}
  + \Bigl(- \frac{55}{3} + 16 \zeta_3 \Bigr) C_F\, T_F\,n_f
  - \frac{16}{27}\,T_F^2\, n_f^2 \Bigr] \,. \nn
\end{align}
The non-cusp anomalous dimension is similarly defined, with coefficients $\gamma^F_n$ that are specific to each function.  
The constants $\kappa_F$ give the proportionality of the cusp part of each function's anomalous dimension to $\Gcusp^q$. The constants $\kappa_F,j_F$ and the $\ord{\as}$ non-cusp anomalous dimensions are
\begin{align}
\label{eq:constantdefs}
j_H &= 1 \,, & \kappa_H &= 4 \,, & \gamma_H^0 &= -12 C_F \,, \nn \\
j_J &= 2-a \,, & \kappa_J &= -\frac{2}{1-a} \,, & \gamma_J^0 &= 6C_F \,, \nn \\
j_S &= 1 \,, & \kappa_S &= \frac{4}{1-a} \,, & \gamma_S^0 &= 0 \,.
\end{align}
For completeness, we include a constant $j_H$ for the hard function in this table, which is implicitly the mass dimension of $Q$ in \eq{gammaH}, and which we henceforth always set equal to 1. Similarly we will always set $j_S=1$ except in equations where we refer to a generic $j_F$.
We define the ``cusp part'' of each function's anomalous dimension by
\be
\label{eq:GammaFdef}
\Gamma_F \equiv -\frac{j_F \kappa_F}{2}\Gcusp^q\,,
\ee
for notational use below. The non-cusp piece of each function's anomalous dimension is given by an expansion
\be
\label{eq:noncuspexpansion}
\gamma_F[\as] = \sum_{n=0}^{\infty} \Bigl( \frac{\as}{4\pi} \Bigr)^{\! n+1} \gamma_F^{n} \,, 
\ee
where $F=H,J,S$. These coefficients are given to $\cO(\as^3)$ for the hard function and the $a=0$ jet and soft functions in \appx{anomalous}. For $a\neq 0$ they are so far known only to $\cO(\as)$ \cite{Hornig:2009vb}.

The convolutions in the RGEs in \eq{anomdimdef} for the jet and soft functions in momentum space can be removed by Fourier or Laplace transforming the functions, in which case the evolution is multiplicative (as in the hard function).  The evolution is performed in the transform space, and the result is transformed back to momentum space (e.g. \cite{Korchemsky:1993uz,Balzereit:1998yf}).  

Two formalisms exist in SCET to write the evolution in momentum space \cite{Becher:2006mr,Becher:2006nr,Ligeti:2008ac}, each with its own advantages.  One method, described in Ref.~\cite{Ligeti:2008ac}, performs the transformation of the jet/soft function solutions in Laplace/position space back to momentum space, and then writes the result for the resummed cross section as a convolution between momentum-space fixed-order jet/soft functions and momentum-space evolution factors.  This method requires calculation of convolutions of plus functions with each other, all explicitly computed in \cite{Ligeti:2008ac}. This is particularly useful for the types of explicit calculations in, e.g., \cite{Abbate:2010xh,Ligeti:2008ac}. The other method, used in Ref.~\cite{Becher:2006mr,Becher:2006nr}, writes the evolution in terms of a derivative operator built out of Laplace-space jet/soft functions acting on a simple $\tau$-dependent function. This formalism turns out to be more transparently relatable to the direct QCD resummation formalism used in, e.g., \cite{Berger:2003iw,Berger:2003pk,Contopanagos:1993xh}, and is somewhat more compact to write down. For the purposes of our study, we find it more convenient to employ the latter formalism. The two methods, as long as formulae are truncated appropriately, yield numerically equivalent results, as illustrated in, e.g., \cite{Kang:2013nha}.

%%%%%%%%%%%%%%%%%%%%%%%%%%%%%%%%%%%%%%%%%%%%%%%%%
\subsubsection{Evolution in SCET}
\label{ssec:SCETevol}
%%%%%%%%%%%%%%%%%%%%%%%%%%%%%%%%%%%%%%%%%%%%%%%%%

Taking the Laplace (similarly, Fourier) transform of the jet function using the variable $x_a$ conjugate to $t_a$ makes the evolution equation multiplicative,
\be \label{eq:Jxevol}
\frac{d}{d\ln\mu} \wt{J} (x_a, \mu) = \wt{\gamma}_J (x_a, \mu) \wt{J} (x_a, \mu) \,,
\ee
and analogously for the soft function.  The Laplace-transformed anomalous dimensions have the form, for $F = J,S$,
\be
\label{eq:gammaFLaplace}
\wt{\gamma}_F (x_a, \mu) = - \kappa_F \Gcusp^q [\as] \ln \Bigl( \mu^{j_F} x_a e^{\gamma_E} \Bigr) + \gamma_F [\as] \,,
\ee
where $x_a$ is dimension $-2+a$ for $F=J$, and dimension $-1$ for $F=S$.
A multiplicative RG equation, as in \eq{Jxevol}, leads to a straightforward evolution equation that allows us to write the function at an arbitrary scale (in terms of its value at another arbitrary scale),
\be \label{eq:Fevolexp}
\wt{F} (x_a, \mu) = \wt{F} (x_a, \mu_0) \, \exp \biggl( \int_{\mu_0}^{\mu} \frac{d\mu'}{\mu'} \, \wt{\gamma}_F (x_a, \mu') \biggr) \,.
\ee
To perform the evolution we define three functions\footnote{Note that these can also be defined as integrals over the running coupling [see \eq{Ketaforms}] by using the defining relation  \eq{RGalpha} of the beta function, $\beta[\alpha_s(\mu)] = \frac{d}{d\ln\mu}\alpha_s(\mu)$.  In this case the large logarithms are expressed as logarithms of ratios of $\alpha_s$ evaluated at parametrically different scales, while in the definitions of \eq{Ketadef} the logarithms appear explicitly.}:
\begin{align} \label{eq:Ketadef}
K_\Gamma (\mu, \mu_0) &\equiv \int_{\mu_0}^{\mu} \frac{d\mu'}{\mu'} \, \Gcusp^q [\alpha_s(\mu')] \ln\frac{\mu'}{\mu_0} \,, \nn \\
\eta_\Gamma (\mu, \mu_0) &\equiv \int_{\mu_0}^{\mu} \frac{d\mu'}{\mu'} \, \Gamma_{\rm cusp}^q [\alpha_s(\mu')] \,, \nn \\
K_{\gamma_F}(\mu, \mu_0) &\equiv \int_{\mu_0}^{\mu} \frac{d\mu'}{\mu'} \, \gamma_F[\alpha_s(\mu')] \,.
\end{align}
These separate the evolution of the function between two arbitrary scales $(\mu_0,\mu)$ from the dependence on the particular scale $x_a^{1/j_F}$ in the anomalous dimension,
\be \label{eq:gammaKomega}
\int_{\mu_0}^{\mu} \frac{d\mu'}{\mu'} \, \wt{\gamma}_F(\mu') = -j_F \kappa_F K_{\Gamma} (\mu, \mu_0) + K_{\gamma_F} (\mu, \mu_0) -\kappa_F \eta_\Gamma (\mu, \mu_0) \ln \bigl( \mu_0^{j_F} e^{\gamma_E} x_a \bigr) \,.
\ee
We define
\begin{equation} \label{eq:Komegadef}
\begin{split}
K_F (\mu, \mu_0) &\equiv -j_F \kappa_F K_{\Gamma} (\mu, \mu_0) + K_{\gamma_F} (\mu, \mu_0) \,, \\
\omega_F (\mu, \mu_0) &\equiv -\kappa_F \, \eta_\Gamma (\mu, \mu_0) \,,
\end{split}
\end{equation}
so that
\be \label{eq:Fevol}
\wt{F} (x_a, \mu) = \wt{F} (x_a, \mu_0) \, \exp \bigl( K_F (\mu, \mu_0) \bigr) \bigl( \mu_0^{j_F} e^{\gamma_E} x_a \bigr)^{\omega_F (\mu, \mu_0)} \,.
\ee
For the hard function, the evolution equation reads
\be
H(\mu) = H(\mu_0) \, \exp \bigl( K_H (\mu, \mu_0) \bigr) \Bigl(\frac{\mu_0}{Q} \Bigr)^{j_H \omega_H (\mu, \mu_0)} \,.
\ee
Using $K_F$ and $\omega_F$ has factored out the $x_a$ dependence in the evolution factor in \eq{Fevol}, and we can perform the inverse Laplace transform to obtain the evolution relation in momentum space:
\be \label{eq:FconvUevol}
F(t_F,\mu) = \int dt_F' U_F(t_F - t_F',\mu,\mu_0) F(t_F', \mu_0) \, ,
\ee
where the evolution kernel $U_F$ is
\be
U_F(t_F,\mu,\mu_0) = \frac{\exp\left(K_F(\mu,\mu_0) + \gamma_E \omega_F(\mu,\mu_0)\right)}{\Gamma(1-\omega_F(\mu,\mu_0))} \biggl[ -\frac{\omega_F}{\mu_0^{j_F}} \cL^{-\omega_F}\Bigl(\frac{t_F}{\mu_0^{j_F}}\Bigr) + \delta(t_F)\biggr] \, ,
\ee
and the distribution $\cL^{-\omega}$ is given by the plus distribution \cite{Ligeti:2008ac}
\be
\cL^{-\omega} (x) = \left[\frac{\theta(x)}{x^{1+\omega}}\right]_+ \,,
\ee
defined in \appx{plus}.
The evolution factor is convolved with the fixed-order expansion of $F(t_F, \mu_0)$ to produce $F (t_F, \mu)$.  These convolutions produce functions of $\omega_F$ multiplying distributions $\cL^{\omega_F}$, and the jet and soft functions in the factorization theorem can then be convolved to produce the total cross section.  The necessary convolutions have been worked out in generality in Ref.~\cite{Ligeti:2008ac}, meaning the procedure is formulaic.

An alternative way to write the evolution uses the following approach illustrated in Ref.~\cite{Becher:2006mr,Becher:2006nr}. This approach turns out to be more transparently relatable to the dQCD literature and is slightly more compact to write down. First one notes that the $x_a$-dependence in the evolution factor in \eq{Fevol} is contained entirely in logs $L_F(\mu_0)$, where
\be
\label{eq:LFdef}
L_F (\mu) \equiv \ln \bigl( \mu^{j_F} e^{\gamma_E} x_a \bigr) = \ln \Bigl[\Bigl(\frac{\mu}{Q}\Bigr)^{j_F} e^{\gamma_E}\nu\Bigr]\,.
\ee
In the last equality we have indicated what the variable $x_a$ gets replaced by when these logs appear in the jet or soft functions in \eq{factorizationx}.
Therefore, we can rewrite the $x$-dependence in terms of logarithms $L_F(\mu)$.  Cleverly, this means we can generate the $L_F$ dependence through partial derivatives with respect to $\omega_F$. Rewriting \eq{Fevol} for $\wt F$ in this way,
\begin{align}
\label{eq:Ftilderewriting}
\wt{F} (L_F (\mu), \mu) &= \wt{F} (L_F (\mu_0), \mu_0) \, \exp \bigl[ K_F (\mu, \mu_0) \bigr] \exp \bigl[ L_F (\mu_0) \, \omega_F (\mu, \mu_0) \bigr] \nn \\
&= \wt{F} (\p_{\omega_F}, \mu_0) \, \exp \bigl[ K_F (\mu, \mu_0) \bigr] \exp \bigl[ L_F (\mu_0) \, \omega_F (\mu, \mu_0) \bigr] \,.
\end{align}
This removes the functional $x_a$-dependence in $\wt{F}$ at $\mu_0$, meaning the inverse transform can be performed completely (without the remaining convolution in \eq{FconvUevol}).  The result is
\be
\label{eq:Fevolderivatives}
F(t_F, \mu) = \wt{F} (\p_{\omega_F}, \mu_0) \, \exp \bigl[ K_F (\mu, \mu_0) \bigr] \Bigl( \frac{\mu_0^{j_F}}{t_F} \Bigr)^{\omega_F(\mu, \mu_0)} \frac{1}{t_F} \, \frac{e^{\gamma_E \omega_F (\mu, \mu_0)}}{\Gamma[-\omega_F (\mu, \mu_0)]} \,.
\ee
This form deserves some remarks.  In the position-space function $\wt{F}$ at the initial scale $\mu_0$, the $\tau$-dependence ($t_F$ dependence) is generated via derivatives acting on the resummed series $\omega_F$ in the evolution factor.  Operationally, the derivatives can be thought of as representing a replacement rule, since we know what functions they are acting on, e.g. \eq{Gbardefs} below.  We will see a similar method used in the QCD formulation of the cross section in \ssec{QCDoverview}.  With jet and soft functions written in the form \eq{Fevolderivatives} convolution over the momentum variables in \eq{factorization} is quite simple to carry out, leading to a compact form for the resummed cross section.

\subsubsection{Resummed Cross Section in SCET}

In position or Laplace space, the factorized cross section \eq{factorizationx} using RG evolved hard, jet, and soft functions takes the form
\begin{align}
\label{eq:SCETresummedLaplace}
\wt\sigma(\nu_a) &=  H_2 (Q^2,\mu_H)  \wt{J} (L_J(\mu_J),\mu_J)^2  \wt{S} (L_S(\mu_S),\mu_S)\\
& \quad \times \exp \bigl( K_H + 2K_J + K_S \bigr) \Bigl( \frac{\mu_H}{Q} \Bigr)^{\omega_H} \Bigl( \frac{\mu_J {(\nu_a e^{\gamma_E})}^{1/j_J}}{Q} \Bigr)^{2j_J\omega_J}\Bigl( \frac{\mu_S \nu_a e^{\gamma_E}}{Q} \Bigr)^{\omega_S}  \,, \nn
\end{align}
Taking the inverse Laplace transform of \eq{SCETresummedLaplace}, or explicitly carrying through the convolutions in \eq{factorization} using the solutions \eq{Fevolderivatives} for the momentum-space jet and soft functions, the momentum-space cross section is
\begin{align}
\label{eq:SCETresummeddist}
\sigma(\tau_a) &=   \exp \bigl( K_H + 2K_J + K_S \bigr) \Bigl( \frac{\mu_H}{Q} \Bigr)^{\omega_H} \Bigl( \frac{\mu_J}{Q \tau_a^{1/j_J}} \Bigr)^{2j_J\omega_J}\Bigl( \frac{\mu_S}{Q \tau_a} \Bigr)^{\omega_S} H_2 (Q^2,\mu_H) \\
& \quad \times  \wt{J} \Bigl( \p_\Omega + \ln \frac{\mu_J^{j_J}}{Q^{j_J}\tau_a} , \mu_J \Bigr)^2 \wt{S} \Bigl( \p_\Omega + \ln \frac{\mu_S}{Q \tau_a} , \mu_S \Bigr)  \frac{1}{\tau_a} \frac{\exp(\gamma_E \Omega)}{\Gamma(-\Omega)} \,, \nn
\end{align}
where
\be
\Omega \equiv 2\omega_J + \omega_S \,.
\ee
The result \eq{SCETresummeddist} for the thrust ($a=0$) appeared in a very similar form in \cite{Becher:2008cf} and for arbitrary $a$ in \cite{Hornig:2009vb}, following the methods in \cite{Becher:2006mr,Becher:2006nr}.
In \eq{SCETresummeddist}, we have pulled through each ratio of factorization scales $\mu_H$, $\mu_J$, and $\mu_S$ to the canonical scales in \eq{SCETnatscales} past the fixed-order functions $\wt{J}$ and $\wt{S}$.  This shifts the derivative in each function by a logarithm of the relevant scale ratio.  Integrating \eq{SCETresummeddist} to yield the cumulant $R(\tau_a)$ defined in \eq{cumulantdef} is simple,
\begin{align}
\label{eq:SCETresummedcumulant}
R(\tau_a) &= \exp \bigl( K_H + 2K_J + K_S \bigr) \Bigl( \frac{\mu_H}{Q} \Bigr)^{\omega_H} \Bigl( \frac{\mu_J}{Q \tau_a^{1/j_J}} \Bigr)^{2j_J\omega_J} \Bigl( \frac{\mu_S}{Q \tau_a} \Bigr)^{\omega_S} H_2 (Q^2,\mu_H)\nn \\
& \quad \times  \wt{J} \Bigl( \p_\Omega + \ln \frac{\mu_J^{j_J}}{Q^{j_J} \tau_a} , \mu_J \Bigr)^2 \wt{S} \Bigl( \p_\Omega + \ln \frac{\mu_S}{Q \tau_a} , \mu_S \Bigr)  \frac{\exp(\gamma_E \Omega)}{\Gamma(1-\Omega)} \,.
\end{align}
The factorization scales $\mu_i$ are arbitrary scales that each function is evolved from, and the cross section is independent of the common scale $\mu$ that all functions are evolved to.  If we kept the hard, jet, and soft functions exact to all orders, then the cross section would be independent of these factorization scales.  Truncating the resummation at a given accuracy introduces dependence on the factorization scales due to the dropped (unknown) higher-order terms. Thus, by varying these scales a theoretical uncertainty due to the omitted higher-order terms can be estimated.\footnote{Typically these scales are varied up and down by a factor of 2. More reliable predictions and estimates of theoretical uncertainties can be obtained by using so-called ``profile scales'' whose functional form varies with $\tau_a$, see e.g. \cite{Abbate:2010xh,Berger:2010xi,Ligeti:2008ac,Kang:2013nha}}
\begin{figure}[t]
\begin{center}
\includegraphics[width=.9\textwidth]{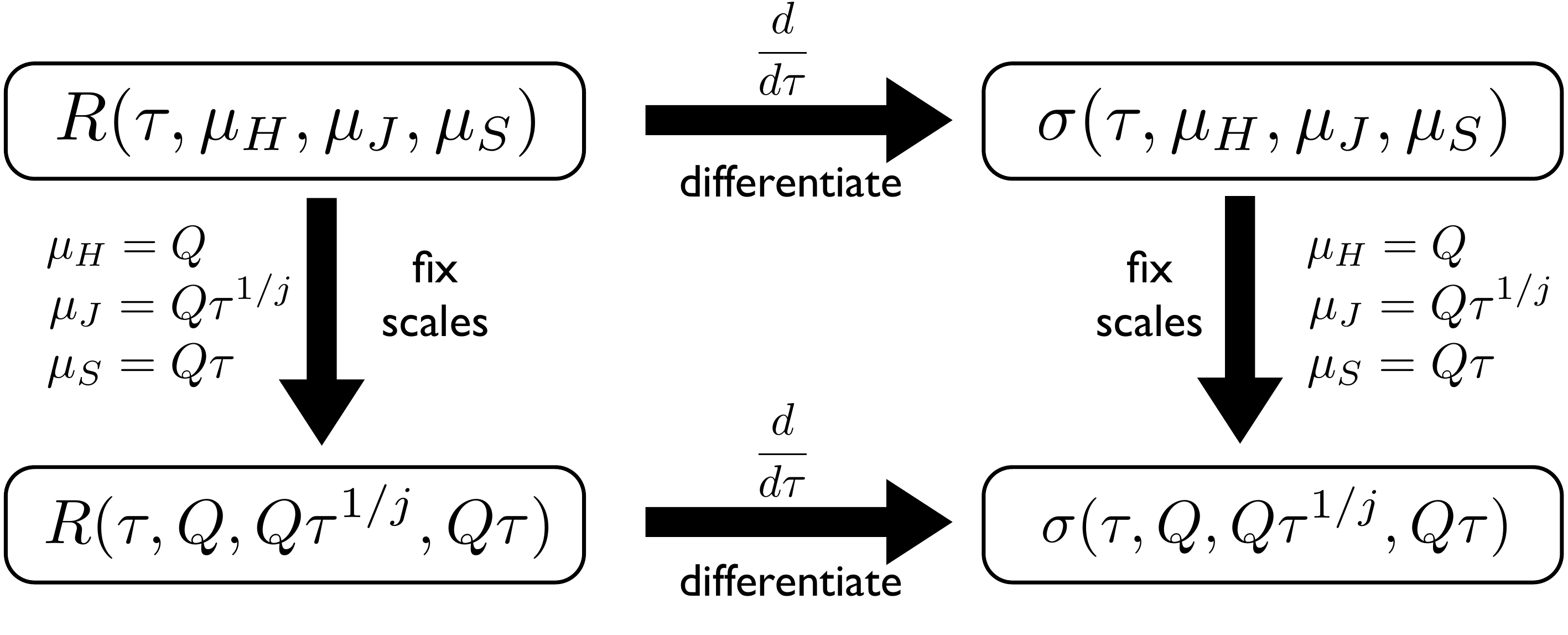}
\vspace{-2em}
\end{center}
\caption{Commutative diagram for obtaining the resummed differential distribution from the resummed cumulant. The resummed distribution can be obtained by first differentiating the cumulant with free scales, and then choosing the scales to be $\tau$-dependent, such as the canonical choices \eq{SCETnatscales} indicated in the figure. Or, it can be obtained by first fixing the scales in the cumulant to be $\tau$-dependent, and then differentiating. Starting with expressions for $R$ truncated at a given \nkll order of accuracy, the latter method will produce more terms in $\sigma$ due to the derivative $d/d\tau$ seeing more $\tau$-dependent terms. However, if expressions for $R,\sigma$ are kept to all orders in $\as$, the two routes produce identical results.
\label{fig:commute}}
\end{figure}

It is worth remarking here that in integrating \eq{SCETresummeddist} to obtain \eq{SCETresummedcumulant} (or differentiating the latter to obtain the former) one assumes that the scales $\mu_{J,S}$ have not yet been chosen and are considered independent of $\tau_a$. Thus the derivative/integral does not act on them. This is represented by the top arrow in the commutative diagram in \fig{commute}. However, eventually, to minimize the logs of $\mu_F/Q\tau_a^{1/j_F}$ in the jet and soft functions we \emph{will} choose them to be functions of $\tau_a$. This is represented by the vertical arrows in \fig{commute}. If we do this first in \eq{SCETresummeddist} or in \eq{SCETresummedcumulant} before integrating/differentiating to obtain the other, then the latter operations are considerably more difficult and yield apparently different results at a truncated order of logarithmic accuracy. If all quantities in \eqs{SCETresummeddist}{SCETresummedcumulant} are kept to all orders in $\as$, then either order of operations will yield exactly the same result---the cross section is independent of the scale choices $\mu_F$. However, at a truncated order, the two operations yield different results. We will remark on this further in \ssec{distributionaccuracy} discussing how to keep $\sigma$ and $R$ as closely to the same accuracy as possible when working to \nkll accuracy.

%%%%%%%%%%%%%%%%%%%%%%%%%%%%%%%%%%%%%%%%%%%%%%%%%
\subsection{Resummed Event Shape Distributions in QCD}
\label{ssec:QCDoverview}
%%%%%%%%%%%%%%%%%%%%%%%%%%%%%%%%%%%%%%%%%%%%%%%%%

The same 2-jet event shape distributions can be calculated directly in QCD, and such techniques were applied before SCET was developed \cite{Catani:1992ua}.  One exploits the universal structure of soft and collinear singularities to write the resummed cross section.  At next-to-leading logarithmic order, for example, the resummation is expressed in terms of an evolution kernel based on splitting functions. We will sketch here the derivation of the resummed angularity cross sections in direct QCD, and then move on to  the final results one finds in the literature.   

The analysis begins by determining the boost properties from the arguments of the soft and jet functions through which logarithmic behavior can occur.  Then, each of the soft, jet, and hard functions obey two equations, one following from renormalization scale variation, the other from boost invariance.  This pattern has been known for some time, and the general solutions derived in \cite{Contopanagos:1996nh} are based on factorization and the strategy for resummations developed in \cite{Collins:1981uk}.  In this case, the renormalization group equations are
\begin{eqnarray}
	\mu \frac{d}{d \mu}\;
	\ln\, S\left(	\frac{Q}{\mu \nu} (\chi_c)^{a-1},
	a, \as(\mu) \right)
	& = & -
	\gamma_s\left(\as(\mu)\right),
	\label{softmu}
	\\
	\mu \frac{d}{d \mu}\;
	\ln\, J_c\left(\frac{ Q\chi_c  }{\mu }, \frac{Q}{\mu \nu} \, (\chi_c)^{a-1} ,
	a,\as(\mu) \right) & = & - \gamma_{J_c}\left(\as(\mu)\right),
	\label{jetmu}
	\\
	\mu \frac{d}{d \mu}\; \ln\,  H\left( \frac{Q}{\mu},\frac{Q\chi_c}{\mu},\as(\mu) \right)
	&=& \gamma_s\left(\alpha_s(\mu)\right)
	+ \sum_{c=1}^2\gamma_{J_c}\left(\alpha_s(\mu)\right)\, ,
	\label{Hmu}
	\end{eqnarray}
	where 
	\begin{eqnarray}
	\chi_c = \frac{Q}{2p_c\cdot {\hat \xi}_c}
	\end{eqnarray}
	with  $\hat \xi_c$ the unit vector introduced in the definition of the QCD jet function, Eq.\ (\ref{eq:QCDjet}), and $\beta_c$ a unit, lightlike vector in the direction of momentum $p_c^\mu = Q\beta_c^\mu/\sqrt{2}$ in the case of two-jet events.   The anomalous dimensions can be defined to be scalars, because any dependence on the vectors $\xi_c$ or $\beta_c$ could be absorbed through a multiplicative redefinition of the hard function.
	
From the behavior under boosts, we derive an equation satisfied by the  jets, which is the only one we need,
\begin{eqnarray}
    \frac{\partial }{\partial \ln \left(\chi_c\right)}
	\ln\, J_c\left(\frac{ Q\chi_c  }{\mu }, \frac{Q}{\mu \nu} \, (\chi_c)^{a-1} ,
	a,\as(\mu) \right)
	& \ & \nonumber
	\\
	&\ & \hspace{-70mm} =\
	    \cK_c\left(\frac{Q}{\mu\,
	\nu}(\chi_c)^{a-1},
	a,\as(\mu)
	\right)   +  \cG_c\left(\frac{Q \chi_c}{\mu},\as(\mu)\right)   \, ,
	    \label{KGend}
	    \end{eqnarray}
	    in terms of perturbative functions $\cK$ and $\cG$, whose low order expansions are given in \cite{Berger:2003iw}.
Using these equations, we can evolve the soft function in the renormalization scale $\mu$, and the jet functions in $\mu$ and $\chi_c$ to organize all logarithms in terms of a limited set of perturbative functions, in much the same manner as for SCET.   We will not give the details here, which for angularities are worked out in Ref.\ \cite{Berger:2003iw}, following \cite{Contopanagos:1996nh}.    The resulting moment-space distribution takes the form
\begin{eqnarray}
\wt\sigma(\nu_a)\ =\  {\cal{N}}(Q) \ \exp \bigl[  E(\ln \nu_a)]\, ,
\label{eq:qcd-exp-nu}
\end{eqnarray}
 where $\ca{N} (Q)$ is a log-free factor with an expansion in $\as$ with coefficients $C_n$ as in \eq{Cexpansion},
\be
\ca{N}(Q) = 1 + \sum_{n=1}^{\infty} \Bigl( \frac{\as(Q)}{4\pi} \Bigr)^n C_n \, ,
\ee
which combines contributions from the hard, soft and jet functions evaluated at unit kinematic ratios at renormalization scale $Q$.
The cumulant cross section is as usual written as the inverse Laplace transform of  (\ref{eq:qcd-exp-nu}),
\begin{align} \label{eq:QCDxsec}
R(\tau_a) = \ca{N}(Q) \frac{1}{2\pi i} \int_{\cC} \frac{d\nu_a}{\nu_a}  \exp \bigl[ \nu_a \tau_a + E(\ln \nu_a) \bigr] \,,
\end{align}
where $\cC$ is the usual contour in the complex $\nu_a$ plane, to the right of singularities.
The QCD exponent $E(\ln \nu_a)$ in Eq.\ (\ref{eq:qcd-exp-nu})) comes from fixed-order calculations of jet and soft functions that bear strong resemblance to those in SCET (although the organization of terms can differ), and is a combined object that describes the perturbative logarithms.  It is built from fixed-order coefficients $A[\alpha_s], B[\alpha_s]$, where $A$ contains the cusp singularity contributions (and, at high orders, can contain additional terms) \cite{Berger:2003iw,Becher:2006nr,Becher:2006mr,Becher:2010tm}.  The form of $E$ for the angularity distribution in \eq{QCDxsec} in terms of $A$ and $B$ can be written
\begin{align}
\label{eq:QCDE}
E(\ln \nu_a) &= 4 \int_{Q (e^{\gamma_E} \nu_a)^{-1/j_J}}^Q \frac{d\mu'}{\mu'} \, A[\as] \ln \frac{\mu'}{Q} \nn \\
& \quad - \frac{4}{1-a} \int_{Q (e^{\gamma_E} \nu_a)^{-1/j_S}}^{Q (e^{\gamma_E} \nu_a)^{-1/j_J}} \frac{d\mu'}{\mu'} \, A[\as] \ln \frac{\mu'}{Q(e^{\gamma_E} \nu_a)^{-1/j_S}} \nn \\
& \quad + 2\int_{Q (e^{\gamma_E} \nu_a)^{-1/j_J}}^Q \frac{d\mu'}{\mu'} \, B_J [\as]  + \int_{Q (e^{\gamma_E} \nu_a)^{-1/j_S}}^Q \frac{d\mu'}{\mu'} \, B_S [\as] \,.
\end{align}
where the integrals $B_S$ and $B_J$ generate non-leading singularities by evolving from the soft and jet scales respectively to the hard scale.  
In fact, it is always possible to absorb the term involving $B_S$ into the other two by following a method originally applied to threshold resummation in Ref.~\cite{Catani:1990rp}.   We may use this freedom to define $A[\alpha_s] = \Gamma_{\rm cusp}[\alpha_s]$, in which case one finds that $B_S[\alpha_s]$ begins only at order $\alpha_s^2$.   In this way, in the QCD angularity analysis of \cite{Berger:2003iw}, the three terms were reduced to two, to check consistency with the with the original CTTW result \cite{Catani:1992ua}.    For the comparison to SCET that we will make in the next subsection, however, it is convenient to keep both of the $B$  terms.

To perform the inverse transform \eq{QCDxsec}, the exponent is expanded about $\ln \nu_a = \ln (1/\tau_a)$ \cite{Catani:1992ua},
\be
\label{eq:ETaylor}
E(\ln \nu_a) = \sum_{n=0}^{\infty} \frac{1}{n!} \ln^n (\nu_a\tau_a) \biggl[ \frac{d^n}{d(\ln \nu_a)^n} E(\ln \nu_a) \Big\rvert_{\nu_a = 1/\tau_a} \biggr] \,.
\ee
Defining $\bar{E} = E(\ln 1/\tau_a)$ and the derivatives
\be
\bar{E}' \equiv \frac{d}{d\ln \nu_a} E(\ln \nu_a) \Big\rvert_{\nu_a = 1/\tau_a} \;\;, \;\; \ldots \;\;, \;\; \bar{E}^{(n)} \equiv \frac{d^n}{d\ln \nu_a^n} E(\ln \nu_a) \bigg\rvert_{\nu_a = 1/\tau_a} \;\;, \;\; \ldots \,,
\ee
we can write
\begin{align}
\exp \bigl[ E(\ln \nu_a) \bigr] &= \exp \biggl[ \, \sum_{n=2}^{\infty} \frac{1}{n!} \bar{E}^{(n)} \p_{\bar{E}'}^n \biggr] \exp \bigl[ \bar{E} + \bar{E}' \ln(\nu_a \tau_a) \bigr] \nn \\
&\equiv \wh{T} (\bar{E}') \exp \bigl[ \bar{E} + \bar{E}' \ln(\nu_a \tau_a) \bigr] \,.
\end{align}
This uses the same kind of derivative operator as in the SCET resummation \eq{SCETresummeddist}, as $\bar{E}'$ is a resummed series and derivatives with respect to it generate additional $\tau_a$ dependence.  As in the SCET case, it allows us to perform the inverse transform \eq{QCDxsec} analytically.  The result is
\be
\label{eq:QCDRderivatives}
R(\tau_a) = \ca{N}(Q) \exp(\bar{E}) \, \wh{T} (\bar{E}') \frac{1}{\Gamma(1-\bar{E}')} \,.
\ee
The operator $\wh{T}$ contains an infinite series of terms (generalizing the result Eq.~(85) in \cite{Catani:1992ua}; a similar form for threshold resummation in Drell-Yan appeared in \cite{Contopanagos:1993xh}).  However, the series is well-ordered, as it is easy to see that the $\bar{E}^{(n)}$ are sequentially higher order ($\bar{E}^{(n)}$ is N$^n$LL, counting in the resummed exponent).  Therefore to work to a given order we simply truncate the series.  As with SCET, since the function that the derivatives are acting on is known, the derivatives with respect to $\bar{E}'$ represent a simple replacement rule.

Note that \eq{QCDRderivatives} contains no free scales like $\mu_{H,J,S}$. As we will see below, they have implicitly been set to the values
\be
\label{eq:QCDscales}
\mu_H = Q\,,\quad \mu_J = \bar\mu_J\equiv Q\bar\tau_a^{1/j_J}\,,\quad \mu_S = \bar\mu_S\equiv Q \bar\tau_a\,,
\ee
where
\be
\label{eq:taubar}
\bar\tau_a \equiv e^{-\gamma_E }\tau_a
\ee
Thus \eq{QCDRderivatives} sits in the lower left-hand box in \fig{commute} (except with $\mu_{J,S}$ rescaled to the values in \eq{QCDscales}). We will show below how to generalize it to have free scales $\mu_{H,J,S}$ as represented in the top left box in \fig{commute}.

Before comparing the cross sections in QCD and SCET, we note that, using the definitions in \eq{constantdefs}, we can write the function $E(\ln \nu_a)$ as
\begin{align}
\label{eq:Efixedscales}
E(\ln \nu_a) &= -2j_J \kappa_J \int_{Q (e^{\gamma_E} \nu_a)^{-1/j_J}}^Q \frac{d\mu'}{\mu'} \, A[\as] \ln \frac{\mu'}{Q(e^{\gamma_E} \nu_a)^{-1/j_J}} \nn \\
& \quad - j_S \kappa_S \int_{Q (e^{\gamma_E} \nu_a)^{-1/j_S}}^Q \frac{d\mu'}{\mu'} \, A[\as] \ln \frac{\mu'}{Q(e^{\gamma_E} \nu_a)^{-1/j_S}} \nn \\
& \quad + 2\int_{Q (e^{\gamma_E} \nu_a)^{-1/j_J}}^Q \frac{d\mu'}{\mu'} \, B_J [\as]  + \int_{Q (e^{\gamma_E} \nu_a)^{-1/j_S}}^Q \frac{d\mu'}{\mu'} \, B_S [\as] \,.
\end{align}
This will make it easier to compare to the SCET resummed form.

%%%%%%%%%%%%%%%%%%%%%%%%%%%%%%%%%%%%%%%%%%%%%%%%%
\subsection{Equivalence Between QCD and SCET Resummation}
\label{ssec:resumequiv}
%%%%%%%%%%%%%%%%%%%%%%%%%%%%%%%%%%%%%%%%%%%%%%%%%

A natural question to ask is, do the SCET and QCD cross sections \eqs{SCETresummedcumulant}{QCDRderivatives} return the same result for the same observable?  One answer is that they had better when working to all orders, or someone is doing something wrong.  But a more pragmatic (and important) question is, if we truncate to a given order in logarithm counting do the two methods still agree?  This question has many possible answers.  QCD and SCET may exactly agree, given some conventions for how to truncate each resummed series at a given order.  They  may agree on the terms up to the required accuracy but naturally include different sets of subleading terms.  They may disagree at the requested level of accuracy (pointing to a fundamental problem in one of the methods). We will show, that written in suitable form, the two formalisms agree exactly. The process of this rewriting will illuminate and improve the results of both methods.

A detailed comparison between the two resummed cross sections above will only make these questions more pointed: the two results \eqs{SCETresummedcumulant}{QCDRderivatives} look \textit{quite} similar in form, and one might believe that appropriate relations between the QCD $A[\as]$ and $B_i[\as]$ functions and the SCET $\Gcusp[\as]$ and $\gamma_i[\as]$ anomalous dimensions would put them in complete agreement.  In fact, this is what we will find in this section.  We start by building the basic correspondence between the two forms, showing that at all orders the two cross sections agree.  Analogous but distinct considerations apply to the analysis  of threshold resummation \cite{Becher:2006mr,Becher:2006nr,Bonvini:2012az,Bonvini:2013td,Sterman:2013nya} when treated in moment space.   In fact the correspondence between SCET and dQCD forms requires a more elaborate analysis for event shapes than for threshold resummation in moment space, but we will see that the resulting physical predictions for event shapes are analytically equivalent, in a way that has not been found for threshold resummed cross sections.

We will work mostly with the SCET formalism, primarily for the explicit dependence on free scales $\mu_F$ that they display. We will massage the SCET form into one easily relatable to the QCD cross section, in particular identifying the scale choices $\mu_F$ that make the SCET and QCD forms transparently equivalent.  To be concrete, we begin first with the form of the resummed jet function in the two formalisms.

From \eq{Fevol} or \eq{Fevolderivatives} we already know how to express the jet (or soft) function  at one scale $\mu$ by RG evolving from another scale $\mu_J$ ($\mu_S$). However, by performing a series of manipulations we can re-express \eqs{Fevol}{Fevolderivatives} in a form that makes the connection with direct QCD in \eq{QCDRderivatives} as transparent as possible and also exponentiates as many logs as possible. For instance, there are series of logs in the prefactor of \eq{Fevolderivatives} in $\wt F(\partial_{\omega_F},\mu_0)$, which are always truncated at some order in $\as$, generated by the action of the derivatives $\partial_{\omega_F}^n$ acting on the gamma function. We will put \eq{Fevolderivatives} in a form that has as many of these logs explicitly exponentiated as possible, leaving the prefactor free of logs.

\begin{figure}[t]
\begin{center}
\hspace{10em}\includegraphics[width=.5\textwidth]{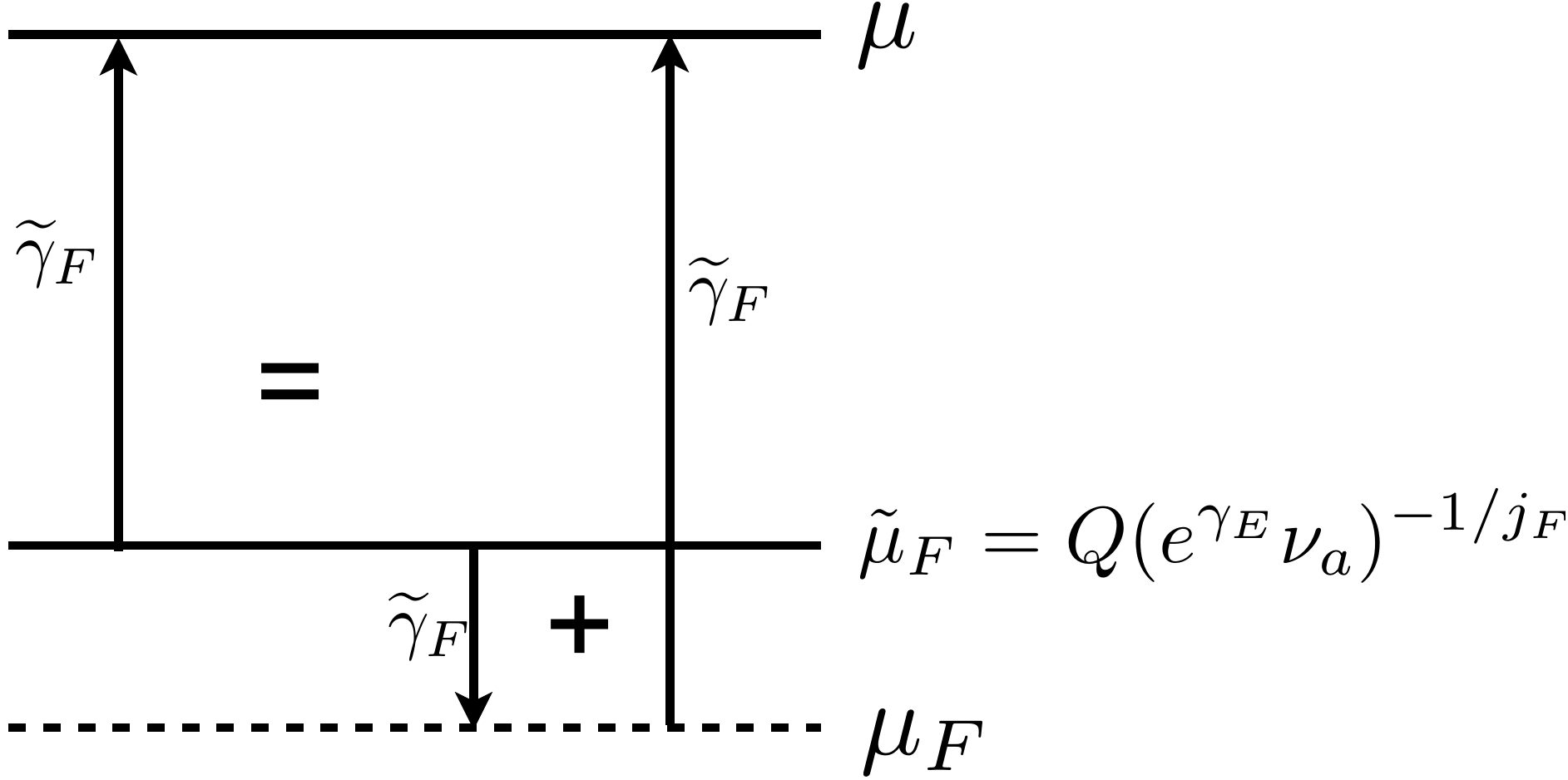}
\vspace{-2em}
\end{center}
\caption{Evolution from the natural scale. In \eq{jetLaplace}, we express a function $\wt F(L_F(\mu),\mu)$ at the factorization scale $\mu$ by performing the RG evolution from its natural scale $\tilde\mu_F$ where all the large logs in the function are minimized. We further break this evolution into two pieces, one from the natural scale $\wt\mu_F$ to another arbitrary scale $\mu_F$, and from $\mu_F$ to $\mu$. We may choose to vary the arbitrary scale $\mu_F$ in order to estimate theoretical uncertainty in a prediction for $F$ at finite resummed accuracy.
\label{fig:Fevolution}}
\end{figure}
We start with the evolution from the natural scale for jet functions in transform space,
\be
\wt{\mu}_J = Q (e^{\gamma_E} \nu_a)^{-1/j_J} \,.
\ee
We then break the evolution into two pieces, that from $\wt{\mu}_J$ to $\mu_J$ and that from $\mu_J$ to $\mu$, illustrated in \fig{Fevolution}. Working from \eqs{Fevol}{Ftilderewriting}, we have,
\be
\label{eq:jetLaplace}
\wt{J} (L_J(\mu), \mu) = \wt{J} (0, \wt{\mu}_J) \exp \biggl( \int_{\tilde{\mu}_J}^{\mu_J} \frac{d\mu'}{\mu'} \, \wt{\gamma}_J (x_a, \mu') \biggr) \exp \biggl( \int_{\mu_J}^{\mu} \frac{d\mu'}{\mu'} \, \wt{\gamma}_J (x_a, \mu') \biggr) \,,
\ee
and similarly for the soft function, for which the discussion below would be entirely parallel.
Note that $L_J(\wt{\mu}_J) = 0$; hence the first argument of $\wt J$ on the right-hand side is 0.  

Our goal is to express each piece in \eq{jetLaplace} in terms of the variable scale $\mu_J$.
The evolution from $\mu_J$ to $\mu$ is straightforward, in the notation of \eqs{Fevol}{LFdef}:
\be
\exp \biggl( \int_{\mu_J}^{\mu} \frac{d\mu'}{\mu'} \, \wt{\gamma}_J (x_a, \mu') \biggr) = \exp \bigl[ K_J (\mu, \mu_J) \bigr] \exp \bigl[ L_J (\mu_J) \omega_J (\mu, \mu_J) \bigr] \,.
\ee
In dealing with the evolution from $\wt{\mu}_J$ to $\mu_J$, we will use derivatives with respect to $\omega_J$ to generate $L_J$ dependence.  We express each scale in terms of $L_J \equiv L_J(\mu_J)$ and $L_J' \equiv L_J (\mu')$ 
\begin{align}
\wt{\mu}_J &= \mu_J e^{ -L_J / j_J} \,, \nn \\
\mu' &= \wt{\mu}_J e^{L_J' / j_J} = \mu_J e^{ (L_J' - L_J) / j_J }\,.
\end{align}
Changing variables from $\mu'$ to $u = L_J' / L_J$, the evolution from $\wt{\mu}_J$ to $\mu_J$ is
\begin{align}
\exp \biggl( \int_{\wt{\mu}_J}^{\mu_J} \frac{d\mu'}{\mu'} \, \wt{\gamma}_J (x_a, \mu') \biggr) &= \exp \biggl( \int_0^1 du \, \biggl\{ - j_J \kappa_J \Gcusp^q \bigl[ \as \bigl(\mu_J e^{ (u-1) L_J / j_J } \bigr) \bigr] \frac{L_J^2}{j_J^2} \, u \nn \\
& \qquad\qquad\qquad\quad + \gamma_J \bigl[ \as \bigl(\mu_J e^{ (u-1) L_J / j_J } \bigr) \bigr]  \frac{L_J}{j_J} \biggr\} \biggr) \,.
\end{align}
This form is convenient to Taylor expand about $u = 1$, that is, $\mu'=\mu_J$, allowing us then to carry out the $u$ integral explicitly.  After some analysis we obtain
\begin{align}
& - j_J \kappa_J \int_0^1 du \, \Gcusp^q \bigl[ \as \bigl(\mu_J e^{ (u-1) L_J / j_J } \bigr) \bigr] \frac{L_J^2}{j_J^2} \, u \\
& \qquad\qquad\qquad\qquad\qquad = -j_J \kappa_J \sum_{n=2}^{\infty} \frac{1}{n!} \biggl( - \frac{L_J}{j_J} \biggr)^n \frac{d^n}{d(\ln\mu_J)^n} \int_{\mu_J}^{\mu} \frac{d\mu'}{\mu'} \Gcusp[\as(\mu')] \ln \frac{\mu'}{\mu_J} \,, \nn \\
& \int_0^1 du \, \gamma_J \bigl[ \as \bigl(\mu_J e^{ (u-1) L_J / j_J } \bigr) \bigr]  \frac{L_J}{j_J} = \sum_{n=1}^{\infty} \frac{1}{n!} \biggl( - \frac{L_J}{j_J} \biggr)^n \frac{d^n}{d(\ln\mu_J)^n} \int_{\mu_J}^{\mu} \frac{d\mu'}{\mu'} \gamma_J [\as(\mu')] \,. \nn
\end{align}
Additionally, one can show by Taylor expansion around $\wt \mu_J = \mu_J$,
\begin{align}
\wt{J} (0, \wt{\mu}_J) &= \wt{J} (0, \mu) \frac{\wt{J} (0, \wt{\mu}_J)}{\wt{J} (0, \mu)} \nn \\
&= \wt{J} (0, \mu) \exp \biggl\{ \sum_{n=0}^\infty \frac{1}{n!} \biggl( - \frac{L_J}{j_J} \biggr)^n \frac{d^n}{d(\ln\mu_J)^n} \biggl[ -\int_{\mu_J}^\mu \frac{d\mu'}{\mu'} \frac{d\ln \wt{J} (0,\mu')}{d\ln\mu'} \biggr] \biggr\} \,.
\end{align}
If we put the various pieces of the evolution together, we obtain the jet function in \eq{jetLaplace} at arbitrary $\mu$,
\be
\label{eq:jetLaplaceexp}
\wt J (L_J(\mu),\mu) = \wt J(0,\mu)\exp \biggl[ E_J(\mu,\mu_J) + \sum_{n=2}^\infty \frac{1}{n!} \Bigl( - \frac{L_J}{j_J}\Bigr)^n \frac{d^nE_J(\mu,\mu_J)}{d(\ln\mu_J)^n} \biggr]  \exp \biggl[ -\frac{L_J}{j_J} \frac{d E_J(\mu,\mu_J)}{d\ln\mu_J}\biggr]\,,
\ee
where we have defined 
\be
\label{eq:EJdef}
E_J(\mu,\mu_J) \equiv K_J (\mu, \mu_J) - \int_{\mu_J}^\mu \frac{d\mu'}{\mu'} \frac{d\ln \wt{J} (0,\mu')}{d\ln\mu'} \,.
\ee
Note that the second term, and thus the difference between $E_J$ and $K_J$, begins at $\cO(\as^2\ln(\mu_J/\mu))$, which is NNLL in \tab{Laplace}.
Defining the derivatives
\be
\label{eq:EJprime}
E_J'(\mu,\mu_J) = \frac{d E_J(\mu,\mu_J)}{d\ln\mu_J} \,,\quad  E_J^{(n)}(\mu_J) =  \frac{d^nE_J(\mu,\mu_J)}{d(\ln\mu_J)^n}\,,
\ee
(note that taking two or more derivatives of $E_J$ with respect to $\mu_J$ removes its dependence on $\mu$) we find that \eq{jetLaplaceexp} can be expressed compactly as
\be
\label{eq:jetLaplacederivatives}
\wt J (L_J(\mu),\mu) = \wt J(0,\mu)\exp \biggl[ E_J(\mu,\mu_J) + \sum_{n=2}^\infty \frac{1}{n!}  E_J^{(n)}(\mu_J) \partial_{E_J'}^n \biggr]  \exp \biggl[ -\frac{L_J(\mu_J)}{j_J} E_J'(\mu,\mu_J)\biggr]\,.
\ee
Now, recalling that $L_J(\mu_J) = \ln(\mu_J^{j_J} e^{\gamma_E} x_a)$, we can inverse Laplace transform \eq{jetLaplacederivatives} from $x_a $ back to $t_a$ and obtain in momentum space:
\be
\label{eq:jetEderivatives}
J(t_a,\mu) = \wt J(0,\mu) \exp\biggl[ E_J(\mu,\mu_J) + \sum_{n=2}^\infty \frac{1}{n!} E_J^{(n)}(\mu_J) \partial_{E_J'}^n\biggr] \frac{1}{t_a} \Bigl( \frac{\mu_J^{j_J} e^{\gamma_E}}{t_a}\Bigr)^{-E_J'(\mu,\mu_J)/j_J} \frac{1}{\Gamma(E_J'/j_J)}\,.
\ee
We have succeeded in expressing $\wt J(L_J(\mu),\mu)$ or $J(t_a,\mu)$ at one scale $\mu$ in terms of RG evolution from any other scale $\mu_J$, with no explicit logs left over in the factor $\wt J(0,\mu)$, which must be truncated at fixed order in $\as$ in \eqs{jetLaplacederivatives}{jetEderivatives}. They are all contained in the exponentiated evolution kernels. (The exponentiated derivative operators must, however, still be truncated at some fixed order in practice; still, the exponentiated form gives an easy prescription that can be evaluated to any desired order.)

Similarly for the soft function, by analogy to \eq{jetEderivatives}
\be
\label{eq:softEderivatives}
S(k,\mu) = \wt S(0,\mu) \exp\biggl[ E_S(\mu,\mu_S) + \sum_{n=2}^\infty \frac{1}{n!} E_S^{(n)}(\mu_S) \partial_{E_S'}^n\biggr] \frac{1}{k}\Bigl( \frac{\mu_S e^{\gamma_E}}{k}\Bigr)^{-E_S'(\mu,\mu_S)} \frac{1}{\Gamma(E_S')}\,,
\ee
where
\begin{subequations}
\label{eq:ESdef}
\begin{gather}
E_S(\mu,\mu_S) \equiv K_S (\mu, \mu_S) - \int_{\mu_S}^\mu \frac{d\mu'}{\mu'} \frac{d\ln \wt{S} (0,\mu')}{d\ln\mu'} \,, \\
E_S'(\mu,\mu_S) = \frac{d E_S(\mu,\mu_S)}{d\ln\mu_S} \,,\quad  E_S^{(n)}(\mu_S) =  \frac{d^nE_S(\mu,\mu_S)}{d(\ln\mu_S)^n}\,.
\end{gather}
\end{subequations}
Finally, we can obtain the total cross section by starting with the Laplace-transformed factorization theorem, obtained by transforming \eq{factorization}:
\be
\wt\sigma (\nu,\mu) = \int_0^\infty d\tau_a e^{-\nu\tau_a} \frac{1}{\sigma_0}\frac{d\sigma}{d\tau_a} =  H_2(Q^2,\mu) \wt J (L_J(\mu),\mu)^2 \wt S(L_S(\mu),\mu)\,,
\ee
plugging in the form \eq{jetLaplaceexp} of the jet function and similarly for the soft function, Laplace transforming all at once from $\nu$ back to $\tau_a$, and integrating over $\tau_a$ to obtain the cumulant:
\begin{align}
\label{eq:SCETcsE}
R(\tau_a) &= H_2(Q^2,\mu_H) \Bigl(\frac{\mu_H}{Q}\Bigr)^{\omega_H(\mu,\mu_H)} e^{K_H(\mu,\mu_H)} \wt J(0,\mu)^2 \wt S(0,\mu) \nn\\
&\quad \times \exp[ 2E_J(\mu,\mu_J) + E_S(\mu,\mu_S)]   \exp\biggl[ \sum_{n=2}^\infty \frac{1}{n!} \Bigl( 2 E_J^{(n)}(\mu_J) \partial_{2E_J'}^n + E_S^{(n)}(\mu_S) \partial_{E_S'}^n \Bigr) \biggr]  \nn \\
&\quad \times \Bigl( \frac{Q^{j_J}\tau_a}{\mu_J^{j_J} e^{\gamma_E}}\Bigr)^{2E_J'(\mu,\mu_J)/j_J} \Bigl( \frac{Q\tau_a}{\mu_S e^{\gamma_E}}\Bigr)^{E_S'(\mu,\mu_S)}  \frac{1}{\Gamma(1+2E_J'/j_j + E_S')}\,.
\end{align}
This generalizes the NLL result Eq.~(85) of \cite{Catani:1992ua} and those of \cite{Berger:2003iw,Berger:2003pk} to arbitrarily high accuracy and restores its dependence on variable hard, jet and soft scales $\mu_{H,J,S}$.

We can show that the SCET cross section now written in the form \eq{SCETcsE} and the QCD cross section \eq{QCDRderivatives} are identical if some identifications and particular scale choices are made.  In \eq{SCETcsE}, let us choose to run the hard, jet, and soft functions from the scales
\be
\label{eq:barcanonicalscales}
\mu_H = Q\,,\quad \bar\mu_J = Q\bar\tau_a^{1/j_J}\,,\quad \bar\mu_S = Q \bar\tau_a\,,
\ee
where $\bar\tau_a = e^{-\gamma_E}\tau_a$ as defined in \eq{taubar}, to the scale $\mu = Q$. Then we have the expression \cite{Contopanagos:1993xh}
\be
\label{eq:SCETcsEfinal}
\begin{split}
R(\tau_a) &= H_2(Q^2,Q) \wt J(0,Q)^2 \wt S(0,Q) \exp\biggl[ \bar E + \sum_{n=2}^\infty \frac{1}{n!}  \bar E^{(n)} \partial_{\bar E'}^n   \biggr]   \frac{1}{\Gamma(1-\bar E')}\,,
\end{split}
\ee
with
\begin{subequations}
\label{eq:Edefs}
\begin{align}
\bar E &\equiv 2\bar E_J + \bar E_S \\
\bar E' &\equiv \frac{d\bar E}{d\ln(1/\tau_a)} \\
\bar E^{(n)} &\equiv \frac{d^n\bar E }{d(\ln(1/\tau_a))^n}
\end{align}
\end{subequations}
where $\bar E_{J,S} \equiv E_{J,S}(Q,\bar \mu_{J,S})$.  Note that the derivatives in \eq{Edefs} are now with respect to $\ln(1/\tau_a)$ instead of $\ln\mu_{J,S}$ as in \eq{EJprime}.

Remarkably, the SCET cross section in \eq{SCETcsEfinal} is now exactly in the form of the QCD cross section \eq{QCDRderivatives}. They are equal if we identify
\begin{align}
\label{eq:AB}
A[\as] &= \Gcusp[\as] \,, \qquad B_F[\as] = \gamma_F [\as] - \frac{d\ln \wt{F} (0, \mu_F)}{d \ln \mu_F} \,, \nn \\
\ca{N} (Q) &= H(Q^2,Q) \wt{J} (0, Q)^2 \wt{S} (0, Q) \,.
\end{align}
Similar identifications were found in Ref.~\cite{Becher:2006mr,Becher:2006nr,Bonvini:2012az,Bonvini:2013td,Sterman:2013nya} for threshold resummation in Drell-Yan production. 

The identification is not without significance for direct QCD: with these identifications, \eq{SCETcsE} also serves to generalize the usual QCD direct resummation formulae by endowing them with free scales $\mu,\mu_{H,J,S}$ and thus the flexibility to vary these scales away from the values in \eq{barcanonicalscales}.  This allows one to match the resummation onto fixed-order perturbation theory using profile scales for $\mu_{J,S}$ to smoothly interpolate between the resummation and fixed-order regimes, and in general allows one to obtain better estimates of theoretical uncertainty.

We note that in \eq{SCETcsEfinal}, 
\be
\label{eq:Eprime}
\bar E' = \bar\Omega + \frac{2B_J[\as(\bar\mu_J)]}{j_J} + B_S[\as(\bar\mu_S)]\,,
\ee
where $\bar\Omega = 2\omega_J(Q,\bar\mu_J) + \omega_S(Q,\bar\mu_S)$. This looks like the argument of the gamma function in the form of the SCET distribution shown in \eq{SCETresummeddist}, shifted by $2B_J/j_J + B_S$. This shift is the result of resumming many of the $\partial_\Omega$-dependent terms in the jet and soft operators in \eq{SCETresummeddist}, which is effectively what has been done by resumming all the $\mu_J$-dependent terms in the jet function into the exponential factors in \eq{jetLaplacederivatives}, and similarly for the soft function.  Also, note that the choices \eq{barcanonicalscales} for jet and soft scales differ from the usual choices in SCET by rescalings of $\tau_a$ to $\bar \tau_a \equiv e^{-\gamma_E}\tau_a$. This just causes the factor of $e^{\gamma_E\Omega}$ in the SCET resummed distribution \eq{SCETresummeddist}  to be absorbed into $K_{J,S}$. Of course, the different scale choices  lead to equivalent results at a given order of accuracy.  

Finally, we note that the resummed form in \eq{SCETcsEfinal} with the non-cusp and $\beta$-function terms resummed into the gamma function is, to our knowledge, new for SCET. Thus it formally resums more terms to all orders in $\as$ than most previous formulae in the SCET literature, capturing more of the non-cusp anomalous dimension and beta function-dependent terms to all orders in $\as$, making it more accurate in the sense of maintaining as closely as possible its equivalence to the inverse Laplace transform of $\widetilde\sigma(\nu)$.

\subsection{Interlude}

In \ssec{resumequiv} we obtained the central result of the paper, \eq{SCETcsEfinal} which puts the SCET resummed cross section into a form transparently equivalent to the dQCD form \eq{QCDRderivatives}. Before moving on, let us take stock of what we have found, and what is yet to come.
\begin{itemize}
\item \eq{SCETcsE} is a rewriting of the usual form of the SCET cumulant cross section \eq{SCETresummedcumulant}. Making the scale choices \eq{barcanonicalscales} in the SCET form \eq{SCETcsE} yields the form \eq{SCETcsEfinal}, which is transparently equivalent to the dQCD form \eq{QCDRderivatives}. \eq{EJdef} for $E_J$ and \eq{ESdef} for $E_S$, along with the definitions \eq{Edefs} and identifications \eq{AB} complete the dictionary for this equivalence.

\item The form \eq{SCETcsE} prior to fixing scales $\mu_{H,J,S}$ shows how to restore the dependence on these scales in the dQCD cross section \eq{QCDRderivatives}, in which the scales are implicitly fixed to the values in \eq{QCDscales}. This is to our knowledge  the first time the final result for the momentum-space distribution has been written with variable scales in dQCD notation, and with all derivative terms $n\geq 2$ written explicitly.

\item The forms \eqs{SCETcsE}{SCETcsEfinal} differ from the typical SCET form \eq{SCETresummedcumulant} due to the additional non-cusp anomalous dimension and beta function dependent terms that are included in $E_{J,S}$ in \eqs{EJdef}{ESdef} and are now all in the exponent or argument of the gamma function in \eqs{SCETcsE}{SCETcsEfinal}.\footnote{A similar exponentiated form for the momentum-space SCET soft function for $t\bar t $ production in hadron collisions was also found in \cite{Ahrens:2011mw}.  Also, Refs.~\cite{Bonvini:2012az,Bonvini:2013td} observed in the context of threshold resummation that inclusion of more terms in the SCET exponent as we did in \eqs{EJdef}{ESdef} would bring SCET and dQCD forms into closer agreement. Our formulae generalize these observations to all orders for the case of event shape distributions.} The prefactors in \eq{SCETcsEfinal} are free of logs, and can be truncated at fixed-order without losing factors containing logs of $\tau_a$. While the extra terms are formally subleading, they help maintain closer equivalence to the Laplace transform to higher orders in $\as$.
\end{itemize}

Now there remains the question of how to evaluate each piece of \eq{SCETcsEfinal} to achieve \nkll or \nkllprime  resummed accuracy. We will show in \sec{schemes}:
\begin{itemize}
\item It is easiest to define resummed accuracy in terms of the Laplace transform \eq{Laplaceexpsimple} since it exponentiates simply in terms of the exponent $\bar E$ for which a straightforward prescription can be given for its computation. We will adopt the common definition of \nkll accuracy in terms of the order to which $\bar E$ is computed in \eq{Laplaceexpsimple}.

\item We will define prescriptions for computing the cumulant $R(\tau)$ or distribution $\sigma(\tau)$ to \nkll accuracy so that they remain at least as accurate as $\wt\sigma(\nu)$ computed to \nkll accuracy.

\item For $R(\tau)$ these prescriptions make it preferable to keep at least the $n=2$ term in the sum of derivative operators in the exponent of \eq{SCETcsEfinal} at NLL accuracy, $n=3$ at NNLL, etc. The exponential of these operators can itself be truncated at $\cO(\as)$ at NLL, $\cO(\as^2)$ at NNLL, etc. This is one higher order than is often kept. We will show why this prescription does better in maintaining equivalent accuracy with the Laplace transform $\wt\sigma(\nu)$. In the standard SCET form \eq{SCETresummedcumulant}, the appropriate prescription requires keeping the differential operator terms in $\wt J,\wt S$ to at least $\cO(\as)$ for NLL, $\cO(\as^2)$ for NNLL, etc.

\item For $\sigma(\tau)$, we will show that the two procedures of computation, 1)  directly from \eq{SCETresummeddist} and then fixing scales $\mu_F$, or  2) by differentiating $R(\tau)$ in \eq{SCETresummedcumulant} after fixing scales $\mu_F$, may give different answers when standard rules for computing to \nkll accuracy are applied. To ensure that procedure 1 gives as accurate an answer as procedure 2, we will show that the differential operator terms in $\wt J,\wt S$ must \emph{not} be truncated according to standard rules too soon, namely when $\sigma(\tau)$ is still written in the form \eq{SCETresummeddist}. Instead, we will show it is expedient to pull a factor of $-\Omega$ through the $\wt J,\wt S$ operators in \eq{SCETresummeddist} first, and then keep a subset of the extra terms generated by the action of the $\partial_\Omega$ derivatives on this factor. At \nkllprime order, though, this is not strictly necessary.
\end{itemize}
The reader who wishes to skip to the final results for formulae for the Laplace transform $\wt\sigma(\nu)$, cumulant $R(\tau)$, and differential distribution $\sigma(\tau)$, and the prescriptions for how to evaluate each properly to \nkll or \nkllprime accuracy can now turn to \sec{summary}. First, in \sec{schemes}, we will develop and justify these prescriptions in detail.

%%%%%%%%%%%%%%%%%%%%%%%%%%%%%%%%%%%%%%%%%%%%%%%%%%%%%%%%%%%%%%%%%%%%%%%%%%%%%%%%%%%%%%%%%%%%%%%%%
\section{Achieving \nkll Logarithmic Accuracy}
\label{sec:schemes}
%%%%%%%%%%%%%%%%%%%%%%%%%%%%%%%%%%%%%%%%%%%%%%%%%%%%%%%%%%%%%%%%%%%%%%%%%%%%%%%%%%%%%%%%%%%%%%%%%%%%%%%%%

In \eq{Laplaceexponent} we defined what \nkll accuracy means, in terms of the number of logs in the exponent of the Laplace transform $\wt\sigma(\nu)$ that are known. In this section, we consider how to actually achieve \nkll accuracy in practice. We review the procedure for the Laplace transform of the distribution $\wt\sigma(\nu)$, and the consider how to compute the cumulant $R(\tau)$ or differential event shape distribution $\sigma(\tau)$ to the equivalent accuracy.  The formalism reviewed in the previous section allows this to be done systematically by calculating anomalous dimensions $\gamma_F$ and hard, jet, and soft functions  $H,J_{n,\bn},S$ order by order in $\as$. We will clarify to what order in $\as$ each of these quantities must be computed to achieve a given order of logarithmic accuracy. The procedure is standard and straightforward for $\wt\sigma(\nu)$. It will turn out that usage of standard formulae for $R(\tau)$ and especially $\sigma(\tau)$ require special care in how they are used (namely, how different parts of them are truncated to finite order in $\as$) to preserve the same level of accuracy as $\wt\sigma(\nu)$. We will explain how to do this properly, and clarify some confusion that can easily arise from a casual reading of the existing literature.

Our aim in this section is to be rather pedagogical and give as clear explanations as possible for the prescriptions we present for evaluating $\wt\sigma(\nu)$, $R(\tau)$, and $\sigma(\tau)$ to \nkll or \nkllprime accuracy. The level of detail is, accordingly, quite high. As we mentioned above, the reader who wishes to skip over it to reach the final results should turn directly to \sec{summary}.

\subsection{Laplace transform}

The RG-evolved Laplace-transformed cross section $\wt\sigma(\nu)$ obtained from the factorization theorem \eq{factorizationx} by using the methods in \ssec{SCETevol} takes the explicit form given in \eq{SCETresummedLaplace},
\begin{equation}
\label{eq:sigmanuresummed}
\begin{split}
\wt\sigma(\nu) &=  e^K H(L_H,\mu_H)\wt J(L_J,\mu_J )^2 \wt S(L_S,\mu_S) \\ 
&\quad \times \Bigl(\frac{\mu_H}{Q}\Bigr)^{\! \omega_H(\mu,\mu_H)} \!\!\biggl(\frac{\mu_J (e^{\gamma_E}\nu)^{1/j_J}}{Q }\biggr)^{\!2j_J\omega_J(\mu,\mu_J)}\!\!\left(\frac{\mu_Se^{\gamma_E}\nu}{Q}\right)^{\!\omega_S(\mu,\mu_S)} \! ,
\end{split}
\end{equation}
or, expressing the logs in the jet and soft functions as the result of derivatives,
\begin{equation}
\label{eq:sigmanuderivative}
\begin{split}
\wt\sigma(\nu) &=  e^K H(L_H,\mu_H)\left(\frac{\mu_H}{Q}\right)^{\omega_H(\mu,\mu_H)} \left(\frac{\mu_J \nu^{1/j_J}}{Q}\right)^{2j_J\omega_J(\mu,\mu_J)}\left(\frac{\mu_S\nu}{Q}\right)^{\omega_S(\mu,\mu_S)-\Omega} \\
&\quad\times \wt J\Bigl(\partial_\Omega + \ln \frac{\mu_J^{j_J}}{Q^{j_J-1} \mu_S},\mu_J \Bigr)^2 \wt S(\partial_{\Omega},\mu_S)  \left(\frac{\mu_S e^{\gamma_E}\nu}{Q}\right)^{\Omega} \,.
\end{split}
\end{equation}
This form will be easier to inverse Laplace transform to the cumulant or distribution in $\tau_a$. In \eqs{sigmanuresummed}{sigmanuderivative}, we have defined the combined RG evolution kernels, 
\begin{subequations}
\label{eq:KOmega}
\begin{align}
K &\equiv  K_H(\mu,\mu_H) + 2K_J(\mu,\mu_J) + K_S(\mu,\mu_S) \\
\Omega &\equiv  2\omega_J(\mu,\mu_J) + \omega_S(\mu,\mu_S)\, ,
\end{align}
\end{subequations}
with the functions $K_F$ and $\omega_F$ defined in Eq.\ (\ref{eq:Komegadef}).
The hard, jet, and soft functions have fixed-order expansions of the form
\be
 F(L_F,\mu_F)  = \sum_{n=0}^\infty \left(\frac{\as(\mu_F)}{4\pi}\right)^n  F_n(L_F,\mu_F)\,,
\ee
where $F=H,\wt J, \wt S$, and to order $\as^2$ the coefficients $F_n$ are given by, using the evolution of the functions $F$ in \eqs{gammaFLaplace}{Fevolexp},
\begin{subequations}
\label{eq:Ffixedorder}
\begin{align}
F_0  &= 1 \\
F_1 &= \frac{\Gamma_F^0}{j_F^2} L_F^2 + \frac{\gamma_F^0}{j_F} L_F + c_F^1 \\
F_2 &= \frac{1}{2j_F^4} (\Gamma_F^0)^2 L_F^4 + \frac{\Gamma_F^0}{j_F^3} \left( \gamma_F^0 + \frac{2}{3}\beta_0\right) L_F^3   \\
&\quad + \frac{1}{j_F^2}\left(\Gamma_F^1 + \frac{1}{2}(\gamma_F^0)^2 + \gamma_F^0\beta_0 + c_F^1 \Gamma_F^0 \right)L_F^2 + \frac{1}{j_F}(\gamma_F^1 + c_F^1\gamma_F^0 + 2c_F^1 \beta_0) L_F + c_F^2 \,, \nn
\end{align}
\end{subequations}
where for $F=H$, $L_H\equiv \ln(\mu_H/Q)$ and for $F=\wt J,\wt S$, $L_F = \ln(\mu_F^{j_F} e^{\gamma_E} \nu/Q^{j_F})$. In \eq{sigmanuderivative}, for the jet and soft functions, each $L_F$ gets replaced by the differential operator shown in the argument of $\wt J,\wt S$.  Recall $j_J = 2-a$ while $j_H=j_S =1$. The quantities $\Gamma_F^n$ are the coefficients in the expansion in $\as$ of $\Gamma_F$ in \eq{GammaFdef}, 
 \be
 \Gamma_F[\as(\mu_F)] = \sum_{n=0}^\infty \left(\frac{\as(\mu_F)}{4\pi}\right)^n  \Gamma_F^n\, ,
 \ee
 and similarly for $\gamma_F$ in Eq.\ (\ref{eq:noncuspexpansion}).

The expression \eq{sigmanuderivative} for $\wt\sigma(\nu)$ contains logarithms of $\mu_H/Q$ and $\mu_F(e^{\gamma_E}\nu)^{1/j_F}/Q$, which can be minimized in $\nu$ space by the choices of scales
\be
\label{eq:nuscalechoices}
\mu=\mu_H = Q\,,\quad \mu_F = \wt \mu_F \equiv Q(e^{\gamma_E}\nu)^{-1/j_F}\,.
\ee
Then $\wt\sigma(\nu)$ takes the particularly simple form,
\be
\label{eq:sigmanusimple}
\wt\sigma(\nu) =  H \wt{J}^2 \wt S e^K \,,
\ee
where $F=H, \wt J, \wt S$ now have the simple log-free expansions
\begin{equation}
F = 1 + \frac{\as(\wt\mu_F)}{4\pi} c_F^1 + \left(\frac{\as(\wt\mu_F)}{4\pi}\right)^2 c_F^2 + \cdots\, .
\end{equation}
\eq{sigmanusimple} organizes terms in $\wt\sigma(\nu)$ into a form much like the CTTW form of the radiator \eq{CTTWradiator}.  The exponent $K$ with the scale choices \eq{nuscalechoices} reduces to 
\be
K = 2K_J(Q,Q s^{1/j_J}) + K_S(Q,Qs),
\ee
where $s\equiv e^{\gamma_E}\nu$. 

\begin{table}
\begin{center}
$
\begin{array}{|c|c|c|c|c|}
\hline
K_{nm} & m=n+1 & m=n & m=n-1 & m=n-2  \\ \hline
n=1 & \frac{1}{2}\Gamma_0 &  & &  \\ \hline
n=2 & \frac{1}{3}\Gamma_0 \beta_0 & \frac{1}{2}\Gamma_1 & & \\ \hline
n=3 & \frac{1}{3}\Gamma_0 \beta_0^2 & (\Gamma_0\beta_1 + 2\Gamma_1\beta_0 ) & \frac12\Gamma_2  & \ \\ \hline
\vdots & \vdots & \vdots & \vdots & \vdots \\ \hline
\text{accuracy} & \text{LL} & \text{NLL} & \text{NNLL} & \text{N$^3$LL} \\ \hline
\end{array}
$
\end{center}
\vspace{-1em}
\caption{Coefficients $K_{nm}$ in the expansion \eq{KFexpansion}  $K_\Gamma = \sum_{n,m} K_{nm} (\as/4\pi)^n \ln^m s^{1/j_F}$ up to $n=3$. A closed form of $K_\Gamma$ to N$^3$LL accuracy is given in \eq{KFclosedform}.}
\label{tab:Ktable}
\end{table}
Each function $K_F$ above is given by \eq{Komegadef} in terms of $K_\Gamma$ and $K_{\gamma_F}$,  whose fixed-order expansions take the form determined by the integrals in Eq.\ (\ref{eq:Ketadef}),
\begin{align}
\label{eq:KFexpansion}
K_\Gamma(Q,Q/s^{1/j_F})  = \  \left(\!\frac{\as(Q)}{4\pi} \!\right) \  \Bigl( K_{12} \ln^2 s^{1/j_F} &+ K_{11} \ln s^{1/j_F} \Bigr) \\
+\left(\!\frac{\as(Q)}{4\pi} \!\right)^2  \Bigl( K_{23} \ln^3 s^{1/j_F} &+ K_{22} \ln^2 s^{1/j_F} + K_{21} \ln s^{1/j_F} \Bigr) \nn \\
+\left(\!\frac{\as(Q)}{4\pi} \!\right)^3  \Bigl( K_{34} \ln^4 s^{1/j_F} &+ K_{33} \ln^3 s^{1/j_F} + K_{32} \ln^2 s^{1/j_F} + K_{31}\ln s^{1/j_F} \Bigr)  \nn \\
+ \cdots \qquad\qquad\qquad\qquad\quad \  & \nn
\end{align}
where the coefficients $K_{nm}$ are given in \tab{Ktable}, and
\begin{align}
\label{eq:kexpansion}
K_{\gamma_F}(Q,Q/s^{1/j_F}) = \biggl[\left(\!\frac{\as(Q)}{4\pi}\!\right) \biggl( & k_{11} \ln s^{1/j_F} \biggr) \\
+\left(\!\frac{\as(Q)}{4\pi}\!\right)^2 \biggl( & k_{22} \ln^2 s^{1/j_F} + k_{21} \ln s^{1/j_F} \biggr) \nn \\
+\left(\!\frac{\as(Q)}{4\pi}\!\right)^3 \biggl( & k_{33} \ln^3 s^{1/j_F} + k_{32} \ln^2 s^{1/j_F} + k_{31} \ln s^{1/j_F}  \biggr) + \cdots \biggr] \,, \nn
\end{align}
where the coefficients $k_{nm}$ are given in \tab{ktable}.
\begin{table}[t]
\begin{center}
$
\begin{array}{|c|c|c|c|}
\hline
k_{nm} & m=n & m=n-1 & m=n-2   \\ \hline
n=1 & \gamma_F^0 &  &   \\ \hline
n=2 & \gamma_F^0 \beta_0 & \gamma_F^1  &  \\ \hline
n=3 & \frac{4}{3}\gamma_F^0 \beta_0^2 & \gamma_F^0\beta_1 + 2\gamma_F^1\beta_0  & \gamma_F^2  \\ \hline
\vdots & \vdots & \vdots & \vdots   \\ \hline
\text{accuracy} & \text{NLL} & \text{NNLL}  & \text{N$^3$LL}   \\ \hline
\end{array}
$
\quad
$
\begin{array}{|c|c|c|c|}
\hline
\eta_{nm} & m=n & m=n-1 & m=n-2   \\ \hline
n=1 & \Gamma_0 &  &   \\ \hline
n=2 & \Gamma_0 \beta_0 & \Gamma_1  &  \\ \hline
n=3 & \frac{4}{3}\Gamma_0 \beta_0^2 & \Gamma_0\beta_1 + 2\Gamma_1\beta_0  & \Gamma_2  \\ \hline
\vdots & \vdots & \vdots & \vdots   \\ \hline
\text{accuracy} & \text{LL} & \text{NLL}  & \text{NNLL}   \\ \hline
\end{array}
$
\end{center}
\vspace{-1em}
\caption{Coefficients $k_{nm}$ and $\eta_{nm}$ in the expansions \eq{kexpansion} for $K_{\gamma_F} = \sum_{n,m} k_{nm} (\as/4\pi)^n \ln^m s^{1/j_F}$ and \eq{omegaFexpansion} for $\eta_\Gamma = \sum_{n,m} \eta_{nm} (\as/4\pi)^n \ln^m s^{1/j_F}$, up to $n=3$. Closed forms for $K_{\gamma_F}$ and $\eta_\Gamma$ up to N$^3$LL accuracy are given by \eqs{etaclosedform}{kFclosedform}.}
\label{tab:ktable}
\end{table}

Similarly the expansion of $\eta_\Gamma$ is,
\begin{align}
\label{eq:omegaFexpansion}
\eta_\Gamma(Q,Q/s^{1/j_F})  =  \biggl[\left(\!\frac{\as(Q)}{4\pi}\!\right) \biggl( & \eta_{11} \ln s^{1/j_F} \biggr) \\
+\left(\!\frac{\as(Q)}{4\pi}\!\right)^2 \biggl( & \eta_{22} \ln^2 s^{1/j_F} + \eta_{21} \ln s^{1/j_F} \biggr) \nn \\
+\left(\!\frac{\as(Q)}{4\pi}\!\right)^3 \biggl( & \eta_{33} \ln^3 s^{1/j_F} + \eta_{32} \ln^2 s^{1/j_F} + \eta_{31} \ln s^{1/j_F}  \biggr) + \cdots \biggr] \,, \nn
\end{align}
with the coefficients $\eta_{nm}$ also given in \tab{ktable}. Although it does not appear in \eq{sigmanusimple}, dependence on $\eta_\Gamma(\mu,\mu_F)$ will reappear if scale choices other than \eq{nuscalechoices} are made in \eq{sigmanuderivative}.  In addition, although \eq{omegaFexpansion} seems to suggest that $\eta_\Gamma$  begins at NLL order rather than LL, we will find that transforming between the Laplace-space $\wt\sigma(\nu)$ and the momentum-space $\sigma(\tau)$ or $R(\tau)$ to a consistent order of accuracy requires keeping anomalous dimensions in $\Omega$ to the same order as in $K$.

In \eqs{KFexpansion}{omegaFexpansion} we have used the three-loop running of the coupling $\as(\mu)$ to perform the fixed-order expansion \cite{Ligeti:2008ac}. The coupling obeys the equation,
\be
\label{eq:RGalpha}
\frac{d\as(\mu)}{d\ln\mu} = \beta[\as(\mu)]\,,
\ee
where the beta function has the perturbative expansion,
\be
\label{eq:betafunction}
\beta[\as] = -2\as\sum_{n=0}^\infty \Bigl(\frac{\as}{4\pi}\Bigr)^{n+1} \beta_n\,.
\ee
The first few coefficients $\beta_n$ are given in \eq{betacoeffs}. The solution of \eq{RGalpha} with the 3-loop beta function gives the 3-loop running coupling,
\be
\label{eq:runningcoupling}
\as(\mu) = \as(Q)\! \left\{X + \as(Q) \frac{\beta_1}{4\pi\beta_0} \ln X + \frac{\as^2(Q)}{16\pi^2} \left[ \frac{\beta_2}{\beta_0}\left(1-\frac{1}{X}\right) + \frac{\beta_1^2}{\beta_0^2} \left(\frac{\ln X}{X} + \frac{1}{X} - 1\right)\!\right]\!\right\}^{\!-1}\,,
\ee
where
\be
X \equiv 1 + \frac{\as(Q)}{2\pi} \beta_0 \ln\frac{\mu}{Q}\,.
\ee
The pattern in \eq{KFexpansion} and Table \ref{tab:Ktable} makes evident that achieving \nkll accuracy in the exponent $K$ of $\wt\sigma(\nu)$ requires the cusp anomalous dimension $\Gamma_\text{cusp}^q[\as]$ to order $\as^{k+1}$,  the non-cusp part $\gamma_F[\as]$ to order $\as^k$, and the $(k+1)$-loop running of $\as$. Counting terms in the prefactors $H,\wt F$ consistently with the convention in \tab{Laplace}, \nkll accuracy can be achieved by calculating the anomalous dimensions, running coupling, and fixed-order functions $H,\wt F$ to the orders given in \tab{LaplaceE}.
At each additional order in log accuracy,  each column is incremented by one power of $\as$. 
\begin{table}[t]
\begin{center}
$
\begin{array}{ | c | c | c | c | c |}
\hline
\text{accuracy} & \Gamma_F & \gamma_F & \beta & H,\wt J,\wt S \\  \hline
\text{LL} & \as & 1 & \as & 1  \\ \hline
\text{NLL} & \as^2 & \as & \as^2 & 1  \\ \hline
\text{NNLL} & \as^3 & \as^2 & \as^3 & \as \\ \hline
\text{N$^3$LL} & \as^4 & \as^3 & \as^4 & \as^2 \\ \hline 
\end{array}
$ 
\quad
$
\begin{array}{ | c | c | c | c | c |}
\hline
\text{accuracy} & \Gamma_F & \gamma_F & \beta & H,\wt J,\wt S \\  \hline
\text{LL} & \as & 1 & \as & 1  \\ \hline
\text{NLL}' & \as^2 & \as & \as^2 & \as  \\ \hline
\text{NNLL}' & \as^3 & \as^2 & \as^3 & \as^2 \\ \hline
\text{N$^3$LL}' & \as^4 & \as^3 & \as^4 & \as^3 \\ \hline 
\end{array}
$
\end{center}
\vspace{-1em}
\caption{Order of anomalous dimensions, beta function, and fixed-order hard, jet, and soft functions required to achieve \nkll and \nkllprime accuracy in the exponent of $\tilde \sigma(\nu)$.}
\label{tab:LaplaceE} 
\end{table}

Equations (\ref{eq:KFexpansion}) and (\ref{eq:omegaFexpansion}) for $K_F,\omega_F$ can be written in closed forms  that resum all the higher order terms in $\as$. This can be done, for example, by changing variables of integration in Eqs.\ (\ref{eq:Ketadef}) and (\ref{eq:gammaKomega}) from $\mu$ to $\alpha_s$ using $d\mu/\mu = d\as/\beta[\as]$, yielding
\begin{subequations} \label{eq:Ketaforms}
\begin{align}
K_\Gamma(\mu,\mu_F) &= \int_{\as(\mu_F)}^{\as(\mu)}\frac{d\alpha}{\beta[\alpha]}\Gamma_{\text{cusp}}^q[\alpha] \int_{\as(\mu_F)}^\alpha \frac{d\alpha'}{\beta[\alpha']} \\
K_{\gamma}(\mu,\mu_F) &= \int_{\as(\mu_F)}^{\as(\mu)}\frac{d\alpha}{\beta[\alpha]} \gamma[\alpha]\,,\qquad \eta_\Gamma(\mu,\mu_F) = \int_{\as(\mu_F)}^{\as(\mu)} \frac{d\alpha}{\beta[\alpha]}\Gamma_{\text{cusp}}^q[\alpha]
\end{align}
\end{subequations} 
Using \tab{LaplaceE} we obtain expressions for $K_{\Gamma},K_{\gamma}^F,\eta_\Gamma$ order-by-order in logarithmic accuracy,
\be
\begin{split}
K &= K^{\text{LL}} + K^{\text{NLL}} + K^{\text{NNLL}} + \cdots \\
\eta & = \eta^{\text{LL}} + \eta^{\text{NLL}} + \eta^{\text{NNLL}} + \cdots\,.
\end{split}
\ee
Thus to N$^3$LL order \cite{Ligeti:2008ac,Abbate:2010xh}:
\begin{subequations}
\label{eq:KFclosedform}
\begin{align}
K_\Gamma^{\text{LL}}(\mu,\mu_F) &=  \frac{\Gamma_0}{4\beta_0^2} \frac{4\pi}{\alpha_s(\mu_F)} \biggl\{  \ln r + \frac{1}{r} - 1  \biggr\} \\
K_\Gamma^{\text{NLL}}(\mu,\mu_F)  &=  \frac{\Gamma_0}{4\beta_0^2} \biggl\{ \left(\frac{\Gamma_1}{\Gamma_0} \minus \frac{\beta_1}{\beta_0}\right) ( r \minus \ln r \minus 1) \minus \frac{\beta_1^2}{2\beta_0} \ln^2 r  \biggr\}\\
K_\Gamma^{\text{NNLL}}(\mu,\mu_F)  &=  \frac{\Gamma_0}{4\beta_0^2} \frac{\as(\mu_F)}{4\pi} \biggl\{  B_2\left( \frac{r^2 \minus 1}{2} - \ln r\right) + \left(\frac{\beta_1 \Gamma_1}{\beta_0\Gamma_0} \minus  \frac{\beta_1^2}{\beta_0^2}\right) ( r \minus r\ln r \minus 1) \nn \\
& \qquad\qquad\qquad + \left(\frac{\Gamma_2}{\Gamma_0} - \frac{\beta_1\Gamma_1}{\beta_0\Gamma_0}\right) \frac{(1-r)^2}{2}  \biggr\} \\
K_\Gamma^{\text{N$^3$LL}}(\mu,\mu_F)  &=  \frac{\Gamma_0}{4\beta_0^2}  \Bigl(\frac{\as(\mu_F)}{4\pi}\Bigr)^2 \biggl\{ \biggl[ \Bigl( \frac{\Gamma_1}{\Gamma_0} - \frac{\beta_1}{\beta_0}\Bigr) B_2 + \frac{B_3}{2}\biggr] \frac{r^2-1}{2} - B_2\Bigl( \frac{\Gamma_1}{\Gamma_0} - \frac{\beta_1}{\beta_0}\Bigr)(r-1) \nn \\
&\qquad\qquad  +  \Bigl(\frac{\Gamma_3}{\Gamma_0} - \frac{\Gamma_2 \beta_1}{\Gamma_0\beta_0} + \frac{B_2\Gamma_1}{\Gamma_0} + B_3\Bigr) \Bigl( \frac{r^3 - 1}{3} - \frac{r^2 -1 }{2} \Bigr) - \frac{B_3}{2}\ln r \nn \\
\label{eq:KFN3LL}
&\qquad\qquad  + \frac{\beta_1}{2\beta_0} \Bigl( \frac{\Gamma_2}{\Gamma_0} - \frac{\Gamma_1 \beta_1}{\Gamma_0 \beta_0} + B_2\Bigr) \Bigl( r^2 \ln r - \frac{r^2-1}{2}\Bigr) \biggr\}\,,
\end{align}
\end{subequations}
where 
\be
r\equiv \frac{\as(\mu)}{\as(\mu_F)}\,, \quad B_2 \equiv \frac{\beta_1^2}{\beta_0^2} - \frac{\beta_2}{\beta_0}\,,\quad B_3 = -\frac{\beta_1^3}{\beta_0^3} + \frac{2\beta_1\beta_2}{\beta_0^2} - \frac{\beta_3}{\beta_0}\,,
\ee
and
\begin{subequations}
\label{eq:etaclosedform}
\begin{align}
\eta_\Gamma^{\text{LL}}(\mu,\mu_F)  &=  -\frac{\Gamma_0}{2\beta_0} \ln r \\
\eta_\Gamma^{\text{NLL}}(\mu,\mu_F)  &=  -\frac{\Gamma_0}{2\beta_0} \frac{\as(\mu_F)}{4\pi} \Bigl( \frac{\Gamma_1}{\Gamma_0} - \frac{\beta_1}{\beta_0}\Bigr) (r-1) \\
\eta_\Gamma^{\text{NNLL}}(\mu,\mu_F)  &=  -\frac{\Gamma_0}{2\beta_0} \Bigl(\frac{\as(\mu_F)}{4\pi}\Bigr)^2 \Bigl( B_2 + \frac{\Gamma_2}{\Gamma_0} - \frac{\Gamma_1\beta_1}{\Gamma_0\beta_0}\Bigr)\frac{r^2-1}{2}   \\
\label{eq:etaN3LL}
\eta_\Gamma^{\text{N$^3$LL}}(\mu,\mu_F)  &=  -\frac{\Gamma_0}{2\beta_0} \Bigl(\frac{\as(\mu_F)}{4\pi}\Bigr)^3 \Bigl( \frac{\Gamma_3}{\Gamma_0} - \frac{\Gamma_2 \beta_1}{\Gamma_0\beta_0} + \frac{\Gamma_1}{\Gamma_0}B_2 + B_3 \Bigr)(r^3-1)  \,,
\end{align}
\end{subequations}
and
\be
\label{eq:kFclosedform}
K_\gamma^{\text{LL}} = 0 \,, \quad K_{\gamma}^{\text{NLL}} = \eta_\gamma^{\text{LL}}\,,\quad K_{\gamma}^{\text{NNLL}} = \eta_\gamma^{\text{NLL}}\,,\quad K_{\gamma}^{\text{N$^3$LL}} = \eta_\gamma^{\text{NNLL}}\,,\dots
\ee
These last relations mean that to get $K_\gamma$ on the left-hand side, use the specified formula for $\eta_\Gamma$ given in \eq{etaclosedform} with the replacement $\Gamma_n\to \gamma_n$. 
The expressions \eqs{KFclosedform}{etaclosedform}  capture terms at all orders in $\as$ required up to N$^3$LL accuracy. Re-expanding \eqs{KFclosedform}{etaclosedform} in powers of $\as(Q)$ using the running coupling \eq{runningcoupling} produces the fixed-order series shown in \eqs{KFexpansion}{omegaFexpansion}. 

Note from \eqs{KFN3LL}{etaN3LL} that N$^3$LL accuracy requires knowing the 4-loop cusp anomalous dimension $\Gamma_3$, which is not known explicitly at the present time. It can however be estimated by Pad\'{e} approximation, yielding approximate N$^3$LL resummed accuracy as in \cite{Abbate:2010xh,Becher:2008cf}, assuming other ingredients such as non-cusp anomalous dimensions and matching coefficients are also known to sufficiently high order.

The way we divided the evolution kernel into $K_F$ and $\omega_F$ in \eqs{KFexpansion}{omegaFexpansion} makes it appear that $\omega_F$ begins at a subleading order in logarithmic accuracy (NLL) compared to $K_F$ (LL). However, it is advisable to keep the cusp anomalous dimension to the same order of accuracy both in $K_F$ and $\omega_F$ at a given order \nkll of logarithmic accuracy, as we indicated in \eq{etaclosedform} or \tab{ktable}. One reason for this is that $K_\Gamma$ and $\eta_\Gamma$ always appear together as in \eq{gammaKomega} in the combination:
\be
\label{eq:Kpluseta}
\begin{split}
& \hspace{-10mm} -j_F\kappa_F K_F(\mu,\mu_0) + (\mu_0^{j_F} e^{\gamma_E} x)^{-\kappa_F\eta_\Gamma(\mu,\mu_0)} \\
&= -j_F\kappa_F \int_{\mu_0}^\mu \frac{d\mu'}{\mu'} \Gamma_{\text{cusp}}^q [\as(\mu')]\Bigl(\ln\frac{\mu'}{\mu_0} + \ln[\mu_0(e^{\gamma_E}x)^{1/j_F}]\Bigr) \\
&= -j_F\kappa_F \int_{\mu_0}^\mu \frac{d\mu'}{\mu'} \Gamma_{\text{cusp}}^q [\as(\mu')] \ln[\mu'(e^{\gamma_E}x)^{1/j_F}]\,,
\end{split}
\ee
whose integrand reproduces the starting form of the anomalous dimension \eq{gammaFLaplace}. Thus keeping $\Gamma_{\text{cusp}}$ to the same order in $K_F$ and $\eta_\Gamma$ is necessary to remove the explicit dependence on the arbitrary scale $\mu_0$ we introduced into the integrand of $K_\Gamma$ in \eq{Ketadef} purely for convenience. In other words, if we had kept the evolution kernel in the form of the last line of \eq{Kpluseta} we would automatically keep $\Gamma_{\text{cusp}}$ to the same order wherever it appears, and this is what we shall do. The remaining dependence on $\mu_0$ in the lower limit of the evolution integrals is cancelled by $\mu_0$-dependent terms in the fixed-order coefficients $H,\wt J, \wt S$. 

\begin{figure}[t]
\begin{center}
\includegraphics[width=.9\textwidth]{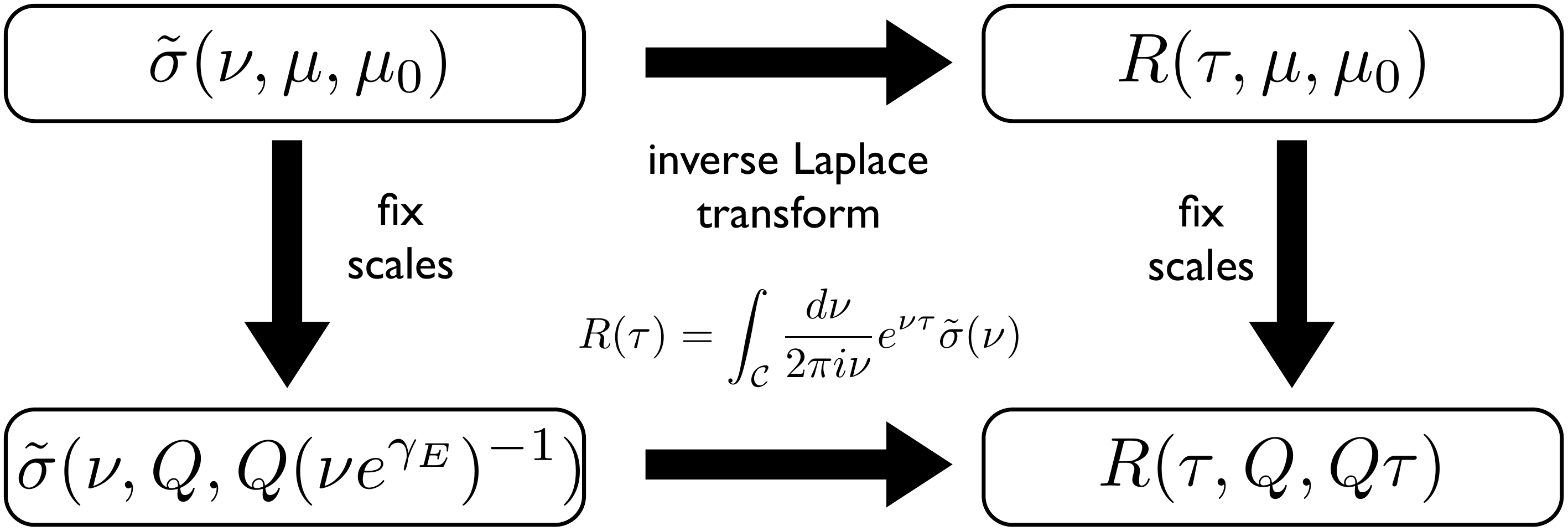}
\end{center}
\caption{Commutative diagram for obtaining the resummed cumulant distribution from the Laplace transform. The resummed cumulant can be obtained by first inverse Laplace transforming $\wt\sigma$ with free scales, and then choosing the scales to be $\tau$-dependent. Or, it can be obtained by first fixing the scales in the Laplace transform to be $\nu$-dependent, and then inverse Laplace transforming. If expressions for $R,\wt\sigma$ are kept to all orders in $\as$, the two routes produce identical results, though care must be exercised at truncated finite orders to maintain equivalent accuracy. (\emph{Nota bene:} When the running of the strong coupling $\as(\mu)$ is taken into account, the bottom route can lead to integration over the Landau pole, which can be avoided by appropriate prescriptions, but in that case the top route is preferable.)
\label{fig:commutenu}}
\end{figure}

Making the scale choices \eq{nuscalechoices} in the Laplace-transformed cross section $\tilde \sigma(\nu)$ in \eq{sigmanuderivative} is appropriate for the purpose of resumming logs of $\nu$ if one wishes to express the distribution directly in terms of this variable. If one is interested instead in summing logs of $\tau_a$ in $\sigma(\tau_a)$ or $R(\tau_a)$, one might think taking the inverse Laplace transform of \eq{sigmanusimple} is the simplest thing to do, but this is actually fairly cumbersome due to the complex $\tau_a$ dependence after scale setting. In addition, in integrating down to $s=0$ one encounters the Landau pole in $\as(Qs)$ and $\as(Qs^{1/j_J})$ whose expansions in $\as(Q)$ produced the $\beta_n$-dependent terms in \eq{KFexpansion}.  Instead, it is simpler to take the inverse Laplace transform of  \eq{sigmanuderivative} before choosing the scales $\mu_F$, and then after the transform to choose $\tau_a$-dependent scales, such as the canonical choice
\be
\label{eq:tauscalechoices}
\mu = \mu_H = Q\,, \quad \mu_F^{\rm nat} = Q\tau_a^{1/j_F}\,,
\ee
in calculating either $\sigma(\tau_a)$ or $R(\tau_a)$.\footnote{Alternatively, one can Taylor expand \eq{sigmanuderivative} around $\nu = 1/\tau_a$ before performing the inverse transform. This is the approach taken in the QCD resummed \cite{Catani:1992ua} cross section given by \eq{QCDxsec}, with the Taylor expansion performed in \eq{ETaylor}.}  Of course, the two orders of operations must be equivalent when the cross section is computed to all orders in $\as$, as illustrated in the commutative diagram \fig{commutenu}. At a truncated order of accuracy, however, results may differ. We carry out these comparisons explicitly in the next subsections.

\subsection{Cumulant}

\subsubsection{Resummed Cumulant in SCET Formalism}
\label{ssec:cumulant}

To obtain the cumulant $R(\tau_a)$, it is simplest to take the inverse Laplace transform of \eq{sigmanuderivative} and integrate with respect to $\tau_a$, or 
\be
\label{eq:radiatorintegral}
R(\tau_a) = \frac{1}{2\pi i}\int_{\cC}\frac{d\nu}{\nu} e^{\nu\tau_a} \wt\sigma (\nu)\,,
\ee
which gives the result in \eq{SCETresummedcumulant}.
Then making the scale choices in \eq{tauscalechoices} leads to the simple form:
\be
\label{eq:radiatorsimple}
R(\tau_a) =  H_2(Q) e^K \wt J(\partial_\Omega,\mu_J^{\rm nat})^2 \wt S(\partial_\Omega,\mu_S^{\rm nat}) \left[\frac{e^{\gamma_E\Omega}}{\Gamma(1-\Omega)}\right]\,,
\ee
where 
\begin{subequations}
\label{eq:canonicalKomega}
\begin{align}
K &= 2K_J(Q,Q\tau_a^{1/j_J}) + K_S(Q,Q\tau_a)\,, \\
\Omega &= 2\omega_J(Q,Q\tau_a^{1/j_J})+ \omega_S(Q,Q\tau_a) \,.
\end{align}
\end{subequations}
The scale choices \eq{tauscalechoices} are the typical ``canonical'' choices made in the SCET literature. Another reasonable set of choices is  \eq{barcanonicalscales}, which lead to a form of the resummed cross section most similar to the typical QCD form \eq{QCDRderivatives}. With the choices \eq{barcanonicalscales}, we obtain for the cumulant \eq{SCETresummedcumulant}:
\be
\label{eq:radiatorbar}
R(\tau_a) =  H_2(Q)e^{\bar K} \wt J(\partial_{\bar\Omega},\bar\mu_J)^2 \wt S(\partial_{\bar \Omega},\bar\mu_S) \left[\frac{1}{\Gamma(1-\bar \Omega)}\right]\,,
\ee
where 
\begin{subequations}
\label{eq:barKomega}
\begin{align}
\bar K &= 2K_J(Q,Q\bar\tau_a^{1/j_J}) + K_S(Q,Q\bar\tau_a)\,, \\
\bar\Omega &= 2\omega_J(Q,Q\bar\tau_a^{1/j_J})+ \omega_S(Q,Q\bar\tau_a) \,,
\end{align}
\end{subequations}
where $\bar\tau_a$ is defined in \eq{taubar}. The two forms \eqs{radiatorsimple}{radiatorbar} are equivalent, when computed to all orders in $\as$. Truncated at a given order of logarithmic accuracy, they may differ in subleading terms but are equal at the given order of accuracy.

We proceed to classify the accuracy of logarithmic resummation for $R(\tau)$ given by \eq{radiatorbar} before considering the distribution $\sigma(\tau_a)$. The discussion for $R$ will turn out to be more straightforward. With the scale choices \eq{barcanonicalscales}, we find from \eq{Komegadef},
\begin{equation}
\bar K_F(Q,Q\bar \tau_a^{1/j_F}) = -j_F\kappa_F K_\Gamma(Q,Q\bar\tau_a^{1/j_F}) + K_{\gamma_F}(Q,Q\bar\tau_a^{1/j_F})\,,
\end{equation}
where $K_\Gamma,K_{\gamma_F}$ have the same fixed-order expansions as in \eqs{KFexpansion}{kexpansion} with the replacement $s\to1/\bar\tau$. Meanwhile, $\bar\Omega = 2\bar\omega_J + \bar\omega_S$, where
\be
\bar\omega_F(Q,Q\bar\tau_a^{1/j_F}) = -\kappa_F \eta_\Gamma(Q,Q\bar\tau_a^{1/j_F})\,,
\ee
with $\eta_\Gamma$ having the same fixed-order expansion as in \eq{omegaFexpansion} with the replacement $s\to 1/\bar\tau$.

Each of the fixed-order jet and soft functions $\wt J,\wt S$ in \eq{radiatorbar} produces a prefactor multiplying the exponentiated series. To calculate the prefactors, we use the relations
\begin{subequations}
\label{eq:Gbardefs}
\begin{align}
\bar\cG(\bar\Omega) &\equiv \Bigl(\frac{\mu}{Q\bar\tau}\Bigr)^{\bar \Omega}\frac{1}{\Gamma(1-\bar\Omega)} \,, \\
\partial_{\bar\Omega}\bar\cG(\bar\Omega) &=  \left[ \ln \frac{\mu}{Q\bar\tau} + \psi(1-\bar\Omega)\right] \cG(\bar\Omega)  \,, \\
\partial_{\bar\Omega}^2\bar\cG(\bar\Omega)  &= \left[\Bigl( \ln\frac{\mu}{Q\bar\tau} + \psi(1-\bar\Omega)\Bigr)^2 - \psi^{(1)}(1-\bar\Omega)\right]\bar\cG(\bar\Omega) \,, \\
\partial_{\bar\Omega}^3\bar\cG(\bar\Omega)  &=  \left[\Bigl(\ln\frac{\mu}{Q\bar\tau} + \psi(1-\bar\Omega)\Bigr)^3 - 3\Bigl( \ln\frac{\mu}{Q\bar\tau} + \psi(1-\bar\Omega)\Bigr)\psi^{(1)}(1-\bar\Omega) + \psi^{(2)}(1-\bar\Omega)\right] \bar\cG(\bar\Omega)\,,
\\ 
\partial_{\bar\Omega}^4\bar\cG(\bar\Omega) &= \biggl[\Bigl( \ln\frac{\mu}{Q\bar\tau} + \psi(1-\bar\Omega)\Bigr)^4 - 6\Bigl( \ln\frac{\mu}{Q\bar\tau} + \psi(1-\bar\Omega)\Bigr)^2\psi^{(1)}(1-\bar\Omega) + 3\psi^{(1)}(1-\bar\Omega)^2 \nn \\
&\qquad\qquad\qquad + 4 \Bigl( \ln\frac{\mu}{Q\bar\tau} + \psi(1-\bar\Omega)\Bigr)\psi^{(2)}(1-\bar\Omega) - \psi^{(3)}(1-\bar\Omega)\biggr] \bar\cG(\bar\Omega) \,,
\end{align}
\end{subequations}
and so on. Here $\psi$ is the digamma function, $\psi(z) = \Gamma'(z)/\Gamma(z)$, and $\psi^{(n)}$ is its $n$th derivative. Note that when applied to \eq{radiatorbar}, the $\mu$ in \eq{Gbardefs} is set equal to $Q\bar\tau$, so the explicit logs of $\mu/Q\bar\tau$ disappear.

To  NLL accuracy, the equation \eq{radiatorsimple} or \eq{radiatorbar} for $R(\tau_a)$ fits into the form \eq{CTTWradiator} given by CTTW, but only to NLL. At this order, the quantities in \eq{CTTWradiator} are given by
\be
\label{eq:CTTWfromSCET}
\ln\Sigma = \bar K - \ln\Gamma(1-\bar \Omega)\,, \quad C(\as) = 1\,,
\ee
but beyond NLL we see that parts of the CTTW ``exponent'' are actually generated by the parts of the fixed-order functions $\wt F$ that contain derivative operators $\partial_{\bar\Omega}$ \cite{Becher:2006nr}. The forms \eqs{QCDRderivatives}{SCETcsEfinal} are properly generalized forms of \eq{CTTWradiator}, where some terms in the $\tau_a$-dependent function $\Sigma(\tau_a,\as)$ are produced by the  differential operator $\wh T(\bar E')$, or equivalently in the forms \eqs{radiatorsimple}{radiatorbar} by the differential operators $\wt F(\partial_{\Omega,\bar\Omega})$.

\subsubsection{Computing Laplace Transform and Cumulant to Consistent Accuracy}
\label{ssec:Laplacecomparison}

\begin{center}
\textbf{\emph{Note:} In this subsection, we will ignore the running of $\as$, treating it as a constant. As a reminder, we will use the shorthand $a_s = \as / 4\pi$.}
\end{center}

Above we have defined \nkll or \nkllprime accuracy by the number of terms in the exponent of $\wt\sigma(\nu)$ in \eq{Laplaceexponent} or \eq{sigmanusimple} that are accurately predicted. In computing the $\tau_a$-distribution in momentum space, the counting of logarithmic accuracy can be a little trickier. For instance, expressions like \eq{SCETcsEfinal}, \eq{radiatorsimple}, \eq{radiatorbar} for the cumulant $R(\tau_a)$ contain a series of derivative operators acting on a gamma function containing large logs. How are the logs generated by these derivatives to be counted when computing to \nkll or \nkllprime accuracy?

Now, the resummed cumulant $R(\tau_a)$ given by \eq{radiatorsimple} or \eq{radiatorbar} is supposed to be equal to the inverse transform \eq{radiatorintegral} of the resummed Laplace-transformed cross section $\wt\sigma(\nu)$ in \eq{sigmanusimple}.  This is exactly true when both formulae are computed to all orders in $\as$. In practice, of course, we have to truncate the accuracy of each at some finite order, so the exact equivalence between the two forms cannot in practice be maintained. Here we will explore this relationship by comparing the results of truncating these formulae for $\wt\sigma(\nu)$ or $R(\tau)$ at a finite logarithmic accuracy. We then prescribe how to compute $R(\tau_a)$ so that the equivalence between it and $\wt\sigma(\nu)$ in \eq{sigmanusimple} computed at a given order of resummed accuracy is maintained as best as possible. We will give a prescription for truncating ingredients of the resummed $R(\tau_a)$ so that it is at least as accurate as $\wt\sigma(\nu)$ at \nkll or \nkllprime accuracy, although subleading terms will differ.

First, note that we defined \nkll accuracy for $\wt\sigma(\nu)$ by the number of terms in the exponent $K$ that are included, given by \tab{Ktable}. \nkll accuracy for $\wt\sigma(\nu)$ is achieved by calculating anomalous dimensions and the fixed-order coefficients to the orders given in \tab{LaplaceE}. Inverse Laplace transforming the \nkll $\wt\sigma(\nu)$ by \eq{radiatorintegral} must then produce the expression \eq{radiatorbar} with $\bar K,\bar\Omega$ and the fixed-order coefficients $H,\wt J,\wt S$ computed to the same orders as given by \tab{LaplaceE}. In particular, this produces an expression for $\bar \Omega$ computed with the cusp anomalous dimension kept to the same order as in $\bar K$ at \nkll accuracy, i.e. to order $\as^{k+1}$, as specified in \tab{LaplaceE}. This is the case even though, by itself, $\omega_F = -\kappa_F\eta_\Gamma$ given by \eq{omegaFexpansion} appears to begin at one lower logarithmic order than $K_\Gamma$ in \eq{KFexpansion}.  Even though it is not \emph{strictly} required for achieving \nkll accuracy, keeping $\Gamma_{\text{cusp}}$ to the same order in $\bar\Omega$ as in $\bar K$ is necessary for keeping the expressions \eq{sigmanusimple} for $\tilde\sigma(\nu)$ and \eq{radiatorbar} for $R(\tau_a)$ exactly equal to each other upon Laplace transformation. [See also comments after \eq{Kpluseta} on benefits of computing $\bar\Omega$ to this order.]

This is not by itself sufficient, however, for we need to consider the contribution of the derivative operator terms in \eq{radiatorbar} as well. To illustrate in a simple example the effect of these terms (being somewhat schematic), consider a distribution $\wt\sigma(\nu,\mu)$ in Laplace space which is made up of a product of transformed functions $\wt{F}$, each of which satisfies an RG equation of the form
\be
\mu\frac{d}{d\mu} \wt{F}(\nu,\mu) = 2\Gamma_F[a_s] \ln\Bigl(\frac{\mu\nu e^{\gamma_E}}{Q}\Bigr) + \gamma_F[a_s]\,,
\ee
where $\Gamma_F,\, \gamma_F$ have the expansions in $a_s$,
\be
\Gamma_F[a_s] = \sum_{n=0}^\infty \Gamma_{F,n} a_s^{n+1}\,,\quad \gamma_F[a_s] = \sum_{n=0}^\infty \gamma_{F,n} a_s^{n+1}\,.
\ee
Then, the resummed $\wt\sigma(\nu,\mu)$ is a product of functions of the form
\be
\wt{F}(\nu,\mu) = \wt{F}(\nu,\mu_F) e^{K_F(\mu,\mu_F)} \biggl(\frac{\mu_F\nu e^{\gamma_E}}{Q}\biggr)^{\omega_F (\mu,\mu_F)}\,,
\ee
where 
\be
K_F (\mu,\mu_F) = \Gamma_F \ln^2\frac{\mu}{\mu_F} + \gamma_F \ln\frac{\mu}{\mu_F}\,,\qquad  \omega_F (\mu,\mu_F) = 2\Gamma_F \ln\frac{\mu}{\mu_F}\,.
\ee
For simplicity, we will write the full cross section $\wt\sigma$ in a schematic form, in terms of only one of these evolution factors,
\be \label{eq:simpleLaplace}
\wt\sigma(\nu,\mu) = \wt\sigma(\nu,\mu_0) e^{K(\mu,\mu_0)} \Bigl( \frac{\mu_0 \nu e^{\gamma_E}}{Q} \Bigr)^{\Omega(\mu,\mu_0)} \,,
\ee
although the number of evolution factors does not affect the conclusions drawn from this discussion.  One can think of $\mu_0$ as representing a set of scales $\mu_i$ for each evolution factor.  Upon choosing the scales $\mu_0 =  Q(\nu e^{\gamma_E})^{-1}$ and $\mu = Q$ yields the expression
\be
\label{eq:simpleLaplacefixed}
\wt\sigma(\nu) = C(a_s) e^{\Gamma \tilde L^2 + \gamma \tilde L} \,,
\ee
where $\tilde L \equiv \ln(\nu e^{\gamma_E})$, and the coefficient $C$ contains no logs,
\begin{align}
C(a_s) &\equiv \wt\sigma(\nu,Q(\nu e^{\gamma_E})^{-1}) = 1 + \sum_{n=1}^\infty C_n a_s^n\,.
\end{align}
Taking the inverse Laplace transform of $\wt\sigma(\nu,\mu)$ in \eq{simpleLaplace} before fixing the scale $\mu_0$, we obtain for the cumulant distribution in momentum space,
\be
\label{eq:simpleR}
R(\tau,\mu) = \int_\mathcal{C}\frac{d\nu}{2\pi i}\frac{e^{\nu\tau}}{\nu} \wt\sigma(\nu) = C(a_s) e^{K(\mu,\mu_0)} e^{\Gamma\partial_{\Omega}^2 + \gamma\partial_{\Omega} } \Bigl(\frac{\mu_0}{Q\tau}\Bigr)^{\Omega(\mu,\mu_0)}\frac{e^{\gamma_E\Omega(\mu,\mu_0)}}{\Gamma(1-\Omega(\mu,\mu_0))}\,.
\ee
The differential operator $e^{\Gamma\partial_\Omega^2 + \gamma\partial_\Omega}$ here is the equivalent of the product of operators $\wt J(\partial_\Omega)^2\wt S(\partial_\Omega)$ in the full SCET expression \eq{SCETresummedcumulant} or the differential operator terms in \eq{SCETcsEfinal}. Thus the order to which we truncated this operator corresponds to the order to which we truncate the jet and soft functions in SCET or the differential operators in the dQCD resummed cumulant.

In \eq{simpleR} we can choose the scale $\mu_0$ to be $Q\tau$ or $Q\bar\tau = Q\tau e^{-\gamma_E}$. Here, we will pick the canonical choice $\mu_0=Q\tau$, and the same scale $\mu=Q$ as in \eq{simpleLaplacefixed}. Then we obtain
\be
\label{eq:simpleRfixed}
R(\tau) =  C e^{\Gamma L^2 + \gamma L} e^{\Gamma\partial_\Omega^2} \frac{e^{\gamma_E(\Omega+\gamma)}}{\Gamma(1-\Omega-\gamma)}\,,
\ee
where 
\be
L\equiv \ln\frac{1}{\tau} \,,\quad \Omega = 2\Gamma L \,,
\ee
and where we have used the translation operator $e^{\gamma\partial_\Omega}$ in \eq{simpleR} to shift the arguments in last factor in \eq{simpleRfixed} from $\Omega$ to $\Omega+\gamma$.

Since the factor $C(a_s)$ is common to both $\wt\sigma(\nu)$ in \eq{simpleLaplacefixed} and $R(\tau)$ in \eq{simpleRfixed}, we will not explicitly include it in comparing the relative accuracy of the two expressions. We focus on comparing how accurately the remaining explicitly log-dependent factors in $R(\tau)$ in \eq{simpleRfixed} reproduce the exponentiated logs in $\wt\sigma(\nu)$ in \eq{simpleLaplacefixed} upon Laplace transformation. It is these terms that determine the shape of the spectrum in $\tau$.

The two formulae \eq{simpleLaplace} for $\wt\sigma(\nu,\mu)$ and \eq{simpleR} are related by Laplace transformation \emph{before} fixing the scale $\mu_0$ to be $\nu$-dependent or $\tau$-dependent. This relation is represented by the top arrow in \fig{commutenu}. Fixing the scales $\mu=Q$ and $\mu_0 = Q(\nu e^{\gamma_E})^{-1}$ or $\mu_0 = Q\tau$ is represented by the vertical arrows. It should follow that transforming between $\wt\sigma(\nu)$ and $R(\tau)$ on the bottom of \fig{commutenu} \emph{after} fixing the scales should also be valid. Evaluating the transform in practice is more involved, however, due to the additional $\nu$ or $\tau$ dependence in the scales (at least once the running of $\as$ is restored), but we can compute at some truncated order in $\as$ to compare the accuracy to which the transform on the bottom of \fig{commutenu} is achieved.

Now, transforming $R(\tau)$ in \eq{simpleRfixed} back to $\tilde \sigma(\nu)$ using
\be
\frac{1}{\nu}\wt\sigma(\nu) =  \int_0^\infty d\tau \, e^{-\nu\tau} R(\tau)
\ee
will yield exactly the same result as \eq{simpleLaplacefixed}, if all objects in both expressions are computed to all orders in $\as$. However, when truncated at a given logarithmic accuracy, the inability to evaluate the infinite series of differential operators in $e^{\Gamma\partial_\Omega^2}$ in \eq{simpleRfixed} prevents this from being exactly realized. An explicit example will serve to illustrate this.

First, consider $\wt\sigma(\nu)$ given, after fixing scales, by \eq{simpleLaplacefixed}, to N$^3$LL accuracy [equation appears in color]:
\be
\label{eq:LaplaceN3LL}
\begin{split}
\wt\sigma(\nu) =  C(a_s) \exp\biggl\{a_s( \red{\Gamma_0 \tilde L^2} + \green{\gamma_0} & \green{\tilde L}) \\
 + a_s^2(\green{ \Gamma_1} & \green{\tilde L^2} + \blue{\gamma_1\tilde L}) \\
&  + a_s^3 ( \blue{\Gamma_2\tilde L^2} + \purple{\gamma_2\tilde L}) + a_s^4\purple{\Gamma_3\tilde L^2 } +\cdots\biggr\}\,,
\end{split}
\ee
where we have colored the terms at LL accuracy in red, NLL in green, NNLL in blue, and N$^3$LL in purple. Now expanding \eq{LaplaceN3LL} in fixed orders in $a_s$ up to $\cO(a_s^3)$ [color]:
\begin{align}
\label{eq:Laplacefixedorder}
\wt\sigma(\nu) = C(a_s)& \biggl\{ \red{1} + a_s (\red{\Gamma_0 \tilde L^2} + \green{\gamma_0\tilde L}) \\
&+ a_s^2 \biggl[\red{\frac{\Gamma_0^2}{2} \tilde L^4} + \green{\Gamma_0\gamma_0 \tilde L^3 + \Bigl(\frac{\gamma_0^2}{2}+\Gamma_1\Bigr)\tilde L^2} + \blue{\gamma_1\tilde L}\biggr] \nn \\
&+ a_s^3 \biggl[\red{\frac{\Gamma_0^3}{6} \tilde L^6} + \green{\frac{\Gamma_0^2\gamma_0}{2} \tilde L^5  + \Bigl(\frac{\Gamma_0\gamma_0^2}{2} \plus \Gamma_1\Gamma_0 \Bigr)\tilde L^4 + \Bigl( \frac{\gamma_0^3}{6} \plus \Gamma_1\gamma_0}  \blue{+ \Gamma_0 \gamma_1\Bigr)\tilde L^3} \nn \\
&\hspace{3.1in} + \blue{(\gamma_1\gamma_0 + \Gamma_2) \tilde L^2} + \purple{\gamma_2 \tilde L}  \biggr]\biggr\}, \nn
\end{align}
where the colors indicate which terms in the fixed-order expansion are generated by expanding the exponent in \eq{LaplaceN3LL} to the corresponding order in logarithmic accuracy. At \nkll order, the first log that is missing in the exponent of \eq{LaplaceN3LL} is $\sim a_s^{k+1}\gamma_k\tilde L$. Thus, in \eq{Laplacefixedorder}, the largest log that is missing at $\cO(\as^{k+n})$ in the fixed-order expansion of the exponentiated logs $e^{\Gamma\tilde L^2 + \gamma\tilde L}$ in $\wt\sigma(\nu)$ is:
\be
\label{eq:missinglogLaplace}
\begin{split}
\text{largest missing log in \nkll $\wt\sigma(\nu)$ at fixed-order $\cO(a_s^{k+n})$: } \\
(a_s^{k+1}\gamma_k \tilde L)\Bigl( a_s^{n-1}\frac{\Gamma_0^{n-1}}{(n-1)!}\tilde L^{2n-2}\Bigr) = a_s^{k+n}\frac{\Gamma_0^{n-1} \gamma_k}{(n-1)!}\tilde L^{2n-1}
\end{split}
\ee
We will use \eqs{Laplacefixedorder}{missinglogLaplace} to compare which terms are accurately predicted by the Laplace transform of the cumulant $R(\tau)$ truncated at a given order. It is reasonable to expect that a definition or prescription for \nkll accuracy in $R(\tau)$ should not mis-predict any logs that are larger than \eq{missinglogLaplace} upon  Laplace transformation.

Now consider the expansion of the cumulant $R(\tau)$ in \eq{simpleRfixed} with the anomalous dimensions kept to the same accuracy. The factor
\be
\label{eq:eKfactor}
e^{K(Q,Q\tau)} = e^{\Gamma L^2 + \gamma L} 
\ee
will take exactly the same form as the expansion in \eq{Laplacefixedorder} for $\wt\sigma(\nu)$, with each $\tilde L \equiv \ln(\nu e^{\gamma_E})$ replaced by $L\equiv \ln(1/\tau)$. However, the powers $\tilde L^n$ and $L^n$ are not exact transforms of each other, as we see from \eqs{LPs}{iLPs}, each contains a series of lower-order logs $L^{n-k}$ or $\tilde L^{n-k}$ for $0\leq k\leq n$.
The extra terms in the transform of each power $\tilde L^n$ are provided by the expansion of the extra factors in \eq{simpleRfixed},
\be
\label{eq:DOmegaseries}
\begin{split}
e^{\Gamma\partial_\Omega^2} \frac{e^{\gamma_E(\Omega+\gamma)}}{\Gamma(1-\Omega-\gamma)} &= \biggl(1 + \Gamma \partial_\Omega^2 + \frac{\Gamma^2}{2} \partial_\Omega^4 + \frac{\Gamma^3}{6}\partial_\Omega^6 + \cdots\biggr) \\
&\quad \times\biggl[ 1 - \frac{\pi^2}{12}(\Omega+\gamma)^2 - \frac{\zeta_3}{3} (\Omega+\gamma)^3 + \frac{\pi^4}{1440} (\Omega+\gamma)^4 \\
&\qquad + \frac{5\pi^2\zeta_3 - 36\zeta_5}{180}(\Omega+\gamma)^5 + \Bigl( \frac{\zeta_3^2}{18} - \frac{\pi^6}{24192}\Bigr)(\Omega+\gamma)^6+\cdots\biggr]\,,
\end{split}
\ee
or, evaluating the derivatives up to terms contributing at $\cO(a_s^3)$,
\be
\label{eq:DOmegaexpansion}
\begin{split}
e^{\Gamma\partial_\Omega^2} \frac{e^{\gamma_E(\Omega+\gamma)}}{\Gamma(1-\Omega-\gamma)} &= 1 - \frac{\pi^2}{12} (\Omega+\gamma)^2 - \frac{\zeta_3}{3}(\Omega+\gamma)^3 \\
&\quad + \Gamma\biggl[- \frac{\pi^2}{6} - 2\zeta_3(\Omega+\gamma) + \frac{\pi^4}{120}(\Omega+\gamma)^2 \biggr] \\
&\quad + \frac{\Gamma^2}{2} \biggl[ \frac{\pi^4}{60} + \Bigl( \frac{10\pi^2}{3}\zeta_3 - 24\zeta_5\Bigr) (\Omega+\gamma)\biggr] \\
&\quad + \frac{\Gamma^3}{6}\Bigl(40\zeta_3^2 - \frac{5\pi^6}{168}\Bigr) + \cdots \,,
\end{split}
\ee
recalling $\Omega=2\Gamma L$ for fixed $a_s$. The derivatives can also be computed from the series expansions of the expressions in \eq{Gbardefs}.

Now to compute $R(\tau_a)$ up to $\cO(a_s^3)$, we multiply together \eqs{eKfactor}{DOmegaexpansion}, and find for the fixed-order expansion of $R(\tau_a)$ [color],
\be
\label{eq:Rfixedorder}
\begin{split}
R(\tau) =\iLP\Biggl( & \frac{C(a_s)}{\nu}  \Biggl\{ \red{1} + a_s (\red{\Gamma_0 \tilde L^2} + \green{\gamma_0\tilde L}) \\
&+ a_s^2 \biggl[\red{\frac{\Gamma_0^2}{2} \tilde L^4} + \green{\Gamma_0\gamma_0 \tilde L^3 + \Bigl(\frac{\gamma_0^2}{2}+\Gamma_1\Bigr)\tilde L^2} + \blue{\gamma_1\tilde L}\biggr] \\
&+ a_s^3 \biggl[\red{\frac{\Gamma_0^3}{6} \tilde L^6} + \green{\frac{\Gamma_0^2\gamma_0}{2} \tilde L^5  + \Bigl(\frac{\Gamma_0\gamma_0^2}{2} \plus \Gamma_1\Gamma_0 \Bigr)\tilde L^4} \\
&\qquad\qquad\qquad\quad    \green{ + \Bigl( \frac{\gamma_0^3}{6} \plus \Gamma_1\gamma_0}  \blue{+ \Gamma_0 \gamma_1\Bigr)\tilde L^3} + \blue{(\gamma_1\gamma_0 + \Gamma_2) \tilde L^2} + \purple{\gamma_2 \tilde L}  \biggr] \! \Biggr\} \! \Biggr)\,.
\end{split}
\ee
where $\iLP(\tilde L^n/\nu)$ is given explicitly by \eq{iLPs}. 
That is, the Laplace transform of $R(\tau)$ is exactly the expansion \eq{Laplacefixedorder} of $\wt\sigma(\nu)$ up to the terms at $\cO(a_s^3)$ that are predicted at N$^3$LL accuracy. However, to achieve this result it was necessary to keep the expansion of the derivative operator $e^{\Gamma\partial_\Omega^2}$ in \eq{DOmegaexpansion} up to the terms of $\cO(a_s^3)$. This operator corresponds to terms in the $\wt F(\partial_\Omega)$ operators in the SCET resummed cumulant \eq{radiatorsimple} or \eq{radiatorbar}, or the $\partial_{\bar E'}$ terms in  the QCD-inspired form \eq{SCETcsEfinal}. This tells us that the Laplace transform of $R(\tau)$ reproduces the expansion \eq{Laplacefixedorder} of $\wt\sigma(\nu)$ exactly only up to the order in $a_s$ to which the differential operators are kept. Thus, to reproduce the \nkll $\wt\sigma(\nu)$ in \eq{Laplacefixedorder} \emph{exactly} to arbitrary order in $a_s$, the differential operators in the expansion of $e^{\Gamma\partial_\Omega^2}$ would have to kept to infinite order. 

Luckily, we will be able to reproduce $\wt\sigma(\nu)$ to \nkll accuracy by keeping only a finite number of these derivative operators, but to higher order than \tab{LaplaceE} might suggest. To see this, let's first start na\"{i}vely, applying the rules in \tab{Laplace} to evaluate $R(\tau)$ given by \eq{simpleRfixed} at NLL order. When \tab{Laplace} tells us to keep the $\wt J,\wt S$ functions to tree level at NLL, interpreted na\"{i}vely we also truncate the operator $e^{\Gamma\partial_\Omega^2}$ in \eq{simpleRfixed} to tree level. Then we would obtain for $R(\tau)$:
\begin{align}
\label{eq:RfixedorderNLL}
R(\tau) &\overset{\text{NLL}}{=}  \exp\Bigl[ a_s (\Gamma_0 L^2 + \gamma_0 L) + a_s^2 \Gamma_1 L^2\Bigr] \frac{\exp\bigl\{ \gamma_E \bigl[ a_s (2 \Gamma_0 L + \gamma_0) + a_s^2 (2 \Gamma_1 L) \bigr] \bigr\}}{\Gamma \bigl[ 1 - a_s (2 \Gamma_0 L + \gamma_0) - a_s^2 (2 \Gamma_1 L) \bigr]} \\
&\ = 1 + a_s (\Gamma_0 L^2 + \gamma_0 L) \nn \\
&\qquad + a_s^2 \biggl[ \frac{\Gamma_0^2}{2} \Bigl(L^4 - \frac{2\pi^2}{3}L^2\Bigr) + \Gamma_0\gamma_0 \Bigl(L^3 - \frac{\pi^2}{3}L\Bigr) + \frac{\gamma_0^2}{2}\Bigl(L^2 - \frac{\pi^2}{6}\Bigr) + \Gamma_1 L^2\biggr] + \cdots \,, \nn
\end{align}
expanded to $\cO(a_s^2)$. Comparing to \eq{iLPs}, the first thing we notice is that \eq{RfixedorderNLL} is missing the $a_s (\pi^2/6)$ term required to accompany $L^2$ for it to Laplace transform properly to $\tilde L^2$. This is fine at NLL accuracy, since the missing $\cO(\as)$ term corresponds to a subleading term in the $\wt\sigma(\nu)$ counting [see \eq{Laplacefixedorder}]. However, we then notice \eq{RfixedorderNLL} also has the wrong $a_s^2\pi^2 L^2$ term required to make $L^4$ transform properly to $\tilde L^4$. This would lead to an incorrect $a_s^2 \tilde L^2$ term in the Laplace transform of \eq{RfixedorderNLL}, which should equal \eq{Laplacefixedorder}. [Continuing to $\cO(a_s^3)$ we would find the $a_s^3 L^4$ term in the Laplace transform of $R(\tau)$ also incorrect.] The mismatch occurs in terms that are completely green (NLL) in \eq{Laplacefixedorder}, thus making the supposedly NLL $R(\tau)$ in \eq{RfixedorderNLL} objectively \emph{less} accurate than $\wt\sigma(\nu)$ in \eq{Laplacefixedorder} at NLL. The similar problem occurs at \nkll order if \tab{LaplaceE} is used to keep the differential operator $e^{\Gamma\partial_\Omega^2}$ to the same order as the fixed-order coefficients $C(a_s)$. To be precise, the largest log at a given fixed order in $a_s$ that is incorrectly predicted at \nkll by using \eq{simpleRfixed} with the differential operator kept to the order specified for the fixed-order coefficients in \tab{LaplaceE} is:
\be
\label{eq:missinglogcumulant}
\begin{split}
&\text{largest incorrect log in \nkll $R(\tau)$ at fixed-order $\cO(\as^{k+n})$:} \\
&\qquad \sim \bigl[a_s^k \Gamma_0^k (\#)\bigr] \Bigl(a_s^{n} \frac{\Gamma_0^n}{n!} L^{2n}\Bigr) = a_s^{k+n}(\#)\frac{\Gamma_0^{n+k}}{n!} L^{2n}\,,
\end{split}
\ee
where $(\#)$ is the constant generated by taking the derivative $\partial_\Omega^{2k}$ of the $(\Omega+\gamma)^{2k}$ term in the series expansion \eq{DOmegaseries}. We see the term in \eq{missinglogcumulant} is \emph{larger} (by one power of a log) than the largest log missing in the \nkll $\wt\sigma(\nu)$ identified in \eq{missinglogLaplace}. This is undesirable for an expression labeled to be \nkll accurate, which should contain the equivalent information as in the \nkll Laplace transform.

These deficiencies are remedied by applying the \nkllprime counting in \tab{LaplaceE} to the differential operator $e^{\Gamma\partial_\Omega^2}$ and thus expanding it to one higher order than at \nkll order. For example, at NLL$'$ order, with the differential operator $e^{\Gamma\partial_\Omega^2}$ expanded to first order, $1+\Gamma\partial_\Omega^2$, and truncating this to $\cO(\as)$, $1+a_s\Gamma_0\partial_\Omega^2$, we find that $R(\tau)$ has the expansion to $\cO(\as^3)$:
\begin{align}
\label{eq:RfixedorderNLLprime}
R(\tau) \overset{\text{NLL$'$}}{=} &C(a_s) \biggl\{ 1 + a_s \Bigl[\Gamma_0\Bigl(L^2 - \frac{\pi^2}{6}\Bigr) + \gamma_0 L\Bigr ] \\
&+ a_s^2 \biggl[ \frac{\Gamma_0^2}{2}( L^4 - \pi^2 L^2 - 8\zeta_3 L)  + \Gamma_0\gamma_0 \Bigl(L^3 - \frac{\pi^2}{2}L - 2\zeta_3\Bigr) + \frac{\gamma_0^2}{2}\Bigl(L^2 - \frac{\pi^2}{6}\Bigr)  + \Gamma_1 L^2 \biggr] \nn \\
&+ a_s^3 \biggl[ \frac{\Gamma_0^3}{6}\Bigl( L^6 - \frac{5\pi^2}{2} L^4 - 40\zeta_3 L^3 \Bigr)  + \frac{\Gamma_0^2 \gamma_0}{2} \Bigl( L^5 - \frac{5\pi^2}{3} L^3 - 20\zeta_3 L^2 \Bigr) \nn \\
&\qquad + \frac{\Gamma_0\gamma_0^2}{2} \Bigl( L^4 - \pi^2 L^2 - 8\zeta_3 L \Bigr)  + \Gamma_0\Gamma_1\Bigl( L^4 - \frac{5\pi^2}{6} L^2 - 4\zeta_3 L \Bigr) \nn \\
&\qquad + \frac{\gamma_0^3}{6}  \Bigl( L^3 - \frac{\pi^2}{2} L - 2\zeta_3\Bigr) + \Gamma_1\gamma_0 \Bigl(L^3 - \frac{\pi^2}{3}L\Bigr)\biggr]\biggr\} \nn
\,.
\end{align}
Note that more of the displayed terms are now the correct inverse Laplace transforms of $\tilde L^n/\nu$, according to \eq{iLPs}, in contrast to \eq{RfixedorderNLL} at unprimed NLL order. Not all the terms in \eq{RfixedorderNLLprime} at $\cO(a_s^2)$ and $\cO(a_s^3)$ satisfy this property---for example, the $a_s^2\Gamma_1 L^2$ term. However, there are enough logs correctly predicted in \eq{RfixedorderNLLprime} that its Laplace transform reproduces all the terms in the fixed-order expansion \eq{Laplacefixedorder} of $\wt\sigma(\nu)$ that are supposed to be correct at NLL$'$ order [the green terms in \eq{Laplacefixedorder}].  That is, at NLL$'$ order,
\begin{align}
R(\tau) = \iLP\Biggl(\frac{1}{\nu} & C(a_s) \biggl\{ 1 + a_s \Bigl[\Gamma_0 \tilde L^2 + \gamma_0 \tilde L\Bigr ] + a_s^2 \biggl[ \frac{\Gamma_0^2}{2}\tilde L^4   + \Gamma_0\gamma_0 \tilde L^3 + \Bigl(\frac{\gamma_0^2}{2}+\Gamma_1\Bigr)\tilde L^2  \biggr]\\
&+ a_s^3 \biggl[ \frac{\Gamma_0^3}{6} \tilde L^6 +  \frac{\Gamma_0^2 \gamma_0}{2} \tilde L^5 + \Bigl( \frac{\Gamma_0\gamma_0^2}{2} + \Gamma_0\Gamma_1\Bigr) \tilde L^4 + \Bigl(\frac{\gamma_0^3}{6} + \Gamma_1\gamma_0\Bigr) \tilde L^3 + \cdots \biggr] \biggr\} \Biggr)
\,,\nn
\end{align}
where the $\cdots$ on the last line indicate terms at $\cO(a_s^2)$ and higher order that are truly subleading at NLL$'$ accuracy in $\wt\sigma(\nu)$. The largest log in the fixed-order expansion of $R(\tau)$ given by \eq{simpleRfixed} that is missing at \nkllprime order is:
\be
\label{missinglogcumulantprime}
\begin{split}
&\text{largest incorrect log in \nkllprime $R(\tau)$ at fixed-order $\cO(\as^{k+n})$:} \\
&\sim \bigl[a_s^{k+1} \Gamma_0^{k+1} (\#)\bigr] \Bigl(a_s^{n-1} \frac{\Gamma_0^{n-1}}{(n-1)!} L^{2n-2}\Bigr) = a_s^{k+n}\frac{\Gamma_0^{n+k}}{(n-1)!} L^{2n-2}\,,
\end{split}
\ee
which is two powers of logs smaller than the corresponding missing log in the \nkll $R(\tau)$ in \eq{missinglogcumulant}, and one power smaller than the corresponding missing log in the \nkll or \nkllprime $\wt\sigma(\nu)$ in \eq{missinglogLaplace}. Thus, the Laplace transform of the \nkllprime cumulant $R(\tau)$ evaluated using \eq{simpleRfixed} \emph{does} reproduce the \nkllprime $\wt\sigma(\nu)$ evaluated from \eq{simpleLaplacefixed}. 

Some of the subleading terms still missing in \eq{RfixedorderNLLprime} to make each term be the exact inverse Laplace transform of the corresponding term in \eq{Laplacefixedorder} can be restored by keeping the $\Gamma$ in the truncated differential operator $1+\Gamma \partial_\Omega^2$ to higher order. Others can be restored by keeping higher-order terms in the Taylor expansion of $e^{\Gamma\partial_\Omega^2}$, as in \eq{DOmegaexpansion}. All these are legitimate options, equivalent at subleading accuracy. The higher the order to which the expansion of the differential operator $e^{\Gamma\partial_\Omega^2}$ is kept, the higher the accuracy to which $R(\tau_a)$ computed using \eq{simpleRfixed} is in fact the correct inverse Laplace transform of $\wt\sigma(\nu)$ computed using \eq{simpleLaplacefixed}. The minimal prescription to maintain equivalence between the accuracy of $R(\tau)$ and $\wt\sigma(\nu)$ at NLL is to keep $e^{\Gamma\partial_\Omega^2}$ up to at least the $\Gamma\partial_\Omega^2$ term of the Taylor expansion, then truncated to $\cO(\as)$ or greater. For \nkll accuracy, one keeps the differential operator up to at least the $\Gamma^k\partial_\Omega^{2k}$ term of the Taylor expansion, then truncated to $\cO(\as^k)$ or greater. This prescription for evaluating $R(\tau)$ to \nkll accuracy is given in \tab{cumulant}. Using \tab{cumulant} ensures that the Laplace transform of $R(\tau)$ reproduces the logs in the fixed-order expansion of $\wt\sigma(\nu)$ \eq{Laplacefixedorder} that are fully predicted by exponentiating \eq{LaplaceN3LL} at the same order of accuracy.
\begin{table}[t]
\begin{center}
$
\begin{array}{ | c | c | c | c | c | c |}
\hline
\text{accuracy} & \Gamma_F & \gamma_F & \beta & \wt J,\wt S (\partial_\Omega\text{ terms})  & c_{H,J,S} \\  \hline
\text{LL} & \as & 1 & \as & 1  & 1 \\ \hline
\text{NLL} & \as^2 & \as & \as^2 & \as & 1  \\ \hline
\text{NNLL} & \as^3 & \as^2 & \as^3 & \as^2 & \as \\ \hline
\text{N$^3$LL} & \as^4 & \as^3 & \as^4 & \as^3 & \as^2 \\ \hline 
\end{array}
\quad
\begin{array}{ | c | c | c | c | c |}
\hline
\text{accuracy} & \Gamma_F & \gamma_F & \beta & H,\wt J,\wt S (\text{full}) \\  \hline
\text{LL} & \as & 1 & \as & 1  \\ \hline
\text{NLL}' & \as^2 & \as & \as^2 & \as  \\ \hline
\text{NNLL}' & \as^3 & \as^2 & \as^3 & \as^2 \\ \hline
\text{N$^3$LL}' & \as^4 & \as^3 & \as^4 & \as^3 \\ \hline 
\end{array}
$
\end{center}
\vspace{-1em}
\caption{Order of anomalous dimensions, beta function, and fixed-order hard, jet, and soft functions required to achieve \nkll and \nkllprime accuracy in the cumulant $R(\tau)$. Using this table instead of \tab{LaplaceE} at \nkll order ensures equivalent accuracy between $R(\tau)$ and $\wt\sigma(\nu)$ upon Laplace transformation of the former.}
\label{tab:cumulant} 
\end{table}

Up until now we have not considered the effect of the $C(a_s)$ factor in front of \eq{simpleLaplacefixed} or \eq{simpleRfixed}. The above considerations apply to maintaining equivalent accuracy between the logs of $\tau$ or $\nu$ predicted by expanding the exponentiated logs in $\wt\sigma(\nu)$ in \eq{simpleLaplacefixed} or the exponentials, derivative operators, and gamma function in $R(\tau)$ in \eq{simpleRfixed}. These terms determine the \emph{shape} of the distribution. If one multiplies through the expansion of $C(a_s)$ in \eq{Laplacefixedorder} (which affects the normalization), there are terms that are missing at \nkll order (with $C(a_s)$ truncated according to \tab{LaplaceE}) that are the same size as those missed by truncating the differential operator $e^{\Gamma\partial_\Omega^2}$ to the same order. For example, at NLL order, one of the leading effects of the missing $\cO(a_s)$ coefficient $C_1$ is a missing $a_s^2 C_1 \Gamma_0 \tilde L^2$ term in \eq{Laplacefixedorder}. Since this is the same size as the $a_s^2 \tilde L^2$ term we pointed out would be mis-predicted by truncating the differential operator $e^{\Gamma\partial_\Omega^2}$ to tree level in \eq{RfixedorderNLL}, it is fair to complain that one should not demand keeping the differential operator to higher order than the order to which the $C_n$ coefficients are known. Thus, it is not \emph{incorrect} to use \tab{LaplaceE} to evaluate all the objects in $R(\tau)$. We are pointing out that, as far as the exponentiation of the logs themselves in \eq{simpleLaplacefixed} is concerned, one \emph{loses} information that is contained in the exponent of \eq{simpleLaplacefixed} by Laplace transforming \eq{simpleRfixed} with the differential operator truncated according to \tab{LaplaceE} instead of \tab{cumulant}. This is information one already has available if one knows the anomalous dimensions needed at \nkll accuracy, and so it need not be thrown away. If we separate the counting of the exponentiated logs in \eq{simpleLaplacefixed} illustrated in \eq{Laplacefixedorder} from the fixed-order non-log coefficient $C(a_s)$, then \tab{cumulant} should be used to evaluate $R(\tau)$ in order to preserve the accuracy of the logs predicted in \eq{Laplacefixedorder} upon Laplace transformation.

For these reasons we say it is \emph{preferable} to compute $R(\tau)$ in momentum space using \tab{cumulant} instead of \tab{LaplaceE} when working to \nkll accuracy. And it is even better to work to full \nkllprime accuracy if possible. Then not only does $R(\tau)$ match the accuracy of the \nkllprime Laplace transform $\wt\sigma(\nu)$, the coefficients $C_n$ and terms in the differential operator $e^{\Gamma\partial_\Omega^2}$ are kept to a consistent accuracy. If the coefficients $C_n$ of the non-logarithmic terms required at \nkllprime accuracy are not available, then keeping just the differential operator terms in the jet and soft functions (or the $\partial_{E'}^n$ terms in the dQCD form) to \nkllprime accuracy in evaluating the cumulant $R(\tau_a)$  is sufficient to maintain equivalent accuracy with the \nkll Laplace transform $\wt\sigma(\nu_a)$, in the manner illustrated in \eqs{Laplacefixedorder}{RfixedorderNLLprime}.

\textbf{\emph{Nota bene:}} In remainder of the paper, we restore the running of $\as$.

\subsection{Distribution}
\label{ssec:countingdistribution}

The result of taking the inverse Laplace transform of \eq{sigmanuderivative} is the  expression for the resummed distribution $\sigma(\tau_a)$ given in \eq{SCETresummeddist}. For the canonical scale choices \eq{tauscalechoices}, this simplifies to:
\be
\label{eq:sigmatausimple}
\sigma(\tau_a) =  H_2(Q)\frac{e^K}{\tau_a} \wt J(\partial_\Omega,\mu_J^{\rm nat})^2 \wt S(\partial_\Omega,\mu_S^{\rm nat}) \left[\frac{e^{\gamma_E\Omega}}{\Gamma(-\Omega)}\right]\,,
\ee
with $K,\Omega$ given by \eq{canonicalKomega}. Alternatively, for the scale choices \eq{barcanonicalscales}, we obtain the form
\be
\label{eq:sigmataubar}
\sigma(\tau_a) =  H_2(Q)\frac{e^{\bar K}}{\tau_a} \wt J(\partial_{\bar\Omega},\bar\mu_J)^2 \wt S(\partial_{\bar\Omega},\bar\mu_S) \left[\frac{1}{\Gamma(-\bar\Omega)}\right]\,,
\ee
where $\bar K,\bar\Omega$ are given by \eq{barKomega}. These forms for $\sigma(\tau_a)$ correspond to making $\tau_a$-dependent scale choices \emph{after} differentiating the cumulant $R(\tau_a)$ given by \eq{SCETresummedcumulant}, which is the clockwise route in \fig{commute}.

For the distribution $\sigma(\tau_a)$, which is the derivative of the cumulant $R(\tau_a)$, one might expect that the accuracy of logarithmic resummation can be classified in the same way as for $R(\tau_a)$ or for $\wt\sigma(\nu)$, and thus achieved by calculating the quantities in \tab{LaplaceE} (or \tab{cumulant}) to the orders specified therein, in particular truncating the fixed-order coefficients $\wt J,\wt S$ to the specified accuracy. This is how one might interpret the prescriptions for computing $\sigma(\tau_a)$ using the form \eq{sigmatausimple} given in, e.g., \cite{Becher:2008cf}. The situation is slightly more subtle, however, and the truncation of $\wt J,\wt S$ according to \tab{LaplaceE} or \tab{cumulant} should not yet be performed in \eq{sigmatausimple} or \eq{sigmataubar}.

Consider the expression in \eq{sigmataubar}. Since the series expansion of  $\Gamma(-\bar\Omega)$ about $\bar\Omega=0$ starts with $-1/\bar\Omega$, it is not expedient to place it directly into the exponent of the cumulant by taking its log as in \eq{CTTWfromSCET}. Instead, let us pull out a factor of $-\bar\Omega$ to turn \eq{sigmatausimple} into as close a form as possible to the cumulant in \eq{radiatorsimple}. This results in
\be
\label{eq:sigmaOmega}
\sigma(\tau_a) = - H_2(Q)\frac{e^{\bar K}}{\tau_a}\wt J(\partial_{\bar\Omega})^2 \wt S(\partial_{\bar\Omega})\left[\frac{\bar\Omega}{\Gamma(1-\bar\Omega)}\right]\,.
\ee
Now pull the factor of $\bar\Omega$ through the functions of $\partial_{\bar\Omega}$ and relate derivatives of the quantity in brackets to those of $\cG(\bar\Omega)$ given  by \eq{Gbardefs}:
\begin{subequations}
\label{eq:Fdefs}
\begin{align}
\cF(\bar\Omega) &\equiv \bar\Omega \cG(\bar\Omega) \,, \\
\partial_{\bar\Omega}\cF(\bar\Omega) &= \bar\Omega\partial_{\bar\Omega}\cG(\bar\Omega) + \cG(\bar\Omega) \,, \\
\partial_{\bar\Omega}^2\cF(\bar\Omega) &= \bar\Omega\partial_{\bar\Omega}^2\cG(\bar\Omega) + 2\partial_{\bar\Omega}\cG(\bar\Omega) \,, \\
\partial_{\bar\Omega}^3\cF(\bar\Omega) &= \bar\Omega\partial_{\bar\Omega}^3\cG(\bar\Omega) + 3\partial_{\bar\Omega}^2\cG(\bar\Omega) \,, \\
\partial_{\bar\Omega}^4\cF(\bar\Omega) &= \bar\Omega\partial_{\bar\Omega}^4\cG(\bar\Omega) + 4\partial_{\bar\Omega}^3\cG(\bar\Omega)\,,
\end{align}
\end{subequations}
and so on. Thus, the distribution \eq{sigmaOmega} can be written very closely to the form of the cumulant \eq{radiatorbar}:
\be
\label{eq:sigmataufromR}
\sigma(\tau_a) = - H_2(Q)\frac{e^{\bar K}}{\tau_a} [ \bar\Omega \wt F(\partial_{\bar\Omega}) + \wt G(\partial_{\bar\Omega}) ] \Bigl[ \frac{1}{\Gamma(1-\bar\Omega)}\Bigr] \,,
\ee
where 
\be
\label{eq:Ftilde}
\wt F(\partial_\Omega) = \wt J(\partial_\Omega)^2\wt S(\partial_\Omega)
\ee
and
\be
\label{eq:Gtilde}
\wt G(L) \equiv \frac{d\wt F(L)}{dL}\,,
\ee
where $L$ is a variable standing in for $\partial_{\bar\Omega}$. The extra terms on each line of \eq{Fdefs} show that $\wt G$ can be constructed from $\wt F$ by differentiation with respect to the $\partial_{\bar\Omega}$ operator itself.

The form \eq{sigmataufromR} is very similar to the cumulant \eq{radiatorsimple}, but presents a conundrum: how do we deal with the prefactors $\wt F,\wt G$ in defining \nkll accuracy for $\sigma(\tau_a)$?  There is now an overall factor of $1/\tau_a$, which should be counted as one log. This promotes the terms in the prefactor $H[\Omega \wt F + \wt G]$ to one higher order in log accuracy than in the cumulant. Applying the power counting that $\as\ln\tau\sim 1$, we would conclude that the tree-level terms in $H,\wt F,\wt G$ are LL order, the $\cO(\as)$ terms are NLL, $\cO(\as^2)$ terms are NNLL, etc.
Strictly applying this power counting, even the constant terms $c_{H,J,S}$ in the hard, jet, and soft functions are promoted to one higher order than in the cumulant alone. This is certainly a possible choice of convention, which corresponds to \nkllprime accuracy in \tab{LaplaceE} and \tab{cumulant}.  As with the cumulant, working to \nkll accuracy in \tab{cumulant} is preferable to \nkll accuracy in \tab{LaplaceE}, and \nkllprime accuracy is even better, as it obtains the correct \nkll terms as well as the complete $\ord{\as^k}$ fixed-order singular terms in the distribution.

We now explore an alternative approach to the distribution, which is to define a prescription so that it equals the derivative of the cumulant at a given order of accuracy, or at least as close to this as possible that we can achieve in practice.

\subsubsection{Keeping Cumulant and Distribution to Consistent Accuracy}
\label{ssec:distributionaccuracy}

We wish to define the \nkll distribution $\sigma(\tau_a)$ by identifying it with the derivative of the \nkll cumulant $R(\tau_a)$ with respect to $\tau_a$, so that integrating the \nkll distribution gives back  the \nkll cumulant.  We will find that first resumming the cumulant in SCET using \eq{SCETresummedcumulant} and then differentiating it (following the counterclockwise route in \fig{commute}) gives a distribution with a closer correspondence with the direct QCD resummation of the cross section than does using \eq{SCETresummeddist} (resulting from the clockwise route in \fig{commute}) directly.  

It is possible to reorganize the result of following the clockwise route in \fig{commute} so that at any truncated \nkll order it more closely matches the result of following the counterclockwise route. Above, we took the formula \eq{sigmataubar} for the SCET distribution obtained by choosing the scales \eq{barcanonicalscales} and plugging into \eq{SCETresummeddist} (clockwise, \fig{commute}), and reorganized its pieces into the form \eq{sigmataufromR}.  This form \eq{sigmataufromR} can be truncated in such a way as to correspond to the derivative of the cumulant \eq{radiatorbar} (counterclockwise, \fig{commute}) at \nkll accuracy, namely, by truncating the terms in $H,\tilde J,\tilde S$ in \eq{sigmataufromR} according to the usual \nkll \tab{LaplaceE}, except for the $\wt G(\partial_{\bar\Omega})$ term produced by pulling $\bar\Omega$ through the $\wt J^2(\partial_{\bar\Omega})\wt S(\partial_{\bar\Omega})$ operators in \eq{sigmaOmega}. We will find that we must keep a subset of terms in $\wt G$ to higher order than we do for $\wt F$ specified in \tab{LaplaceE} in order for the \nkll distribution to equal the derivative of the \nkll cumulant. Truncating $\tilde J,\tilde S$ in \eq{sigmatausimple} according to \tab{LaplaceE} \emph{before} pulling $\bar\Omega$ through the differential operators in \eq{sigmaOmega} will \emph{not} produce the derivative of the full \nkll radiator but will be missing terms that are in $\wt G$ in \eq{sigmataufromR}.  Some of these missing terms are in fact required at unprimed \nkll accuracy; at \nkllprime accuracy, they are formally subleading, but ensure closer numerical correspondence with the derivative of the cumulant. If full \nkllprime results are not available, the truncation rules for \nkll accuracy in \tab{cumulant} also suffice to make \eq{sigmataufromR} match the accuracy of the \nkll cumulant.

Since remembering which terms in $\wt G$ are strictly necessary can be cumbersome, the most straightforward way to compute the differential distribution so that it matches the accuracy of the \nkll cumulant is simply to follow the counterclockwise route in \fig{commute} and differentiate after choosing appropriate $\tau_a$-dependent scales. We will derive below a generic formula for the result of this procedure.  This procedure guarantees that the resulting distribution $\sigma(\tau_a)$ matches the accuracy of the cumulant $R(\tau_a)$ that we started with. 

We note that this approach of differentiating the resummed cumulant to obtain the distribution has been previously considered, and two potential (related) issues have been noted~\cite{Abbate:2010xh}.  First, the uncertainties from scale variation may be unreasonably small in the tail region of the distribution, and second, the accuracy of the tail region when matching onto fixed order may be compromised.  These complications arise because the tail region of the distribution has large cancellations between singular terms (those participating in the resummation) and nonsingular terms.  Care is required when turning off the resummation in the tail region to preserve this delicate cancellation and also produce reliable uncertainties.  While these problems will not be solved here, we hope that the approach taken here offers ways to address these issues in the future.

Following the counterclockwise route in \fig{commute}, we take the derivative of the cumulant \emph{after} setting the scales $\mu_{J,S}$ to be $\tau_a$-dependent.  For generality, we will allow for arbitrary functional dependence $\mu_{J,S}=\mu_{J,S}(\tau_a)$ (generalizing the canonical choices in \fig{commute}). Starting from \eq{SCETresummedcumulant} and taking the $\tau_a$ derivative:
\begin{align} \label{eq:dRdtau}
\frac{dR}{d\tau_a} &= \sigma_n (\tau_a) + \frac{1}{\tau_a} \exp(K_H + 2K_J + K_S) \Bigl( \frac{\mu_H}{Q} \Bigr)^{\omega_H} \Bigl( \frac{\mu_J}{Q \tau_a^{1/j_J}} \Bigr)^{2j_J \omega_J} \Bigl( \frac{\mu_S}{Q\tau_a} \Bigr)^{\omega_S} H_2 (Q^2, \mu_H) \nn \\
& \quad \times \biggl\{ \sum_{F = J_n,J_{\bn},S} \frac{d\ln \mu_F}{d\ln \tau_a} \biggl[ \frac{dK}{d\ln\mu_F} + j_F \omega_F + j_F \frac{d\omega_F}{d\ln \mu_F} \ln \frac{\mu_F}{Q \tau_a^{1/j_F}} \\
& \qquad\qquad\qquad\qquad\qquad\quad + \frac{d\Omega}{d\ln\mu_F} \p_\Omega + \frac{d}{d\ln \mu_F} \ln \wt{F} \Bigl( \p_\Omega + \ln \frac{\mu_F^{j_F}}{Q^{j_F} \tau_a} , \mu_F \Bigr) \biggr]  \biggr\} \nn \\
& \quad \times \wt{J}_n \Bigl( \p_\Omega + \ln \frac{\mu_J^{j_J}}{Q^{j_J} \tau_a} , \mu_J \Bigr) \wt{J}_\bn \Bigl( \p_\Omega + \ln \frac{\mu_J^{j_J}}{Q^{j_J} \tau_a} , \mu_J \Bigr) \wt{S} \Bigl( \p_\Omega + \ln \frac{\mu_S}{Q\tau_a}, \mu_S \Bigr) \frac{\exp(\gamma_E \Omega)}{\Gamma(1 - \Omega)} \,. \nn
\end{align}
The spectrum $\sigma_n (\tau_a)$ in the first term is the ``natural'' SCET spectrum, given in \eq{SCETresummeddist}.  The jet functions $J_n$ and $J_{\bn}$ and their profiles $\mu_{J_n}$ and $\mu_{J_\bn}$ are treated separately to avoid the combinatoric factors from a common jet function label.  Using the relations
\begin{align}
\frac{dK}{d\ln\mu_F} &= -j_F \omega_F - \gamma_F [\as(\mu_F)] \,, \nn \\
\frac{d\omega_F}{d\ln\mu_F} &= - \frac{2}{j_F} \Gamma_F [\as(\mu_F)] \,, \nn \\
\frac{d}{d\ln\mu_F} \wt{F} \Bigl( \p_\Omega + \ln \frac{\mu_F^{j_F}}{Q^{j_F} \tau_a} , \mu_F \Bigr) &= \Biggl[ \beta[\as(\mu_F)] \frac{\p \wt{F} (\mathbf{L}_F, \mu_F)}{\p \alpha_s} + j_F \frac{d\wt{F} (\mathbf{L}_F, \mu_F)}{d\mathbf{L}_F} \Biggr] \,, \nn \\
\textrm{with} & \;\; {\mathbf{L}_F = \p_\Omega + \ln \mu_F^{j_F} / Q^{j_F} \tau_a} \,,
\end{align}
we can write the spectrum $\sigma_R$, which is the derivative of the cumulant when working to all orders:
\begin{align} \label{eq:sigmaR}
\sigma_R (\tau_a) &= \frac{dR}{d\tau_a} = \sigma_n (\tau_a) + \delta\sigma_R (\tau_a) \,, \\
\delta\sigma_R (\tau_a) &= \frac{1}{\tau_a}  \exp( K_H + 2K_J + K_S ) \Bigl( \frac{\mu_H}{Q} \Bigr)^{\omega_H} \Bigl( \frac{\mu_J}{Q \tau_a^{1/j_J}} \Bigr)^{2j_J \omega_J} \Bigl( \frac{\mu_S}{Q \tau_a} \Bigr)^{\omega_S} H_2 (Q^2, \mu_H)  \nn \\
& \quad \times\Biggl\{ \sum_{F = J,J,S} \wt{P}_F (\mathbf{L}_F, \mu_F) \frac{1}{\wt{F} (\mathbf{L}_F, \mu_F)} \Biggr\} \wt{J} ( \mathbf{L}_J, \mu_J)^2 \wt{S} ( \mathbf{L}_S, \mu_S) \frac{\exp(\gamma_E \Omega)}{\Gamma(1 - \Omega)} \,, \nn \\
\textrm{with} &\quad \mathbf{L}_J = \partial_\Omega + \ln \mu_J^{j_J} / Q^{j_J} \tau_a \,, \quad \mathbf{L}_S = \partial_\Omega + \ln \mu_S / Q \tau_a \,. \nn
\end{align}
The function $\wt{P}_F$ is defined
\begin{align} \label{eq:PF}
\wt{P}_F (L_F, \mu_F) &= \frac{d \ln \mu_F}{d \ln \tau_a} \biggl\{ j_F \frac{\partial \wt{F} (L_F, \mu_F)}{\partial L_F} + \beta[\as(\mu_F)] \frac{\partial \wt{F} (L_F, \mu_F)}{\partial \as} - \wt{\gamma}_F (L_F, \mu_F) \wt{F} (L_F, \mu_F) \biggr\} \,.
\end{align}
The $\wt \gamma_F$ in the last term of \eq{PF} takes the same form as the usual anomalous dimension in \eq{gammaFLaplace},
\be
\wt{\gamma}_F (L_F, \mu) \equiv - \kappa_F \Gcusp^q [\as] L_F + \gamma_F [\as] \,,
\ee
but in \eq{sigmaR} it becomes a differential operator upon making the replacements $L_F\to \mathbf{L}_F$ given in the last line of \eq{sigmaR}.
In \eq{sigmaR}, the factor of $1/\wt{F}$ is present to cancel the factor of $\wt{F}$ present in $\wt{J}^2 \wt{S}$.  Note that in this form there is no need to separately label the two jet functions [as there was in \eq{dRdtau}].  

Since \tab{LaplaceE} (or \tab{cumulant} if using the prescription discussed in \ssec{Laplacecomparison}) specifies to what accuracy the various ingredients in the cumulant are kept when working to a given order, for $\sigma_R$ \tab{LaplaceE} or \tab{cumulant} should also be used to define the orders of resummed accuracy.  The fact that the spectrum $\sigma_n$ and the derivative of the cumulant are equivalent when working to all orders implies that $\wt{P}_F$ vanishes if all orders of accuracy are kept.  In fact, working to N$^k$LL$'$, $\wt{P}_F$ starts at $\ord{\as^{k+1}}$. [However, working to \nkll\!\!, $\wt{P}_F$ starts at $\cO(\as^k)$.] Although we have derived $\sigma_R$ by differentiating the cumulant, the way in which the truncation occurs at a given finite order of accuracy is different between the cumulant and $\sigma_R$.  For example, the all-orders  evolution kernel $\eta_\Gamma$ in \eq{Ketadef} and cusp anomalous dimension $\Gamma_{\text{cusp}}$ obey the exact relation  
\be
\label{eq:detadmu}
\frac{d\eta_\Gamma(\mu,\mu_F)}{d\ln \mu_F} = -\Gamma_{\text{cusp}}^q [\as(\mu_F)]
\ee
but at a finite order of accuracy the function defining $\eta_\Gamma$ is truncated in such a way that its derivative differs at the level of higher order terms beyond the requested order of accuracy. For example, at NLL, using \eqs{Ketaforms}{etaclosedform} and the counting of anomalous dimensions and beta functions terms in \tab{LaplaceE}, we find
\be
\frac{d(\eta_\Gamma^{\text{LL}} \plus \eta_\Gamma^{\text{NLL}})}{d\ln\mu_F} = -\frac{\as(\mu_F)}{4\pi} \Gamma_0 - \Bigl(\frac{\as(\mu_F)}{4\pi}\Bigr)^{\!2} \Gamma_1 - \Bigl(\frac{\as(\mu_F)}{4\pi}\Bigr)^{\!3} \Bigl(\frac{\Gamma_1\beta_1}{\beta_0} - \frac{\Gamma_0\beta_1^2}{\beta_0^2}\Bigr)\,,
\ee
thus with both sides of \eq{detadmu} truncated at NLL accuracy, the relation holds only up to subleading terms [here $\cO(\as^3)$]. These kinds of mismatches, formally subleading, lead to small numerical differences in the values of $\sigma_R$ in \eq{sigmaR} and $dR/d\tau_a$ computed by differentiating \eq{SCETresummedcumulant} when truncated to a finite accuracy.

The form in \eq{sigmaR} is fully generic, and simplifies significantly with simple scale choices such that we can compare to the form in \eq{sigmataufromR}.  For the scale choice in \eq{barcanonicalscales},
\be
\label{eq:radiatorderivative2}
\sigma_R(\tau_a) =   H_2(Q)e^{\bar K} \biggl[- \frac{1}{\tau_a}\Bigl( \bar\Omega + \sum_F \frac{1}{j_F}\bar\gamma_F(\dOmegabar)\Bigr) \wt F(\dOmegabar)  + \frac{d\wt F(\dOmegabar)}{d\tau_a} \biggr] \frac{1}{\Gamma(1-\bar\Omega)}\,,
\ee
where the operator $\bar\gamma_F(\dOmegabar)$ is defined
\be
\label{eq:gammaFoperator}
\bar\gamma_F(\dOmegabar) = -\kappa_F \Gamma_\text{cusp}^q[\as(\bar\mu_F)] \dOmegabar + \gamma_F[\as(\bar\mu_F)]\,,
\ee
which is formed from the anomalous dimension \eq{gammaFLaplace} replacing $L_F$ with $\dOmegabar$. 

If we start with the cumulant $R$ at a given order of \nkll accuracy with $\gamma_F,\Gamma_F,\wt F$ computed to the orders given in \tab{LaplaceE} or \tab{cumulant}, then \eq{radiatorderivative2} is the differential distribution at the equivalent order of \nkll accuracy, where each piece of \eq{radiatorderivative2} is truncated to the same order as in $R(\tau_a)$ itself. Suppose, however, we start directly with the formula \eq{sigmataufromR} for the distribution $\sigma(\tau_a)$. Note that the first term in brackets of each of \eqs{sigmataufromR}{radiatorderivative2} are the same. Thus the remaining terms in \eq{radiatorderivative2} must correspond to the terms in the $\wt G(\dOmegabar)$ operator in \eq{sigmataufromR}. Note that we must keep the terms in $\wt G$ to higher order than $\wt F$ in \tab{LaplaceE} or \tab{cumulant} to keep the two formulae \eqs{sigmataufromR}{radiatorderivative2} equivalent. For example, using the \tab{LaplaceE} prescription for computing to NLL accuracy, we start with the formula \eq{radiatorderivative2}, and keep $\wt F$ to $\cO(1)$, $\gamma_F$ to $\cO(\as)$ and $\Gamma_\text{cusp}^q$ to $\cO(\as^2)$. Then we obtain the result
\begin{align}
\label{eq:dRdtNLL}
\frac{dR_{\text{NLL}}}{d\tau_a} = &-  \frac{e^{\bar K_{\text{NLL}}}}{\tau_a}\biggl(\bar\Omega_{\text{NLL}} + \sum_{F}\frac{\as(\bar\mu_F)}{4\pi}\frac{1}{j_F} \biggl\{\gamma_F^0 - \kappa_F\biggl[ \Gamma_\text{cusp}^0 + \Gamma_\text{cusp}^1\Bigl(\frac{\as(\bar\mu_F)}{4\pi}\Bigr) \biggr]\dOmegabar\biggr\} \biggr) \nn \\
&\times  \frac{1}{\Gamma(1-\bar\Omega_{\text{NLL}})}\,.
\end{align}
Meanwhile, from the formula \eq{sigmataufromR}, applying the rules in \tab{LaplaceE} na\"ively to $\wt F$ and $\wt G$ [that is, to $\wt F$ already in \eq{sigmaOmega}], we would obtain the formula
\be
\sigma(\tau_a) = - \frac{1}{\tau_a} e^{\bar K} \bar\Omega_{\text{NLL}} \frac{1}{\Gamma(1-\bar\Omega_{\text{NLL}})}\,,
\ee
which contains only the first term in parentheses in \eq{dRdtNLL}. The extra terms in \eq{dRdtNLL} are contained as a subset of the terms in the $\wt G$ operator in \eq{sigmataufromR}:
\begin{align}
\label{eq:G2}
\wt G(\dOmegabar) &= \sum_F  \biggl\{ \frac{\as(\bar\mu_F)}{4\pi} \frac{1}{j_F} \mbox{\boldmath$\Bigl( -\kappa_F\Gamma_0 \dOmegabar + \gamma_F^0\Bigr)$}  + \Bigl(\frac{\as(\bar\mu_F)}{4\pi} \Bigr)^2 \biggl[ \frac{(\kappa_F\Gamma_0)^2}{2j_F^2} \dOmegabar^3 \\ 
&\qquad - \frac{\kappa_F\Gamma_0}{2j_F^2} (3\gamma_F^0 + 2\beta_0)\dOmegabar^2 
+ \frac{1}{j_F^2} \Bigl(\mbox{\boldmath$-\kappa_Fj_F\Gamma_1 $} + (\gamma_F^0)^2 + 2\gamma_F^0\beta_0 - j_F\kappa_F c_F^1 \Gamma_0\Bigr) \dOmegabar \nn \\
&\qquad + \frac{1}{j_F} (\gamma_F^1 + c_F^1\gamma_F^0 + 2 c_F^1\beta_0)\biggr] + \cdots \biggr\}\,, \nn
\end{align}
which is computed using \eqs{Ftilde}{Gtilde} and the expansions of $\wt J,\wt S$ to $\cO(\as^2)$ given by \eq{Ffixedorder}. The dots indicate additional $\cO(\as^2)$ cross terms and higher order terms in $\as$, and we have indicated in bold where the extra terms in \eq{dRdtNLL} are found, namely the $\cO(\as)$ part and the $\Gamma_1$ term of the $\cO(\as^2)$ part of \eq{G2}. Thus, we see that at least a subset of terms in $\wt G$ are needed to higher order than those in $\wt F$ in \eq{sigmataufromR} in order to reproduce the result of differentiating the cumulant in \eq{dRdtNLL}. And these terms are contained in $\wt P_F$ in \eq{PF} at NLL accuracy. 

This pattern continues at higher orders. At unprimed \nkll order, the extra terms in $\wt G$ that are contained in $\wt P_F$ in \eq{PF} are required to be kept to maintain the same accuracy as the \nkll cumulant. At primed \nkllprime orders, the extra terms are subleading---the leading ones are actually captured by the $\sigma_n$ term in \eq{sigmaR}, but the additional terms in $\delta\sigma_R$ ensure closer numerical equivalence with the derivative of the cumulant.

To recap, our proposed formula \eq{sigmaR} for computing the differential distribution $\sigma(\tau_a)$ adds to the usual form terms that are necessary to maintain equivalent  \nkll accuracy with the cumulant or Laplace transform. These terms are not strictly needed at \nkllprime accuracy, where the modifications are subleading. Thus our proposed formula does not differ at leading accuracy from any \nkllprime results for differential event shape distributions in the literature, e.g. those in \cite{Hornig:2009vb,Abbate:2010xh}. The extra terms are required when evaluating the distribution directly at unprimed \nkll accuracy, using the standard formula (equivalent to $\sigma_n$) that appears in, e.g., \cite{Becher:2008cf,Becher:2009th,Hornig:2009vb}. Thus they should be added to, e.g., the NLL results in \cite{Hornig:2009vb}. We note that distributions obtained by differentiation of \nkll cumulants do retain the correct accuracy even at unprimed order, as in, e.g., \cite{Schwartz:2007ib,Becher:2008cf,Chien:2010kc,Kang:2013nha}.
To ensure that the resummed differential distribution given by \eq{SCETresummeddist} integrates to the \nkll cumulant computed from \eq{SCETresummedcumulant}, we compute the distribution $\sigma_R$ in \eq{sigmaR}. In that formula, each piece including the fixed-order functions $\wt F$ can be truncated according to the rules in \tab{LaplaceE} or \tab{cumulant}, and then $\sigma_R$ will maintain the same accuracy as the corresponding cumulant with the pieces truncated according to the same rules.

\subsubsection{Counting Accuracy of Cumulant and Distribution in dQCD Formalism}
\label{ssec:dQCDcounting}

We can consider also how to compute the distribution and cumulant from the QCD-inspired formalism in \ssec{resumequiv} to consistent accuracy. We can differentiate the radiator in \eq{SCETcsEfinal} [after the scale choices in \eq{barcanonicalscales}], or we can differentiate \eq{SCETcsE} first and then plug in the scale choices \eq{barcanonicalscales}. The first procedure gives:
\be
\label{eq:sigmatauE1}
\begin{split}
\sigma(\tau_a) &= -\frac{1}{\tau_a} \frac{dR}{d\ln(1/\tau_a)} \\
&= -\frac{1}{\tau_a} H_2(Q) \wt J(0,Q)^2 \wt S(0,Q)   \\
&\quad\times \Biggl[ \sum_{m=0}^\infty  \frac{1}{m!} \bar E^{(m+1)} \partial_{\bar E'}^m\Biggr] \exp\biggl[ \bar E + \sum_{n=2}^\infty \frac{1}{n!}  \bar E^{(n)} \partial_{\bar E'}^n   \biggr]\frac{1}{\Gamma(1-\bar E')}\,,
\end{split}
\ee
where the operator in the first set of brackets is formed by differentiating the argument of the exponential. Meanwhile the second procedure gives:
\be
\label{eq:sigmatauE2}
\begin{split}
\sigma(\tau_a)  &=  H_2(Q) \wt J(0,Q)^2 \wt S(0,Q) \exp\biggl[ \bar E + \sum_{n=2}^\infty \frac{1}{n!}  \bar E^{(n)} \partial_{\bar E'}^n   \biggr]  \frac{1}{\tau_a} \frac{-\bar E'}{\Gamma(1-\bar E')} \\
&= -\frac{1}{\tau_a} H_2(Q) \wt J(0,Q)^2 \wt S(0,Q)   \\
&\quad \times \biggl[ \bar E' + \sum_{m=2}^\infty \frac{1}{(m-1)!} \bar E^{(m)} \partial_{\bar E'}^{m-1}\biggr] \exp\biggl[ \bar E + \sum_{n=2}^\infty \frac{1}{n!}  \bar E^{(n)} \partial_{\bar E'}^n   \biggr] \frac{1}{\Gamma(1-\bar E')}\,,
\end{split}
\ee
where the terms in the first set of brackets are formed by moving the $\bar E'$ in the numerator on the first line through the differential operator in front of it. The operators summed over $m$ on the last line turn out to be the derivatives of the operators summed over $n$ on the first line with respect to $\partial_{\bar E'}$ itself. With a change of index ($m\to m+1$), we note that \eq{sigmatauE2} is precisely equal to \eq{sigmatauE1}. 

\begin{table}[t]
\begin{center}
$
\begin{array}{ | c | c | c | c | c | c |}
\hline
\text{accuracy} & \Gamma_F & \gamma_F & \beta &  \bar E^{(j)} & H,\wt J,\wt S \\  \hline
\text{LL} & \as & 1 & \as & j=1,\as  & 1 \\ \hline
\text{NLL} & \as^2 & \as & \as^2 & j=2, \as & 1  \\ \hline
\text{NNLL} & \as^3 & \as^2 & \as^3 & j=3, \as^2 & \as \\ \hline
\text{N$^3$LL} & \as^4 & \as^3 & \as^4 & j=4, \as^3 & \as^2 \\ \hline 
\end{array}
\quad
\begin{array}{ | c | c | c | c | c | c |}
\hline
\text{accuracy} & \Gamma_F & \gamma_F & \beta &  \bar E^{(j)} & H,\wt J,\wt S \\  \hline
\text{LL} & \as & 1 & \as & j=1,\as  & 1 \\ \hline
\text{NLL}' & \as^2 & \as & \as^2 & j=2,\as & \as  \\ \hline
\text{NNLL}' & \as^3 & \as^2 & \as^3 & j=3,\as^2 & \as^2 \\ \hline
\text{N$^3$LL}' & \as^4 & \as^3 & \as^4 & j=4,\as^3 & \as^3 \\ \hline 
\end{array}
$
\end{center}
\caption{Order of anomalous dimensions, beta function, coefficients $\bar E^{(n)}$ of derivative operators, and non-log coefficients in $H,\wt J,\wt S$ in dQCD form \eq{SCETcsE} of cumulant $R(\tau)$ required to achieve \nkll and \nkllprime accuracy. These tables are also applicable to the form for the distribution $\sigma(\tau_a)$ in the \emph{last} line of \eq{sigmatauE2}.}
\label{tab:Ecounting} 
\end{table}

In the resummed cumulant \eq{SCETcsEfinal}, the number of terms we keep in the differential operator is very simple: keep only $\bar E$ at LL, keep up to the $\bar E''$ term (truncating the exponential derivative at $\cO(\as)$) at NLL, keep up to $\bar E^{(3)}$ (truncated at $\cO(\as^2)$) at NNLL, up to $\bar E^{(4)}$ (truncated at $\cO(\as^3)$) at N$^3$LL, etc. (This counting, summarized in \tab{Ecounting}, corresponds to the rules in \tab{cumulant}.) The distribution \eq{sigmatauE1} given by differentiating that formula then has terms in the sum in the prefactor in brackets up to at $\bar E''$ at LL, $\bar E^{(3)}$ at NLL, $\bar E^{(4)}$ at NNLL, etc. To get \eq{sigmatauE2} to agree exactly with this result, it does not suffice to truncate the differential operator in the first line according to this same scheme. At NLL, if we truncate the sum starting at $\bar E^{(3)}$ entirely from the start, we will be missing the $\bar E^{(3)}$ term in the last line of \eq{sigmatauE2} which is present in the last line of \eq{sigmatauE1} if we start with the NLL $R(\tau_a)$ in the first line. The similar mismatch occurs at higher \nkll orders. The mismatched terms are formally of subleading order. But if one is interested in keeping the derivative of the cumulant and the directly computed differential distribution in \eqs{sigmatauE1}{sigmatauE2} numerically equal to each other, then, similarly to the SCET formalism in the section above, one should keep an extra term in the sum over derivatives in the first set of brackets in the last line of \eq{sigmatauE2} than specified by \tab{Ecounting}.

\section{Final Formulae and Prescriptions for \nkll\!\!($'$) Accuracy}
\label{sec:summary}

In this section, we collect in summary form the formulae that can be used to obtain the Laplace transform, cumulant, and distribution at a consistent order of \nkll accuracy, in both SCET and QCD-inspired forms. These results follow from the detailed discussion in \sec{schemes}.

The standard counting rules in \tab{LaplaceE} for achieving \nkll or \nkllprime accuracy apply to the formulae \eqs{Laplacevariablescales}{LaplacevariableE} for $\wt\sigma(\nu)$ below. The terms resummed in perturbation theory by computing these formulae according to the rules in these tables form our baseline definition for these orders of accuracy. Our prescriptions for computing $R(\tau_a)$ or $\sigma(\tau_a)$ are motivated by the requirement that they reproduce the accuracy of $\wt\sigma(\nu)$ computed in this way, upon Laplace transformation.

For the cumulant we give two forms, in SCET and dQCD-like notation, in \eqs{cumulantvariable}{cumulantvariableE}. Our proof in \sec{compare} of the equivalence of these forms is a central technical result of this paper. Thus the dQCD-inspired form \eq{cumulantvariableE} below can actually be viewed as a SCET form as well, using the definitions \eqs{EJdef}{ESdef} for the exponent. The rules in \tab{cumulant}, in which derivative operator terms in $R(\tau_a)$ are kept to one higher order in $\as$ than implied by \tab{LaplaceE}, ensure that the Laplace transform of $R(\tau_a)$ reproduces the result of computing $\tilde\sigma(\nu)$ in \eq{Laplacevariablescales} or \eq{LaplacevariableE} according to the rules in \tab{LaplaceE}, order by order in $\as$ to the accuracy illustrated in \ssec{Laplacecomparison}. \eq{cumulantvariableE} is the first time to our knowledge that a dQCD form for $R(\tau_a)$ with variable jet and soft scales and with derivative operator terms up to arbitrarily high order has been given (see \cite{Contopanagos:1993xh} for a similar form with fixed scales in the context of threshold resummation). 

For the differential distribution $\sigma(\tau_a)$, our final result, which we call $\sigma_R$, is \eq{sigmaRfinal}. In this form, it is safe to truncate the ingredients according to the same rules as the cumulant $R(\tau_a)$ in \tab{cumulant}. We write $\sigma_R$ as a sum of two terms $\sigma_n + \delta\sigma_R$, where the first piece $\sigma_n$ is the usual way \eq{SCETresummeddist} the SCET differential distribution is written, as in \cite{Becher:2008cf,Becher:2009th,Hornig:2009vb}. We illustrated in \ssec{countingdistribution} that applying the truncation rules for \nkll accuracy in \tab{LaplaceE} or \tab{cumulant} directly to $\sigma_n$ would yield a result less accurate than taking the derivative of $R(\tau_a)$ computed at the same accuracy. At \nkllprime order, the accuracy of $\sigma_n$ or $\sigma_R$ is formally the same, but the integral of $\sigma_R$ numerically matches the cumulant better. (We note the final results in \cite{Becher:2008cf} were computed in terms of the cumulant, not $\sigma(\tau)$, so those results do not suffer from these issues.) The extra term $\delta\sigma_R$ restores the missing pieces. Our formula \eq{sigmaRfinal} is to our knowledge the first time a generic formula for the resummed differential distribution $\sigma(\tau_a)$ that possesses this automatic equivalence to $dR/d\tau_a$ has been written down.

It would be instructive to perform similar analyses and comparisons for the event shape resummation performed in \cite{Abbate:2010xh} using the formalism of \cite{Ligeti:2008ac}; an exercise that nevertheless lies outside the scope of the present paper.

In the following subsections, those equations which are \boxed{\text{boxed}} represent our final forms for the Laplace transform, cumulant, and differential distribution, which exhibit full dependence on the hard, jet, and soft scales $\mu_{H,J,S}$ and which are also written in a form where truncation of the ingredients according to the appropriate table in the text will preserve the resummed logarithmic accuracy of the expression, as defined by equivalence to the accuracy of $\tilde\sigma(\nu)$. 

\subsection{Laplace Transform}
For the Laplace transform $\wt\sigma(\nu)$, \eq{sigmanuderivative} gives with variable scales, in standard SCET form,
\begin{equation}
\label{eq:Laplacevariablescales}
\boxed{
\begin{split}
\wt\sigma(\nu) &=  e^K H_2(Q^2,\mu_H)\left(\frac{\mu_H}{Q}\right)^{\omega_H(\mu,\mu_H)} \left(\frac{\mu_J \nu^{1/j_J}}{Q}\right)^{2j_J\omega_J(\mu,\mu_J)}\left(\frac{\mu_S\nu}{Q}\right)^{\omega_S(\mu,\mu_S)-\Omega} \\
&\quad\times \wt J\Bigl(\partial_\Omega + \ln \frac{\mu_J^{j_J}}{Q^{j_J-1} \mu_S},\mu_J \Bigr)^2 \wt S(\partial_{\Omega},\mu_S)  \left(\frac{\mu_S e^{\gamma_E}\nu}{Q}\right)^{\Omega} \,,
\end{split}
}
\end{equation}
where $K,\Omega$ are defined in \eq{KOmega} and $\omega_F$ in \eq{Komegadef}. $H_2$ is the hard function appearing in the factorization theorem \eq{factorization}, and $\wt J,\wt S$ are the Laplace transforms of the jet and soft functions $J,S$ appearing therein. $H_2$ is given to $\cO(\as)$ in SCET by \eq{hardoneloop}, and the momentum space jet and soft functions by \eqs{jetoneloop}{softoneloop}. The generic definitions of $J,S$ are given in \eqs{SCETjet}{SCETsoft}.

In the dQCD-inspired form that was shown in \ssec{resumequiv} to be equivalent to SCET with the identification in \eqs{EJdef}{ESdef}, we have, by transforming \eq{SCETcsE},
\be
\label{eq:LaplacevariableE}
\boxed{
\begin{split}
\wt\sigma(\nu) &=  H_2(Q^2,\mu_H) \Bigl( \frac{\mu_H}{Q}\Bigr)^{\omega_H(\mu,\mu_H)}e^{K_H(\mu,\mu_H)} \wt J(0,\mu)^2 \wt S(0,\mu) \\
&\quad \times e^{2E_J(\mu,\mu_J) + E_S(\mu,\mu_S)} \exp\biggl[ \sum_{n=2}^\infty \frac{1}{n!} \Bigl( 2 E_J^{(n)}(\mu_J) \partial_{2E_J'}^n + E_S^{(n)}(\mu_S) \partial_{E_S'}^n \Bigr) \biggr] \\
&\quad \times \Bigl(\frac{Q}{\mu_J(e^{\gamma_E}\nu)^{1/j_J}}\Bigr)^{2E_J'(\mu,\mu_J)} \Bigl(\frac{Q}{\mu_S e^{\gamma_E}\nu}\Bigr)^{E_S'(\mu,\mu_S)}\,.
\end{split}
}
\ee

With the scale choices $\mu=\mu_H=Q, \mu_F =\wt\mu_F\equiv  Q(e^{\gamma_E}\nu)^{-1/j_F}$ given in \eq{nuscalechoices}, the above expressions for $\wt\sigma(\nu)$ simplify considerably,
\be
\label{eq:Laplacefinal}
\begin{split}
\wt\sigma(\nu) &=  H_2(Q) \wt J(0,\wt\mu_J)^2\wt S(0,\wt\mu_S) e^K  \\ 
&=  H_2(Q)\wt J(0,Q)^2\wt S(0,Q)e^{\bar E} \,,
\end{split}
\ee
where here $K = 2K_J(Q,\wt\mu_J) + K_S(Q,\wt\mu_S)$, and $\bar E = 2\bar E_J + \bar E_S$ is defined in \eq{Edefs}. The derivatives $E_{J,S}^{(n)}$ are defined in \eqs{EJprime}{ESdef}. As noted after \eq{EJdef}, the difference between the exponents $K$ and $\bar E$ is NNLL. These differences at \nkll accuracy, $k\geq 2$, are made up by differences due the scales in the jet and soft functions, which differ in the two lines of \eq{Laplacefinal}, calculated to the order appropriate to the accuracy in question (see \tab{LaplaceE}).

\emph{Nota bene:} The primary definition of \nkll or \nkllprime accuracy is based on the accuracy of the exponents in the simple exponentiated forms in \eq{Laplacefinal}. These accuracies can be achieved by computing ingredients according to \tab{LaplaceE}. The theoretical uncertainty at finite resummed \nkll accuracy can be estimated by varying the scales in \eqs{Laplacevariablescales}{LaplacevariableE}.

\subsection{Cumulant}
For the cumulant, the usual resummed form given in SCET with variable scales is given by \eq{SCETresummedcumulant},
\be
\label{eq:cumulantvariable}
\boxed{
\begin{split}
R(\tau_a) &= e^K \Bigl( \frac{\mu_H}{Q} \Bigr)^{\omega_H} \Bigl( \frac{\mu_J}{Q \tau_a^{1/j_J}} \Bigr)^{2j_J\omega_J} \Bigl( \frac{\mu_S}{Q \tau_a} \Bigr)^{\omega_S} \\
& \quad \times H_2 (Q^2,\mu_H) \wt{J} \Bigl( \p_\Omega + \ln \frac{\mu_J^{j_J}}{Q^{j_J} \tau_a} , \mu_J \Bigr)^2 \wt{S} \Bigl( \p_\Omega + \ln \frac{\mu_S}{Q \tau_a} , \mu_S \Bigr)  \frac{\exp(\gamma_E \Omega)}{\Gamma(1-\Omega)} \,,
\end{split}
}
\ee
where $K,\Omega$ are defined in \eq{KOmega} and $\omega_F$ in \eq{Komegadef}. The counting rules in \tab{cumulant} for computing to \nkll or \nkllprime accuracy apply to \eq{cumulantvariable}. See also the note below.

Meanwhile the dQCD-inspired form, shown in \ssec{resumequiv} to be equivalent to SCET, is given by \eq{SCETcsE},
\be
\label{eq:cumulantvariableE}
\boxed{
\begin{split}
R(\tau_a) &=  H_2(Q^2,\mu_H) \Bigl(\frac{\mu_H}{Q}\Bigr)^{\omega_H(\mu,\mu_H)} e^{K_H(\mu,\mu_H)} \wt J(0,\mu)^2 \wt S(0,\mu) \\
&\quad \times e^{ 2E_J(\mu,\mu_J) + E_S(\mu,\mu_S)}   \exp\biggl[ \sum_{n=2}^\infty \frac{1}{n!} \Bigl( 2 E_J^{(n)}(\mu_J) \partial_{2E_J'}^n + E_S^{(n)}(\mu_S) \partial_{E_S'}^n \Bigr) \biggr]  \\
&\quad \times \Bigl( \frac{Q^{j_J}\tau_a}{\mu_J^{j_J} e^{\gamma_E}}\Bigr)^{2E_J'(\mu,\mu_J)/j_J} \Bigl( \frac{Q\tau_a}{\mu_S e^{\gamma_E}}\Bigr)^{E_S'(\mu,\mu_S)}  \frac{1}{\Gamma(1+2E_J'/j_j + E_S')}\,.
\end{split}
}
\ee
A closely related form was derived in the context of threshold resummation in Ref.~\cite{Contopanagos:1993xh}. The counting rules in \tab{Ecounting} apply to \eq{cumulantvariableE}. Again the functions $E_{J,S}$ are defined in \eqs{EJdef}{ESdef}, and the derivatives $E_{J,S}^{(n)}$ in \eqs{EJprime}{ESdef}.

Choosing the scales in \eq{barcanonicalscales}, $\mu=\mu_H=Q,\mu_F=\bar\mu_F\equiv Q(e^{-\gamma_E}\tau_a)^{1/j_F}$, we obtained for the above forms,
\be
\label{eq:cumulantfinalbar}
\begin{split}
R(\tau_a) &=  H_2(Q)e^{\bar K} \wt J(\partial_{\bar\Omega},\bar\mu_J)^2 \wt S(\partial_{\bar \Omega},\bar\mu_S) \left[\frac{1}{\Gamma(1-\bar \Omega)}\right] \\
&=  H_2(Q) \wt J(0,Q)^2 \wt S(0,Q) \exp\biggl[ \bar E + \sum_{n=2}^\infty \frac{1}{n!}  \bar E^{(n)} \partial_{\bar E'}^n   \biggr]   \frac{1}{\Gamma(1-\bar E')} \,,
\end{split}
\ee
where $\bar K,\bar\Omega$ are defined in \eq{barKomega}, and $\bar E,\bar E^{(n)}$ in \eq{Edefs}. The first line is the SCET form, \eq{radiatorbar}, and the second line follows directly from \eq{cumulantvariableE}. The sign change in the argument of the gamma function between \eqs{cumulantvariableE}{cumulantfinalbar} is due to switching taking derivatives with respect to $\ln\mu_F$ in the former and $\ln(1/\tau_a)$ in the latter. These scale choices make the parallel between SCET and QCD forms, as given in previous literature, most transparent. As we will discuss in \sec{numerical}, however, it is often preferable to use the ``canonical'' scales $\mu=\mu_H=Q,\mu_F^{\rm nat} \equiv Q\tau_a^{1/j_F}$. In this case, the resummed cumulant takes the form
\be
\label{eq:cumulantfinal}
\begin{split}
R(\tau_a) &=  H_2(Q)e^{K} \wt J(\partial_{\Omega},\mu_J^{\rm nat})^2 \wt S(\partial_{ \Omega},\mu_S^{\rm nat}) \left[\frac{e^{\gamma_E\Omega}}{\Gamma(1- \Omega)}\right] \\
&=  H_2(Q) \wt J(0,Q)^2 \wt S(0,Q) \exp\biggl[  E + \sum_{n=2}^\infty \frac{1}{n!}   E^{(n)} \partial_{ E'}^n   \biggr]   \frac{e^{\gamma_E E'}}{\Gamma(1- E')} \,,
\end{split}
\ee
where the $K,\Omega$ here are defined in \eq{canonicalKomega}, and $E,E^{(n)}$ are defined by
\be
\label{eq:canonicalEdefs}
E = 2E_J(Q,Q\tau_a^{1/j_J}) + E_S(Q,Q\tau_a)\,,\quad E^{(n)} = \frac{d^n E}{d(\ln(1/\tau_a))^n}\,,
\ee
$E_{J,S}$ being defined by \eqs{EJdef}{ESdef}.

The dQCD-inspired ``$E$'' forms in \eqs{cumulantfinalbar}{cumulantfinal} in fact have SCET definitions by way of the relations in \eqss{EJdef}{ESdef}{canonicalEdefs} (see also \eqs{AB}{Eprime}). They resum a larger set of terms than the $\Omega$ terms thanks to more logs generated by $\wt J^2\wt S$ being put in the exponent and gamma functions.

\emph{Nota bene:}  While \tab{LaplaceE} could be used to evaluate the ingredients in the above formulae for $R(\tau_a)$ to \nkll accuracy, as explained in \ssec{Laplacecomparison} it is preferable to keep the differential operator terms in $\wt J(\partial_{\bar\Omega}),\wt S(\partial_{\bar\Omega})$ or the $\partial_{\bar E'}$ operators to the order corresponding to \nkllprime accuracy, as described in \tab{cumulant}. This maintains better equivalence with the accuracy of the \nkll $\wt\sigma(\nu)$. Similarly in \eq{cumulantfinal}. Working to full \nkllprime accuracy as given by \tab{LaplaceE} or \tab{cumulant} is the ideal. These rules for $\partial_\Omega$ terms also apply to evaluation and truncation of the exponential of $\partial_{E'}^n$ operators in the dQCD forms above, which is summarized in \tab{Ecounting}.

\subsection{Differential Distribution}

For the cumulant, a common prescription is the form in \eq{SCETresummeddist}, which we label $\sigma_n$:
\begin{align}
\label{eq:sigman}
\sigma_n (\tau_a) &=   \exp \bigl( K_H + 2K_J + K_S \bigr) \Bigl( \frac{\mu_H}{Q} \Bigr)^{\omega_H} \Bigl( \frac{\mu_J}{Q \tau_a^{1/j_J}} \Bigr)^{2j_J\omega_J}\Bigl( \frac{\mu_S}{Q \tau_a} \Bigr)^{\omega_S} H_2 (Q^2,\mu_H) \\
& \quad \times  \wt{J} \Bigl( \p_\Omega + \ln \frac{\mu_J^{j_J}}{Q^{j_J}\tau_a} , \mu_J \Bigr)^2 \wt{S} \Bigl( \p_\Omega + \ln \frac{\mu_S}{Q \tau_a} , \mu_S \Bigr)  \frac{1}{\tau_a} \frac{\exp(\gamma_E \Omega)}{\Gamma(-\Omega)} \,. \nn
\end{align}
Often, the rules in \tab{LaplaceE} are used to compute $\sigma_n$ to a given accuracy.  To achieve the same accuracy as the Laplace transform or cumulant at \nkll order, it is necessary to use the rules in \tab{cumulant}. At \nkllprime order the two tables are the same, and $\sigma_n(\tau_a)$ matches the accuracy of $\wt\sigma(\nu)$.

An alternate approach is to define a distribution that will reproduce the derivative of the cumulant at any accuracy.  This distribution is labeled $\sigma_R (\tau_a)$, with [\eq{sigmaR}]:
\be \label{eq:sigmaRfinal}
\boxed{
\begin{split}
&\sigma_R (\tau_a) = \sigma_n (\tau_a) + \delta\sigma_R (\tau_a) \,, \\
&\delta\sigma_R (\tau_a) =   \frac{1}{\tau_a} \exp( K_H + 2K_J + K_S ) \Bigl( \frac{\mu_H}{Q} \Bigr)^{\omega_H} \Bigl( \frac{\mu_J}{Q \tau_a^{1/j_J}} \Bigr)^{2j_J \omega_J} \Bigl( \frac{\mu_S}{Q \tau_a} \Bigr)^{\omega_S} \\
& \times H_2 (Q^2, \mu_H) \Biggl\{ \sum_{F = J,J,S} \wt{P}_F (\mathbf{L}_F, \mu_F) \frac{1}{\wt{F} (\mathbf{L}_F, \mu_F)} \Biggr\} \wt{J} ( \mathbf{L}_J, \mu_J)^2\wt{S} ( \mathbf{L}_S, \mu_S) \frac{\exp(\gamma_E \Omega)}{\Gamma(1 - \Omega)} \,, \\
& \qquad \textrm{with} \quad \mathbf{L}_J = \partial_\Omega + \ln \mu_J^{j_J} / Q^{j_J} \tau_a \,, \quad \mathbf{L}_S = \partial_\Omega + \ln \mu_S / Q \tau_a \,,
\end{split}
}
\ee
and [\eq{PF}]:
\begin{align} \label{eq:PFfinal}
\wt{P}_F (L_F, \mu_F) &= \frac{d \ln \mu_F}{d \ln \tau_a} \biggl\{ j_F \frac{\partial \wt{F} (L_F, \mu_F)}{\partial L_F} + \beta[\as(\mu_F)] \frac{\partial \wt{F} (L_F, \mu_F)}{\partial \as} - \wt{\gamma}_F (L_F, \mu_F) \wt{F} (L_F, \mu_F) \biggr\} \,.
\end{align}
In the form \eq{sigmaRfinal}, $\sigma_R(\tau_a)$ can be evaluated using the rules in \tab{LaplaceE} or \tab{cumulant} to achieve \nkll or \nkllprime accuracy. Using the form \eq{sigmaRfinal} will guarantee that equivalent accuracy is maintained with differentiating the \nkll or \nkllprime cumulant $R(\tau_a)$, evaluated according the same rules.

The dQCD-inspired differential cross section, found by taking the derivative of \eq{cumulantvariableE}, can be written with free scales in the form,
\be
\label{eq:distvariableE}
\begin{split}
\sigma (\tau_a) &= \frac{1}{\tau_a}H_2(Q^2,\mu_H) \Bigl(\frac{\mu_H}{Q}\Bigr)^{\omega_H(\mu,\mu_H)} e^{K_H(\mu,\mu_H)} \wt J(0,\mu)^2 \wt S(0,\mu) \\
&\quad \times e^{ 2E_J(\mu,\mu_J) + E_S(\mu,\mu_S)}   \exp\biggl[ \sum_{n=2}^\infty \frac{1}{n!} \Bigl( 2 E_J^{(n)}(\mu_J) \partial_{2E_J'}^n + E_S^{(n)}(\mu_S) \partial_{E_S'}^n \Bigr) \biggr]  \\
&\quad \times \Bigl( \frac{Q^{j_J}\tau_a}{\mu_J^{j_J} e^{\gamma_E}}\Bigr)^{2E_J'(\mu,\mu_J)/j_J} \Bigl( \frac{Q\tau_a}{\mu_S e^{\gamma_E}}\Bigr)^{E_S'(\mu,\mu_S)}  \frac{1}{\Gamma(2E_J'/j_j + E_S')}\,.
\end{split}
\ee
Alternatively one may also differentiate \eq{cumulantvariableE} after choosing $\tau_a$-dependent scales, which allows one to apply the counting rules in \tab{Ecounting} directly.

Upon making the particular choices of scales in \eq{barcanonicalscales}, we obtained in SCET and dQCD-inspired forms [\eqs{radiatorderivative2}{sigmatauE1}, respectively],
\be
\label{eq:sigmataufinalbar}
\begin{split}
\sigma(\tau_a) &=   H_2(Q)e^{\bar K} \biggl[-\frac{1}{\tau_a} \Bigl( \bar\Omega + \sum_F \frac{1}{j_F}\bar\gamma_F(\dOmegabar)\Bigr) \wt F(\dOmegabar)  + \frac{d\wt F(\dOmegabar)}{d\tau_a} \biggr] \frac{1}{\Gamma(1-\bar\Omega)} \\
&= -\frac{1}{\tau_a} H_2(Q^2,Q) \wt J(0,Q)^2 \wt S(0,Q)   \\
&\quad\times \Biggl[ \sum_{n=0}^\infty  \frac{1}{n!} \bar E^{(n+1)} \partial_{\bar E'}^n\Biggr] \exp\biggl[ \bar E + \sum_{n=2}^\infty \frac{1}{n!}  \bar E^{(n)} \partial_{\bar E'}^n   \biggr]\frac{1}{\Gamma(1-\bar E')}\,. 
\end{split}
\ee
and similarly with the canonical scale choices \eq{tauscalechoices},
\be
\label{eq:sigmataufinal}
\begin{split}
\sigma(\tau_a) &=  H_2(Q)e^{ K} \biggl[-\frac{1}{\tau_a} \Bigl( \Omega + \sum_F \frac{1}{j_F}\hat\gamma_F(\dOmega)\Bigr) \wt F(\dOmega) + \frac{d\wt F(\dOmega)}{d\tau_a} \biggr] \frac{e^{\gamma_E\Omega}}{\Gamma(1-\Omega)} \\
&= -\frac{1}{\tau_a} H_2(Q^2,Q) \wt J(0,Q)^2 \wt S(0,Q)   \\
&\quad\times \Biggl[ \sum_{n=0}^\infty  \frac{1}{n!}  E^{(n+1)} \partial_{ E'}^n\Biggr] \exp\biggl[  E + \sum_{n=2}^\infty \frac{1}{n!}   E^{(n)} \partial_{ E'}^n   \biggr]\frac{e^{\gamma_E E'}}{\Gamma(1- E')}\,.
\end{split}
\ee
In these expressions the operator $\bar\gamma_F$ was defined in \eq{gammaFoperator}, and $\hat\gamma_F$ is given by the same formula with scales $\mu_F$ in \eq{tauscalechoices}. The exponents $\bar K,\bar\Omega$ are defined in \eq{barKomega}, and $K,\Omega$ by \eq{canonicalKomega}. The exponents $\bar E,\bar E^{(n)}$ are given by \eq{Edefs}, and $E,E^{(n)}$ by \eq{canonicalEdefs}.  In these forms, one may directly apply the counting rules in \tab{LaplaceE} or \tab{cumulant} for the SCET forms, and \tab{Ecounting} for the dQCD-inspired forms.

\emph{Nota bene:}  Use the forms in \eq{sigmataufinalbar} or \eq{sigmataufinal} keeping all objects to orders specified in \tab{LaplaceE} or \tab{cumulant} at \nkll or \nkllprime accuracy to maintain equivalence with accuracy of $R(\tau_a)$. \emph{Do not} apply \nkll rules in  \tab{LaplaceE} or \tab{cumulant} directly to \eq{sigmaOmega}.

%%%%%%%%%%%%%%%%%%%%%%%%%%%%%%%%%%%%%%%%%%%%%%%%%%%%%%%%%%%%%%%%%%%%%%%%%%%%%%%%%%%%%%%%%%%%%%%%%%%%%%%%%
\section{Numerical Comparison of Angularity Distributions}
\label{sec:numerical}
%%%%%%%%%%%%%%%%%%%%%%%%%%%%%%%%%%%%%%%%%%%%%%%%%%%%%%%%%%%%%%%%%%%%%%%%%%%%%%%%%%%%%%%%%%%%%%%%%%%%%%%%%

A numerical study is useful to compare the various resummation prescriptions and the relationship between the dQCD and SCET formalisms.  For different values of the angularity parameter $a$, we will study the effect of different prescriptions on NLL and NLL$'$ distributions in both SCET and dQCD formalisms, for both the resummed cumulant and spectrum.  These prescriptions lead to notable differences between the resummed cross sections, and we will find that the best agreement between SCET and dQCD results are in the most consistent versions of the resummation formulae.

For the resummed cross sections in SCET and dQCD, we use the distributions in \eqss{sigman}{sigmaRfinal}{distvariableE} to NLL and NLL$'$ accuracy.  We label the dQCD-inspired form in \eq{distvariableE} as $\sigma_Q (\tau_a)$ for clarity.  This study will allow us to contrast the standard SCET resummed distribution, $\sigma_n$ [\eq{sigman}], with the SCET resummed form that matches closely with the derivative of the cumulant, $\sigma_R$ [\eq{sigmaRfinal}], as well as the dQCD-inspired form $\sigma_Q$ [\eq{distvariableE}].  The NLL and NLL$'$ distributions for the SCET resummed forms $\sigma_n$ and $\sigma_R$ are obtained from \eqs{sigman}{sigmaRfinal} by applying the counting rules\footnote{We use \tab{LaplaceE} instead of \tab{cumulant} to define the NLL spectrum for two reasons.  First, since \tab{LaplaceE} is a prescription commonly used, it provides a useful point of comparison, especially for the results in \cite{Hornig:2009vb}.  Second, the difference between the NLL and NLL$'$ distributions is larger when using \tab{LaplaceE}, making the orders of accuracy more distinct.  Using \tab{cumulant}, we find that the NLL distributions are more similar to NLL$'$.} in \tab{LaplaceE}.  The NLL and NLL$'$ distributions for the dQCD-inspired form $\sigma_Q$ is obtained from \eq{distvariableE} using the counting prescription described in \ssec{dQCDcounting}.  For the case of NLL or NLL$'$ accuracy, this means that \eq{distvariableE} takes the form
\begin{align} \label{eq:dQCDNLL}
\sigma_Q (\tau_a) &= -\frac{1}{\tau_a}   e^{K_H + 2E_J + E_S} H_2 (Q^2, \mu_H) \wt{J} (0, \mu)^2 \wt{S} (0, \mu) \Bigl( \frac{\mu_H}{Q} \Bigr)^{\omega_H} \Bigl( \frac{Q^{j_J} \tau_a}{\mu_J^{j_J}} \Bigr)^{2E_J' / j_J} \Bigl( \frac{Q \tau_a}{\mu_S} \Bigr)^{E_S'} \nn \\
& \quad \times \biggl\{ E' + 2 E_J'' \Bigl( \p_{E'} + \ln \frac{\mu_J^{j_J}}{Q^{j_J} \tau_a} \Bigr) + E_S'' \Bigl( \p_{E'} + \ln \frac{\mu_S}{Q \tau_a} \Bigr) \nn \\
& \qquad \quad + E' E_J'' \Bigl( \p_{E'} + \ln \frac{\mu_J^{j_J}}{Q^{j_J} \tau_a} \Bigr)^2 + \frac12 E' E_S'' \Bigl( \p_{E'} + \ln \frac{\mu_S}{Q \tau_a} \Bigr)^2 \biggr\} \, \frac{\exp(\gamma_E E')}{\Gamma( 1 - E')} \,.
\end{align}
where $E' = - 2E_J' / j_J - E_S'$.  This form arises from \eq{distvariableE} by expanding the exponentiated derivative operator to $\ord{\as}$, which involves keeping only the leading nontrivial $E_J''$ and $E_S''$ terms in the expansion.  We also pulled several factors through the derivative operator, as in previous forms.

The numerical study is performed at a center of mass energy of $Q = 100$ GeV, and we make canonical scale choices, \eq{tauscalechoices}, for the central scales of each distribution.  Uncertainty estimates are made through scale variation of $\mu_H = \mu$, $\mu_J$, and $\mu_S$ each up and down by a factor of 2.  The envelope of these scale variations determines the overall uncertainty.  The uncertainty estimates here should be taken as nominal; as we will see, the uncertainties for $\sigma_R$ and $\sigma_Q$ are not robust (being either overestimated or underestimated in certain regions of $\tau_a$ with the scale variations used here) and require further study. Ideally more refined scale variations using parameters in the profile functions themselves should be performed, as in e.g. \cite{Abbate:2010xh,Berger:2010xi,Ligeti:2008ac,Kang:2013nha}. We leave such an improved study of the uncertainties for future work. Here, our focus is on the change in behavior of the central values of the curves amongst $\sigma_{n,R,Q}$ and how well they agree with one another.  Since there is no matching of the far tail to fixed-order perturbation theory, we only study the behavior of the distributions in the resummation region.

\begin{figure}[!t]
\vspace{-1em}
\begin{center}
\includegraphics[width=.96\textwidth]{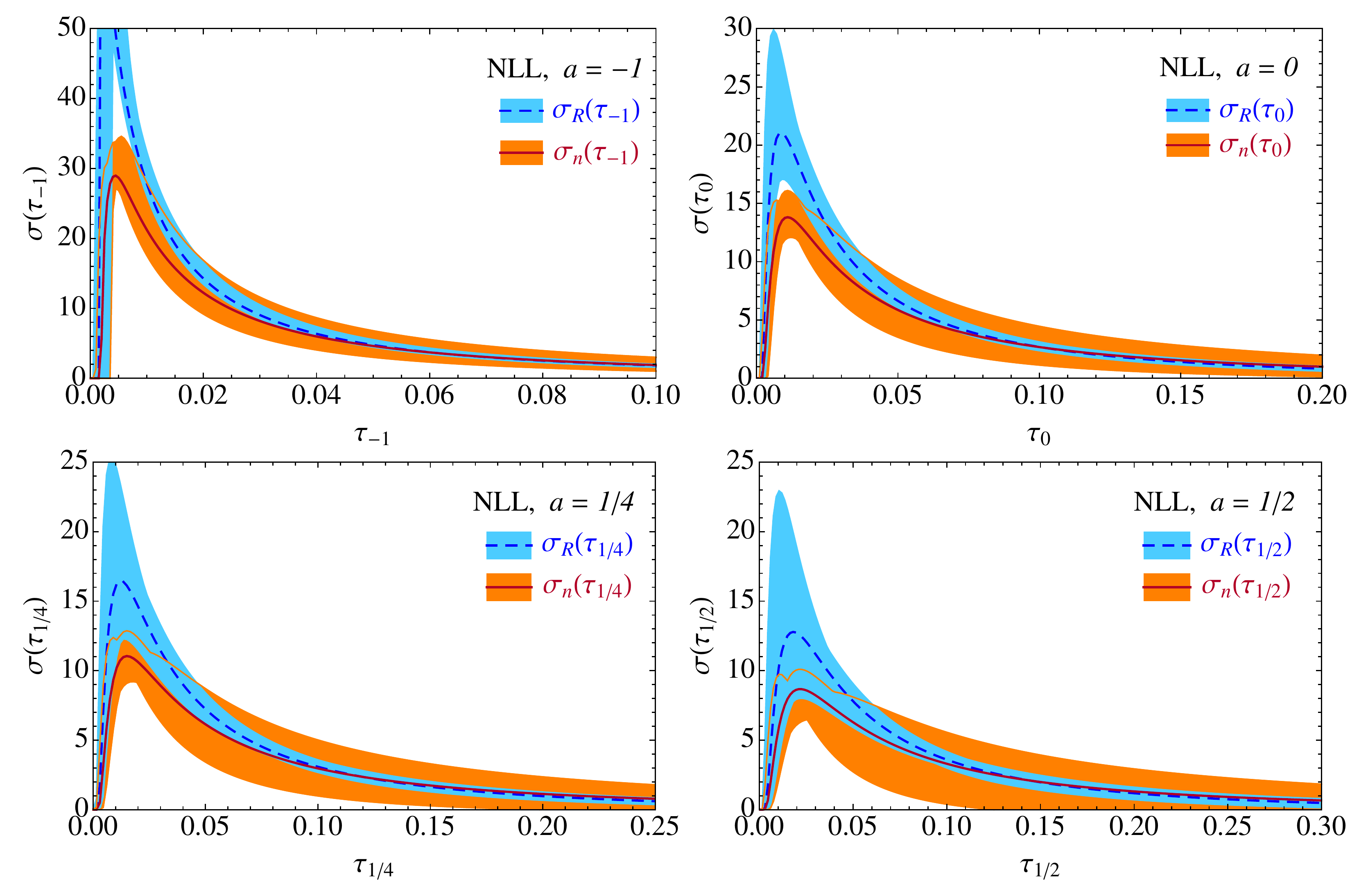}
\vspace{-2em}
\end{center}
\caption{The NLL distributions for $\sigma_n (\tau_a)$ and $\sigma_R (\tau_a)$, for $a = -1,\, 0,\, 1/4,$ and $1/2$ at $Q = 100$ GeV.  These compare the natural SCET resummation ($\sigma_n$) with the resummed form that closely matches the derivative of the cumulant ($\sigma_R$).
\label{fig:RnNLLcomp}}
\end{figure}

\begin{figure}[!b]
\vspace{-4em}
\begin{center}
\includegraphics[width=.96\textwidth]{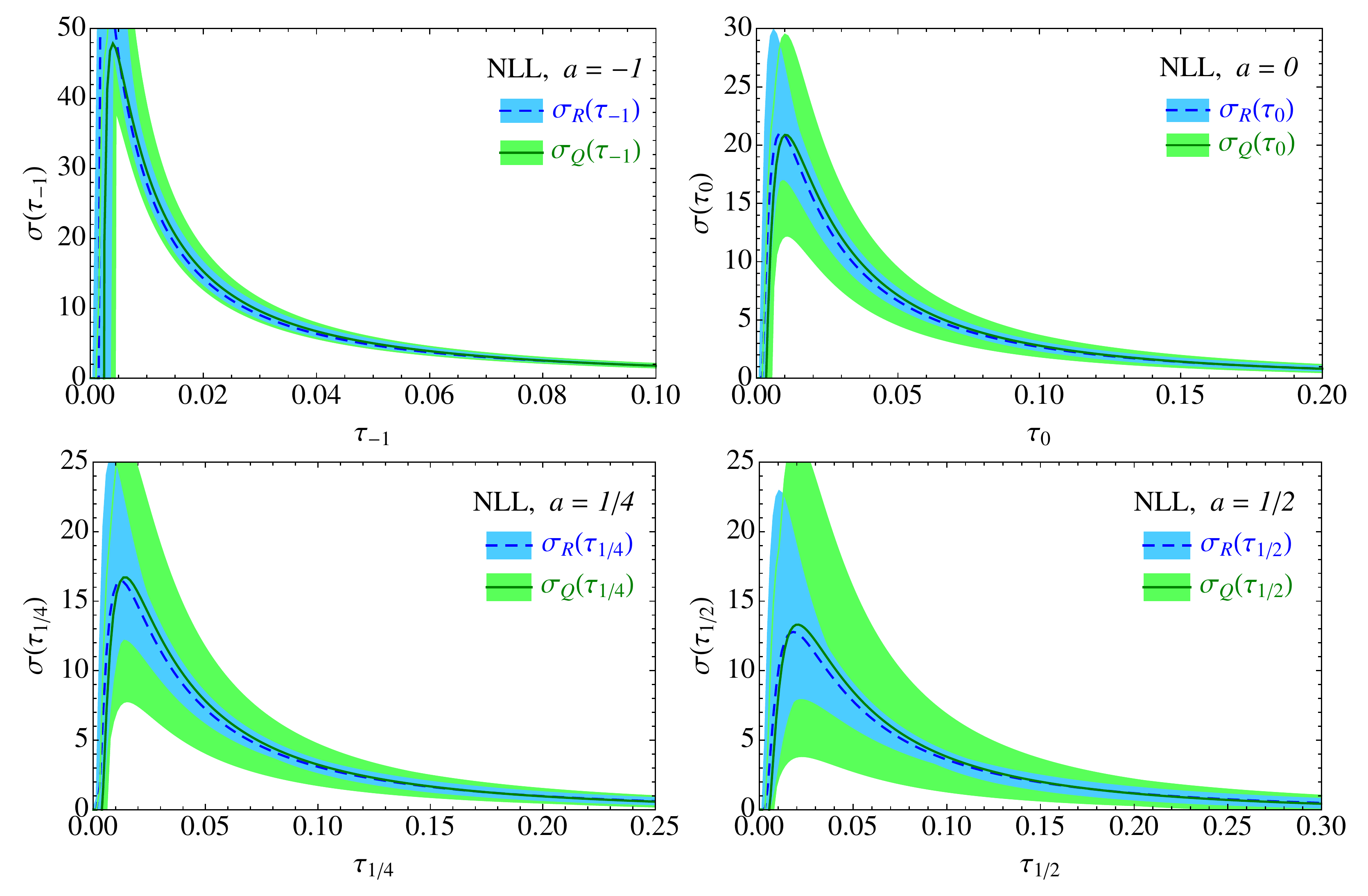}
\vspace{-2em}
\end{center}
\caption{The NLL distributions for $\sigma_Q (\tau_a)$ and $\sigma_R (\tau_a)$, for $a = -1,\, 0,\, 1/4,$ and $1/2$ at $Q = 100$ GeV.  These compare the dQCD-like resummation ($\sigma_Q$) with the resummed form that closely matches the derivative of the cumulant ($\sigma_R$).  Note that these forms are in much better agreement than the comparison in \fig{RnNLLcomp}.
\label{fig:RQNLLcomp}}
\end{figure}

In \fig{RnNLLcomp}, we plot the NLL resummed distributions for $\sigma_n (\tau_a)$ and $\sigma_R (\tau_a)$ for $a = -1,\, 0,\, 1/4,$ and $1/2$.  In \fig{RQNLLcomp} we plot the NLL resummed distributions for the same values of $a$, comparing $\sigma_R (\tau_a)$ and $\sigma_Q (\tau_a)$.
\figs{RnNLLcomp}{RQNLLcomp} show a much better agreement between the dQCD distribution $\sigma_Q$ and the SCET distribution $\sigma_R$ (which is close to the derivative of the cumulant) than $\sigma_Q$ and the standard SCET distribution $\sigma_n$.  In \cite{Hornig:2009vb}, angularity distributions in dQCD and SCET were compared using resummation formulae very similar to $\sigma_n$ and $\sigma_Q$.  A discrepancy was observed which is very similar to the discrepancy seen here in \fig{RnNLLcomp}, and we see that it is ameliorated by including the additional resummed terms present in $\sigma_R$.  The good agreement between $\sigma_R$ and $\sigma_Q$ in \fig{RQNLLcomp} provides strong evidence to support the analytic arguments that the dQCD and SCET resummations are in close correspondence.  We note when $\tau_a \lesssim 0.01$, the central soft scale $\mu_S = Q \tau_a \lesssim 1$ GeV becomes nonperturbative and the predictions become unreliable without additional care.

The uncertainty bands on each distribution are determined by the envelope of scale variations, as described above.  The relative size of uncertainty bands on $\sigma_n$ and $\sigma_R$ arise because of the relationship of $\sigma_R$ to the cumulant.  In the large $\tau_a$ regime, the distribution $\sigma_n$ produces reasonable uncertainties.  The cumulant has scale uncertainties whose $\tau_a$ dependence is very similar between different variations, meaning the derivative of the cumulant (which matches $\sigma_R$ closely) has very small scale uncertainties at large $\tau_a$.  This is observed in \fig{RnNLLcomp}.  We also note that at large $\tau_a$, $\sigma_Q$ has scale variation that is smaller than $\sigma_n$ but larger than $\sigma_R$, suggesting that more careful scale variations (with profiles) are needed when using $\sigma_Q$ in phenomenological applications.  This is also evident in the fact that $\sigma_Q$ has very large uncertainties in the peak region.  Finally, the increase in uncertainties with $a$ is due to the logarithmic structure of the resummation and the gradual breakdown of the resummation framework for both dQCD and SCET as $a\to1$.  Consistent uncertainties across $a$ values may be achieved with more careful scale variation and profile functions, and further study on how to obtain robust uncertainties for $\sigma_R$ and $\sigma_Q$ is needed.

\begin{figure}[!t]
\vspace{-1em}
\begin{center}
\includegraphics[width=.96\textwidth]{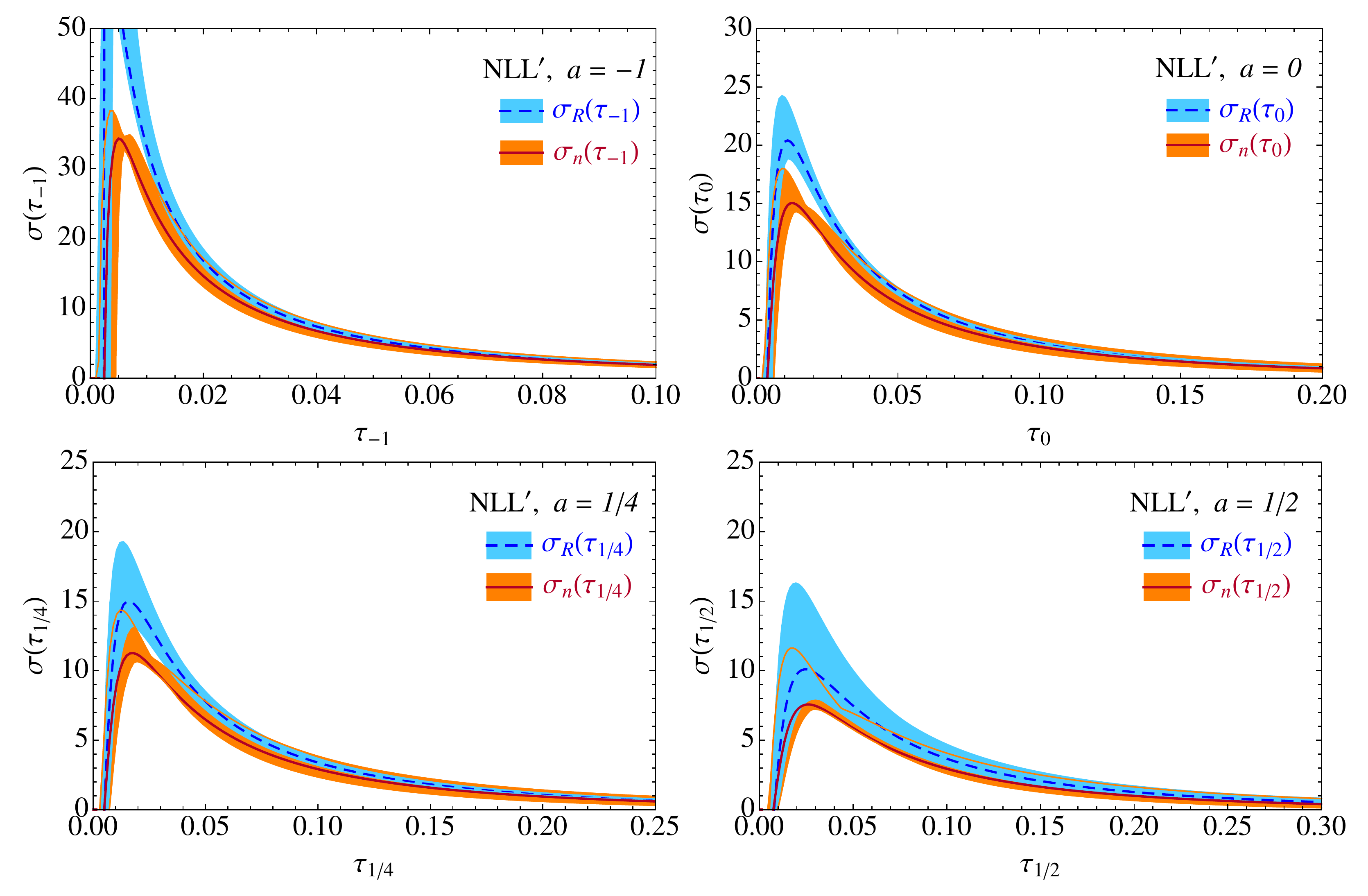}
\vspace{-2em}
\end{center}
\caption{The NLL$'$ distributions for $\sigma_n (\tau_a)$ and $\sigma_R (\tau_a)$, for $a = -1,\, 0,\, 1/4,$ and $1/2$ at $Q = 100$ GeV.  These compare the natural SCET resummation ($\sigma_n$) with the resummed form that closely matches the derivative of the cumulant ($\sigma_R$).
\label{fig:RnNLLPcomp}}
\end{figure}

\begin{figure}[!b]
\vspace{-4em}
\begin{center}
\includegraphics[width=.96\textwidth]{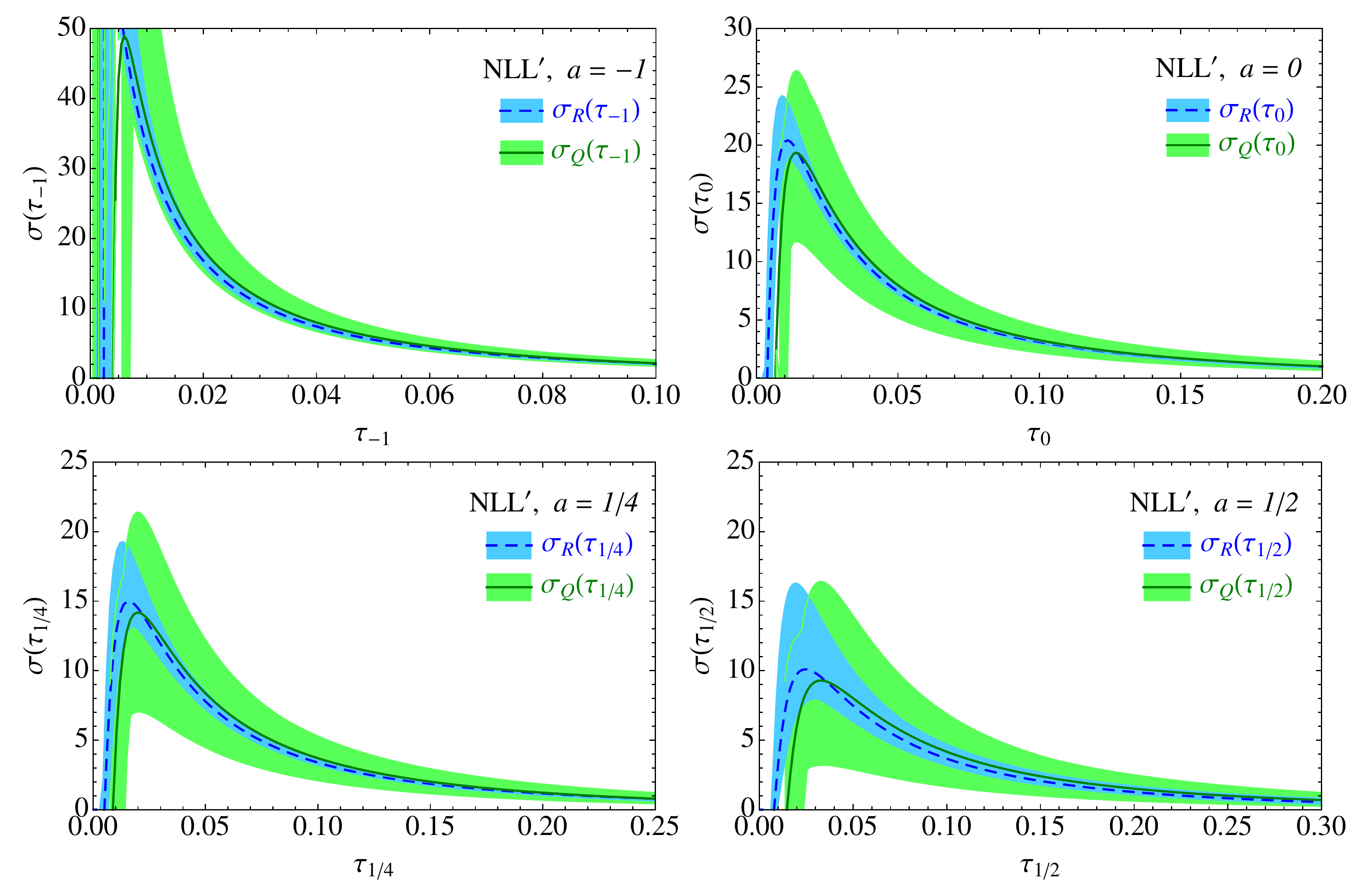}
\vspace{-2em}
\end{center}
\caption{The NLL$'$ distributions for $\sigma_Q (\tau_a)$ and $\sigma_R (\tau_a)$, for $a = -1,\, 0,\, 1/4,$ and $1/2$ at $Q = 100$ GeV.  These compare the dQCD-like resummation ($\sigma_Q$) with the resummed form that closely matches the derivative of the cumulant ($\sigma_R$).  Note that these forms are in better agreement than the comparison in \fig{RnNLLPcomp}.
\label{fig:RQNLLPcomp}}
\end{figure}

In \fig{RnNLLPcomp}, we plot the NLL$'$ resummed distributions for $\sigma_n (\tau_a)$ and $\sigma_R (\tau_a)$ for $a = -1,\, 0,\, 1/4,$ and $1/2$.  In \fig{RQNLLPcomp} we compare $\sigma_R (\tau_a)$ and $\sigma_Q (\tau_a)$ at NLL$'$ accuracy at the same values of $a$.
These figures can be contrasted with \figs{RnNLLcomp}{RQNLLcomp}.


The agreement between the central values of $\sigma_R$ and $\sigma_Q$ improves from NLL to NLL$'$ in the region of $\tau_a$ past the peak.  The moderate difference between them in the peak of the distribution observed in \fig{RQNLLPcomp} arises from the different treatment in $\sigma_R$ and $\sigma_Q$ of the $\ord{\as}$ nonlogarithmic singular terms that are included in the fixed-order functions at NLL$'$.  We note this difference remains small compared to the uncertainties in the $\sigma_Q$ distribution.

The uncertainties in $\sigma_n$, $\sigma_R$, and $\sigma_Q$ at NLL$'$ follow the same general pattern as the NLL case.  An exception is that the uncertainties in the $\sigma_n$ distribution decrease noticeably in working to NLL$'$; this occurs because $\sigma_n$ at NLL contains the hard, jet, and soft functions only at tree level, while they are taken to $\ord{\as}$ in the NLL$'$ spectrum. In $\sigma_R$ and $\sigma_Q$,  however, the NLL spectrum has contributions from the hard, jet, and soft functions beyond tree level. Had we used the prescription in \tab{cumulant} to define the NLL spectrum for $\sigma_n$, we would find the uncertainties in that case are much closer to the NLL$'$ spectrum in \fig{RnNLLPcomp}.

Further study of the relationship between the SCET and dQCD resummed angularity distributions is of interest to more deeply probe the numerical effect of the different approaches to resummation.  As these approaches are formally consistent when working to the same order of accuracy, the relative agreement gives insight into the uncertainties that accompany each resummation scheme.  Of particular interest would be a careful study that includes matching to fixed-order perturbation theory in the large $\tau_a$ tail region of the distribution and a systematic study of the profile scales and variations needed to obtain reliable uncertainties across the spectrum.

%%%%%%%%%%%%%%%%%%%%%%%%%%%%%%%%%%%%%%%%%%%%%%%%%%%%%%%%%%%%%%%%%%%%%%%%%%%%%%%%%%%%%%%%%%%%%%%%%%%%%%%%%
\section{Conclusions}
\label{sec:conclusions}
%%%%%%%%%%%%%%%%%%%%%%%%%%%%%%%%%%%%%%%%%%%%%%%%%%%%%%%%%%%%%%%%%%%%%%%%%%%%%%%%%%%%%%%%%%%%%%%%%%%%%%%%%

In this paper we have performed a detailed study of resummed $e^+e^-$ event shape distributions in both SCET and dQCD, using angularities as a generic example.  This study contains several parts, examining three different ways to express the cross sections: the cumulant $R(\tau_a)$, the spectrum $\sigma (\tau_a)$, and its Laplace transform $\wt{\sigma}(\nu)$.

In \sec{orders} we reviewed standard logarithmic counting schemes.  As the cross section in Laplace space directly exponentiates, it is simplest to define logarithmic accuracy by counting in the exponent of $\wt{\sigma} (\nu)$.  We also discussed the original CTTW convention that determines logarithmic accuracy by counting in the exponent of the cumulant $R(\tau_a)$.  While either definition of logarithmic accuracy is valid, and one can translate between the two, the advantage of defining accuracy in terms of the Laplace-transformed cross section is that a closed algebraic form for $\wt{\sigma} (\nu)$ is easily obtained.  A major goal of \sec{schemes} is to define prescriptions for $R(\tau_a)$ and $\sigma(\tau_a)$ that are correct to a given order of accuracy when these cross sections are transformed to $\wt{\sigma} (\nu)$.

In \sec{compare}, we reviewed resummation techniques in SCET and dQCD and showed the equivalence between resummed forms, culminating in the equivalence relations \eqss{EJdef}{ESdef}{AB}.  This equivalence reveals alternative forms of the resummed cross section in SCET that more closely correspond with the resummed cross section in dQCD, as well as a form of the dQCD cross section with dependence on arbitrary soft and collinear factorization scales, results which generalize both approaches to resummation.

In \sec{schemes} we discussed the precise prescriptions needed to obtain a given order of logarithmic accuracy. We explained how to compute the resummed $R(\tau_a)$ and distribution $\sigma(\tau_a)$ so that their accuracies match that of $\wt\sigma(\nu)$ after Laplace transformation. We also studied a way of defining the resummed spectrum $\sigma (\tau_a)$ in terms of the resummed cumulant $R(\tau_a)$ [or equivalently the Laplace-transformed cross section $\wt{\sigma} (\nu)$].  This leads to a novel way, \eq{sigmaR}, of writing the resummed spectrum that matches onto the derivative of the cumulant.  

A compact summary of the dQCD and SCET resummed forms for the cross section, its Laplace transform, and the cumulant is given in \sec{summary}.

Finally, in \sec{numerical} we have performed a short numerical study of the angularity distributions.  We have shown that numerical discrepancies between the dQCD forms (given in \cite{Berger:2003iw,Berger:2003pk}) and the SCET forms (given in \cite{Hornig:2009vb}), originally observed in the latter study, are resolved by using versions of the resummed cross sections that arise from the equivalence of the dQCD and SCET forms.  This gives confidence that the novel resummed forms derived in this paper may be applied in phenomenological studies. For example, we are now in a position to perform a robust comparison to LEP data in \cite{Achard:2011zz}.  Further work is warranted to study the resummed forms introduced here at higher resummed orders and to determine techniques to achieve robust uncertainty estimates. It would also be informative to extend our study to resummation of angularities with $a\geq 1$ \cite{Dokshitzer:1998kz,Becher:2011pf,Becher:2012qc,Chiu:2011qc,Chiu:2012ir} or to the recently introduced ``recoil-free'' observables \cite{Larkoski:2014uqa}.

Although the comparisons and lessons in this paper are formulated in terms of event shape distributions in $e^+e^-$ collisions, the observations about how to compute different ways of writing a cross section to consistent accuracy are applicable generally to any cross section computed in dQCD. In the paper \cite{Sterman:2013nya} which appeared recently, similar comparisons are performed for threshold resummation in hadron collisions.

\acknowledgments
We thank the organizers of the 2010 ``Joint Theoretical-Experimental Workshop on Jets and Jet Substructure at the LHC'' at the University of Washington (supported by DOE Contract DE-FG02-96ER40956) where this project was germinated, and the organizers of the 2011 INT Program on ``Frontiers of QCD'', the 2012 Boston Jet Physics Workshop, and the 2013 ESI Program on ``Jets and Quantum Fields for LHC and Future Colliders'' in Vienna, Austria, where it progressed. We thank S. Rappoccio for asking the question that instigated this project.  GS thanks M. Zeng, and CL and JRW thank I. Stewart and F. Tackmann, for many useful discussions. We also thank the CNYITP at Stony Brook (CL and JRW) and Los Alamos National Laboratory (JRW) for hospitality during portions of this work.  The work of LGA was partially supported by the P2IO Labex. The work of SDE was supported by DOE Contract DE-FG02-96ER40956. The work of CL is supported by DOE Contract DE-AC52-06NA25396 and by the LDRD office at Los Alamos. The work of GS, IS, and LGA was supported by the National Science Foundation,  grants  PHY-0653342, PHY-0969739 and PHY-1316617. The work of IS was supported by DOE Contract DE-FG02-93ER40762.  The work of JRW was supported by DOE Contract DE-AC02-05CH11231, and the LHC Theory Initiative, US National Science Foundation grant NSF-PHY-0705682.

\appendix

\section{Plus Distributions}
\label{app:plus}

In this Appendix we collect definitions and properties of plus distributions used in the main text.

In general for a function $q$, the plus distribution is defined by (see, e.g., \cite{Ligeti:2008ac})
\be
\label{eq:plusdef}
\begin{split}
[q(x)]_+ &= \lim_{\epsilon\to 0} \frac{d}{dx} \bigl [\theta(x-\epsilon)Q(x)\bigr] \\
&= \lim_{\epsilon\to 0} \bigl[\theta(x-\epsilon)q(x) + \delta(x-\epsilon)Q(x)\bigr]\,,
\end{split} 
\ee
where
\be
\label{eq:Qdef}
Q(x) = \int_1^x dx' q(x')\,.
\ee
With the choice of lower limit in \eq{Qdef}, the definition in \eq{plusdef} satisfies $\int_0^1 dx [q(x)]_+ = 0$. 
The result of integrating a plus distribution against a suitable test function $f(x)$ is 
\be
\int_{-\infty}^{x_{\text{max}}} \!\! dx\, [\theta(x)q(x)]_+ f(x) = \int_0^{x_{\text{max}}} \!\! dx \, q(x) [f(x) - f(0)] \ + \ f(0) Q(x_{\text{max}})\,.
\ee
We define two special plus distributions which commonly appear,
\be
\cL_n(x) \equiv \left[ \frac{\theta(x)\ln^n x}{x}\right]_+ ,\  (n\geq 0) \,,\qquad \cL^a(x) \equiv \left[\frac{\theta(x)}{x^{1-a}}\right]_+\,.
\ee
For the case $n=-1$, we define
\be
\cL_{-1} (x) \equiv \delta(x) \,.
\ee
The plus function $\cL_n$ obeys the rescaling relation,
\be
\lambda\cL_n(\lambda x) = \sum_{k=0}^n {n\choose k} \ln^k \lambda \,\cL_{n-k}(x) + \frac{\ln^{n+1}\lambda}{n+1} \delta(x)
\ee

\section{Laplace Transforms}
\label{app:Laplace}

In this Appendix we collect results for the Laplace transforms and inverse Laplace transforms ($\iLP$) between the logs
\be
L\equiv \ln \frac{1}{\tau} \,,\qquad \tilde L \equiv \ln(\nu e^{\gamma_E})
\ee
The Laplace transforms, defined by
\be
\wt F(\nu) \equiv \LP \{ F\} (\nu) = \int_0^\infty d\tau e^{-\nu\tau} F(\tau)
\ee
are given by
\begin{subequations}
\label{eq:LPs}
\begin{align}
\LP\bigl\{ 1\bigr\} &= \frac{1}{\nu} \\
\LP\bigl\{ L\bigr\} &= \frac{1}{\nu} \, \tilde L \\
\LP\bigl\{ L^2\bigr\}&= \frac{1}{\nu} \biggl\{ \tilde L^2 + \frac{\pi^2}{6} \biggr\} \\
\LP\bigl\{ L^3\bigr\}  &= \frac{1}{\nu} \biggl\{ \tilde L^3 + \frac{\pi^2}{2} \tilde L + 2\zeta_3 \biggr\} \\
\LP\bigl\{ L^4\bigr\} &= \frac{1}{\nu} \biggl\{ \tilde L^4 + \pi^2 \tilde L^2 + 8\zeta_3 \tilde L  + \frac{3\pi^4}{20} \biggr\} \\
\LP\bigl\{ L^5\bigr\} &= \frac{1}{\nu} \biggl\{ \tilde L^5 + \frac{5\pi^2}{3} \tilde L^3 + 20\zeta_3 \tilde L^2 + \frac{3\pi^4}{4} \tilde L + \frac{10\pi^2}{3}\zeta_3 + 24\zeta_5 \biggr\} \\
\LP\bigl\{ L^6\bigr\} &= \frac{1}{\nu} \biggl\{ \tilde L^6 + \frac{5\pi^2}{2} \tilde L^4 + 40\zeta_3 \tilde L^3 + \frac{9\pi^4}{4}\tilde L^2 + (20\pi^2 \zeta_3 + 144\zeta_5)\tilde L + 40\zeta_3^2 + \frac{61\pi^6}{168} \biggr\} \,,
\end{align}
\end{subequations}
and so on. 
The inverse Laplace transforms, defined by
\be
\iLP \{ \wt F\}(\tau) = \int_{\gamma-i\infty}^{\gamma+i\infty}\frac{d\nu}{2\pi i} e^{\nu\tau} \wt F(\nu)\,,
\ee
where $\gamma$ lies to the right of all the poles of $\wt F$ in the complex plane,
are given by
\begin{subequations}
\label{eq:iLPs}
\begin{align}
\iLP\Bigl\{\frac{1}{\nu} \Bigr\} &= 1 \\
\iLP\Bigl\{\frac{1}{\nu}\tilde L\Bigr\} &= L \\
\iLP\Bigl\{\frac{1}{\nu}\tilde L^2\Bigr\} &= L^2 - \frac{\pi^2}{6} \\
\iLP\Bigl\{\frac{1}{\nu}\tilde L^3\Bigr\}  &= L^3 - \frac{\pi^2}{2} L - 2\zeta_3 \\
\iLP\Bigl\{\frac{1}{\nu}\tilde L^4\Bigr\}  &= L^4 - \pi^2 L^2 - 8\zeta_3 L  + \frac{\pi^4}{60} \\
\iLP\Bigl\{\frac{1}{\nu}\tilde L^5\Bigr\}  &= L^5 - \frac{5\pi^2}{3} L^3 - 20\zeta_3 L^2 + \frac{\pi^4}{12} L + \frac{10\pi^2}{3}\zeta_3 - 24\zeta_5 \\
\iLP\Bigl\{\frac{1}{\nu}\tilde L^6\Bigr\} &= L^6 - \frac{5\pi^2}{2} L^4 - 40\zeta_3 L^3 + \frac{\pi^4}{4} L^2 + (20\pi^2 \zeta_3 - 144\zeta_5) L + 40\zeta_3^2 - \frac{5\pi^6}{168}\,,
\end{align}
\end{subequations}
and so on. The results explicitly tabulated in \eqs{LPs}{iLPs} are needed to transform logs in the fixed-order expansions of event shape distributions in QCD up to $\cO(\as^3)$.

\section{Anomalous Dimensions}
\label{app:anomalous}
The coefficients of the beta function up to three-loop order in $\overline{\mathrm{MS}}$  are given by~\cite{Tarasov:1980au, Larin:1993tp}
%%%
\begin{align} \label{eq:betacoeffs}
\beta_0 &= \frac{11}{3}\,C_A -\frac{4}{3}\,T_F\,n_f
\,,\\
\beta_1 &= \frac{34}{3}\,C_A^2  - \Bigl(\frac{20}{3}\,C_A\, + 4 C_F\Bigr)\, T_F\,n_f
\,, \nn\\
\beta_2 &=
\frac{2857}{54}\,C_A^3 + \Bigl(C_F^2 - \frac{205}{18}\,C_F C_A
 - \frac{1415}{54}\,C_A^2 \Bigr)\, 2T_F\,n_f
 + \Bigl(\frac{11}{9}\, C_F + \frac{79}{54}\, C_A \Bigr)\, 4T_F^2\,n_f^2
\,,\nn
\end{align}
and the cusp anomalous dimension coefficients by~\cite{Korchemsky:1987wg, Moch:2004pa}:
\begin{align}\label{eq:Gacuspexp}
\Gamma^q_0 &= 4C_F
\,,\nn\\
\Gamma^q_1 &= 4C_F \Bigl[\Bigl( \frac{67}{9} -\frac{\pi^2}{3} \Bigr)\,C_A  -
   \frac{20}{9}\,T_F\, n_f \Bigr]
\,,\nn\\
\Gamma^q_2 &= 4C_F \Bigl[
\Bigl(\frac{245}{6} -\frac{134 \pi^2}{27} + \frac{11 \pi ^4}{45}
  + \frac{22 \zeta_3}{3}\Bigr)C_A^2 
  + \Bigl(- \frac{418}{27} + \frac{40 \pi^2}{27}  - \frac{56 \zeta_3}{3} \Bigr)C_A\, T_F\,n_f
\nn\\* & \hspace{8ex}
  + \Bigl(- \frac{55}{3} + 16 \zeta_3 \Bigr) C_F\, T_F\,n_f
  - \frac{16}{27}\,T_F^2\, n_f^2 \Bigr]
\,.\end{align}
%%%
The $\overline{\mathrm{MS}}$ non-cusp anomalous dimension $\gamma_H = 2\gamma_C$ for the hard function $H$ can be obtained~\cite{Idilbi:2006dg, Becher:2006mr} from the IR divergences of the on-shell massless quark form factor $C(q^2,\mu)$ which are known to three loops~\cite{Moch:2005id},
%%%
\begin{align} \label{eq:gaHexp}
\gamma_H^0 &= -12 C_F
\,,\nn\\
\gamma_H^1
&= - 2C_F 
\Bigl[
  \Bigl(\frac{82}{9} - 52 \zeta_3\Bigr) C_A
+ (3 - 4 \pi^2 + 48 \zeta_3) C_F
+ \Bigl(\frac{65}{9} + \pi^2 \Bigr) \beta_0 \Bigr]
\,,\nn\\
\gamma_H^2
&= -4C_F \Bigl[
  \Bigl(\frac{66167}{324} - \frac{686 \pi^2}{81} - \frac{302 \pi^4}{135} - \frac{782 \zeta_3}{9} + \frac{44\pi^2 \zeta_3}{9} + 136 \zeta_5\Bigr) C_A^2
\nn\\ & 
\quad + \Bigl(\frac{151}{4} - \frac{205 \pi^2}{9} - \frac{247 \pi^4}{135} + \frac{844 \zeta_3}{3} + \frac{8 \pi^2 \zeta_3}{3} + 120 \zeta_5\Bigr) C_F C_A
\nn\\ & 
\quad + \Bigl(\frac{29}{2} + 3 \pi^2 + \frac{8\pi^4}{5} + 68 \zeta_3 - \frac{16\pi^2 \zeta_3}{3} - 240 \zeta_5\Bigr) C_F^2 \nn \\
&\quad + \Bigl(-\frac{10781}{108} + \frac{446 \pi^2}{81} + \frac{449 \pi^4}{270} - \frac{1166 \zeta_3}{9} \Bigr) C_A \beta_0
\nn\\ & 
\quad + \Bigl(\frac{2953}{108} - \frac{13 \pi^2}{18} - \frac{7 \pi^4 }{27} + \frac{128 \zeta_3}{9}\Bigr)\beta_1
+ \Bigl(-\frac{2417}{324} + \frac{5 \pi^2}{6} + \frac{2 \zeta_3}{3}\Bigr)\beta_0^2
\Bigr]
\,.\end{align}
%%%
The non-cusp three-loop anomalous dimension for the $a=0$ quark jet function is  given by \cite{Becher:2006mr},
%%%
\begin{align}\label{gaBexp}
 \gamma_J^0 &= 6 C_F
\,,\nn\\
\gamma_J^1 
&= C_F 
\Bigl[
  \Bigl(\frac{146}{9} - 80 \zeta_3\Bigr) C_A
+ (3 - 4 \pi^2 + 48 \zeta_3) C_F
+ \Bigl(\frac{121}{9} + \frac{2\pi^2}{3} \Bigr) \beta_0 \Bigr]
\,,\nn\\
\gamma_J^2 
&= 2 C_F\Bigl[
  \Bigl(\frac{52019}{162} - \frac{841\pi^2}{81} - \frac{82\pi^4}{27} -\frac{2056\zeta_3}{9} + \frac{88\pi^2 \zeta_3}{9} + 232 \zeta_5\Bigr)C_A^2
\nn\\ & 
\quad + \Bigl(\frac{151}{4} - \frac{205\pi^2}{9} - \frac{247\pi^4}{135} + \frac{844\zeta_3}{3} + \frac{8\pi^2 \zeta_3}{3} + 120 \zeta_5\Bigr) C_A C_F
\nn\\ &
\quad + \Bigl(\frac{29}{2} + 3 \pi^2 + \frac{8\pi^4}{5} + 68 \zeta_3 - \frac{16\pi^2 \zeta_3}{3} - 240 \zeta_5\Bigr) C_F^2 \nn \\
&\quad + \Bigl(-\frac{7739}{54} + \frac{325}{81} \pi^2 + \frac{617 \pi^4}{270} - \frac{1276\zeta_3}{9} \Bigr) C_A\beta_0
\nn\\ &
\quad + \Bigl(-\frac{3457}{324} + \frac{5\pi^2}{9} + \frac{16 \zeta_3}{3} \Bigr) \beta_0^2
+ \Bigl(\frac{1166}{27} - \frac{8 \pi^2}{9} - \frac{41 \pi^4}{135} + \frac{52 \zeta_3}{9}\Bigr) \beta_1
\Bigr]
\,.\end{align}
%%%
The anomalous dimension for the soft function is obtained from $\gamma_S = -\gamma_H -2 \gamma_J$.

%%%%%%%%%%%%%%%%%%%%%%%%%%%%%%%%%%%%%%%%%%%%%%%%%%%%%%%%%%%%%%%%%%%%%%%%%%%%%%%%%%%%%%%%%%%%%%%%%%%%%%%%%

\bibliography{resum}

\end{document}